\documentclass[reprint,preprintnumbers,amsmath,amssymb,superscriptaddress,nofootinbib,aps]{revtex4-1}
\usepackage{lipsum}
\usepackage{appendix}
\usepackage{diagbox}
\usepackage{physics}
\usepackage{amsmath}
\usepackage{amssymb}
\usepackage{mathrsfs}
\usepackage[nolist,nohyperlinks]{acronym}
\usepackage[caption=false]{subfig}
\usepackage{url}
\usepackage{multirow}
\usepackage{booktabs}
\usepackage{xcolor}
\definecolor{dblue}{RGB}{25,25,125}
\usepackage[%
  colorlinks = true,
  urlcolor= dblue,
  linkcolor= dblue,
  citecolor= dblue
]{hyperref}
\usepackage{float}
\usepackage{graphicx}
\usepackage{capt-of}
\usepackage{tikz-cd}
\usepackage{tikz}
\usetikzlibrary{shapes,arrows}
\usetikzlibrary{shapes.geometric, arrows}
\usetikzlibrary{graphs}
\usetikzlibrary{cd}
\definecolor{dblue}{RGB}{25,25,125}
\tikzstyle{startstop} = [rectangle,rounded corners, minimum width=3cm,minimum height=1cm,text centered, draw=black]
\tikzstyle{io} = [trapezium, trapezium left angle = 70,trapezium right angle=110,minimum width=3cm,minimum height=1cm,text centered,draw=black]
\tikzstyle{process} = [rectangle,minimum width=2.5cm,minimum height=1cm,text centered,text width =2.5cm,draw=black]
\tikzstyle{decision} = [diamond,minimum width=2.5cm,minimum height=1cm,text centered,draw=black]
\tikzstyle{arrow} = [thick,->,>=stealth]
\usepackage{dcolumn}
\usepackage{bm}

\usepackage[commentmarkup=uwave]{changes}
\definechangesauthor[name=Haowen, color=red]{HZ}

\definecolor{mpl_red}{HTML}{D62728}

\newcommand{\D}{\text{d}}

\newcommand{\feature}{\bm\Lambda\bm}

\newcommand{\data}{\{\bm{d}_i\}_{i=1}^{N_\mathrm{obs}}}
\newcommand{\orcid}[1]{\begingroup
  \hypersetup{hidelinks}\href{https://orcid.org/#1}{\includegraphics[width=10pt]{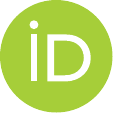}} \endgroup}

\begin{document}

\title{Two-Step Procedure to Detect Cosmological Gravitational Wave Backgrounds with Next-Generation Terrestrial Gravitational-Wave Detectors}

\author{Haowen Zhong \orcid{0000-0001-8324-5158}\,}
\email{zhong461@umn.edu}
\affiliation{
 School of Physics and Astronomy, University of Minnesota, Minneapolis, MN 55455, USA
}

\author{Luca Reali \orcid{0000-0002-8143-6767}\,}
\email{lreali1@jhu.edu}
\affiliation{William H. Miller III Department of Physics and Astronomy, Johns Hopkins University, Baltimore, Maryland 21218, USA
}

\author{Bei Zhou \orcid{0000-0003-1600-8835}\,}
\email{beizhou@fnal.gov}
\affiliation{Theory division, Fermi National Accelerator Laboratory, Batavia, Illinois 60510, USA}
\affiliation{Kavli Institute for Cosmological Physics, University of Chicago, Chicago, Illinois 60637, USA}

\author{Emanuele Berti \orcid{0000-0003-0751-5130}\,}
\email{berti@jhu.edu}
\affiliation{William H. Miller III Department of Physics and Astronomy, Johns Hopkins University, Baltimore, Maryland 21218, USA
}

\author{Vuk Mandic \orcid{0000-0001-6333-8621}\,}
\email{vuk@umn.edu}
\affiliation{
School of Physics and Astronomy, University of Minnesota, Minneapolis, MN 55455, USA
}

\preprint{FERMILAB-PUB-24-0871-T}

\date{\today}
\begin{abstract}

  Cosmological gravitational-wave backgrounds are an exciting science target for next-generation ground-based detectors, as they encode invaluable information about the primordial Universe. However, any such background is expected to be obscured by the astrophysical foreground from compact-binary coalescences. We propose a novel framework to detect a cosmological gravitational-wave background in the presence of binary black holes and binary neutron star signals with next-generation ground-based detectors, including Cosmic Explorer and the Einstein Telescope. Our procedure involves first removing all the individually resolved binary black hole signals by notching them out in the time-frequency domain. Then, we perform joint Bayesian inference on the individually resolved binary neutron star signals, the unresolved binary neutron star foreground, and the cosmological background. For a flat cosmological background, we find that we can claim detection at $5\,\sigma$ level when $\Omega_\mathrm{ref}\geqslant 2.7\times 10^{-12}/\sqrt{T_\mathrm{obs}/\mathrm{yr}}$, where $T_\mathrm{obs}$ is the observation time (in years), which is within a factor of $\lesssim2$ from the sensitivity reached in absence of these astrophysical foregrounds. 
\end{abstract}
\maketitle
\begin{acronym}
    \acro{GW}{gravitational-wave}
    \acro{PSD}{power spectral density}
    \acro{GR}{general relativity}
    \acro{CBC}{compact binary coalescence}
    \acro{BH}{black hole}
    \acro{BBH}{binary black hole}
    \acro{BNS}{binary neutron star}
    \acro{NSBH}{neutron star-black hole}
    \acro{SFR}{star formation rate}
    \acro{LVK}{LIGO-Virgo-KAGRA}
    \acro{ET}{Einstein Telescope}
    \acro{CE}{Cosmic Explorer}
    \acro{PE}{parameter estimation}
    \acro{SGWB}{stochastic gravitational wave background}
    \acro{CGWB}{cosmological gravitational wave background}
    \acro{XG}{next generation}
    \acro{ML}{Machine Learning}
    \acro{DL}{Deep Learning}
    \acro{NN}{Neural Network}
    \acro{SNR}{signal-to-noise ratio}
    \acro{PDF}{probability density function}
    \acro{CDF}{cumulative density function}
\end{acronym}

\noindent
\textbf{\em Introduction.} The \ac{XG} of ground-based \ac{GW} detectors, such as \ac{CE}~\cite{CE, CEHorizon, CE2_PSD} and the \ac{ET}~\cite{ET, ET_PSD}, will see an unprecedented increase in sensitivity both to individual \ac{CBC} events and to stochastic signals. This will allow us to individually resolve nearly all of the \ac{BBH} mergers and a significant fraction of \ac{BNS} and \ac{NSBH} events in the entire Universe~\cite{Gupta:2023lga, Branchesi:2023mws}. Furthermore, we will probe the \acp{SGWB} across several orders of magnitude in amplitude~\cite{TaniaCE,Renzini:2022alw}.  
In particular, the detection of a \ac{CGWB} would open up a unique window to observe the earliest moments of the Universe and probe physics at energies close to the Planck scale~\cite{Grishchuk:1974ny, Starobinsky:1979ty, Grishchuk:1993te, Barnaby:2011qe, Damour:2004kw, Siemens:2006yp, Olmez:2010bi, Regimbau:2011bm}. This detection is challenging because of the simultaneous presence of the \acp{SGWB} from \acp{CBC}, which effectively act as a foreground, limiting the sensitivity to other subdominant \acp{SGWB}~\cite{Callister:2016ewt}.

Several methods have been proposed to bypass the \ac{CBC} foreground. Given the high fraction of individually resolved \ac{CBC} signals with \ac{XG} detectors, one could simply fit and subtract them from the data in the frequency domain to lower the foreground~\cite{TaniaCE, subtractionSurabhi}. However, the recovery of detected signals is never perfect, and the pile-up of residuals from imperfect subtraction can produce an effective foreground that is comparable to the original \ac{SGWB} before removal~\cite{Zhou:2022nmt, Zhou:2022otw, Song:2024pnk}. Techniques have been proposed to further reduce this effective foreground in the frequency domain, from applying a projection method~\cite{Cutler:2005qq, Sharma:2020btq} to estimating and subtracting the expected residual power~\cite{Pan:2023naq}. Ref.~\cite{Belgacem:2024ntv} showed that subtraction results in the frequency domain are actually significantly more optimistic if one abandons the linear-signal approximation in favor of more realistic estimates of the residuals. Moreover, the authors derive an optimal filter specifically targeted to the cosmological background search. They find that applying this match filtering procedure significantly lowers the impact of the \ac{CBC} foreground in the search for a cosmological background, introducing a dependence on the specific spectral shape of such background. On the other hand, Refs.~\cite{Zhong:2022ylh,Zhong:2024} proposed a time-frequency approach to notch-out the detected signals by masking the associated pixels. They showed that this method could reduce the \ac{CBC} foreground by
one order of magnitude even when residuals from imperfect recovery are taken into account~\cite{Zhong:2024}.

All of the aforementioned techniques can only lower the \ac{CBC} foreground by removing individually resolved CBC signals, but the sensitivity to subdominant \acp{SGWB} would still be limited by the \ac{BNS} signals that are too weak to be individually detected, where the unresolvable \ac{BNS} foreground is about at the order of $\mathcal{O}(10^{-11})$ at $f_\mathrm{ref}=25$Hz~\cite{Zhong:2024, Bellie:2023jlq}. Refs.~\cite{Smith:2017vfk,Smith:2020lkj,Biscoveanu:2020gds} proposed a Bayesian framework to simultaneously fit all of the \ac{CBC} signals without distinguishing between resolved and unresolved. Ref.~\cite{Li:2024iua} suggested using the information on the \ac{CBC} population from detected events to predict the foreground from unresolved signals and remove it from the data. While promising, the computational feasibility of these methods with a realistic number of events in the context of \ac{XG} detectors has yet to be proven.

In this Letter, we propose a novel method that combines the notching procedure of Refs.~\cite{Zhong:2022ylh, Zhong:2024} with the joint hierarchical inference of Ref.~\cite{Callister:2020arv}. First, we notch out all the detected \ac{BBH} signals, showing that the residual \ac{BBH} foreground is negligible. Then, we jointly fit within a unified Bayesian framework the resolved \ac{BNS} events, the foreground from unresolved \ac{BNS} events, and a \ac{CGWB}. For illustration, we apply this procedure to mock data including realistic \ac{BBH} and \ac{BNS} populations and a flat \ac{CGWB}.


\noindent
\textbf{\em Astrophysical population and setup.} For both \acp{BBH} and \acp{BNS}, we assume the same population models as in Refs.~\cite{Zhong:2022ylh,Zhong:2024}. We employ the
\texttt{PowerLaw+Peak} model from the latest \ac{LVK} catalog~\cite{KAGRA:2021duu} to characterize the \ac{BBH} mass distribution, while we assume that the \ac{BNS} masses are uniformly distributed within the range $[1,2]\,\rm{M_\odot}$~\cite{KAGRA:2021duu,Landry:2021hvl}.We assume zero spins and an isotropic distribution of the orbital orientation and sky position for all the binaries. To obtain the redshift distribution, we convolve the \ac{SFR}~\cite{Finkel:2021zgf}
\begin{equation}
    R_f(z)=\mathcal{N}\frac{ae^{b(z-z_p)}}{a-b+be^{a(z-z_p)}}
    \label{eq:SFR}
\end{equation}
with a power-law time delay distribution $p(t_d) \propto t_d^p$ with $p=-1$. The source-frame merger rate $R_m(z)$ then reads
\begin{equation}
R_m(z)=\int_{t_{\min}}^{t_{\max}}R_f(z_f)p(t_d)\D t_d\,,
\label{eq:Rm}
\end{equation}
where $z_f=z[t(z)-t_d]$ is the binary formation time, we set $t_{\min}=20$ Myr for \acp{BNS} and $50$ Myr for \acp{BBH}, and we set $t_{\max}$ to be the Hubble time in both cases. We fix the parameters in the \ac{SFR} to $z_p=2.00$, $a=2.37$ and $b=1.80$~\cite{Finkel:2021zgf}, while $\mathcal{N}$ is a normalization factor chosen so that the local merger rate $\mathcal{R}_0$ is consistent with the \ac{LVK} results~\cite{KAGRA:2021duu}.
In particular, for \acp{BNS} we set $\mathcal{R}_0=320 \, \mathrm{Gpc}^{-3} \,\mathrm{yr}^{-1}$; for \acp{BBH}, we choose a value of $\mathcal{R}_0$ such that $\mathcal{R}_m(z=0.2)=28.1\, \mathrm{Gpc}^{-3} \,\mathrm{yr}^{-1}$~\cite{KAGRA:2021duu}.

Besides the \ac{CBC} signals, we also simulate a power-law \ac{CGWB} 
\begin{equation}
\Omega_\mathrm{GW}(f)=\Omega_\mathrm{ref}\left(\frac{f}{f_\mathrm{ref}}\right)^\alpha\,,    
\end{equation}
where we choose $\alpha=0$ and $~\Omega_\mathrm{ref}=2.0\times10^{-11}$ at $f_\mathrm{ref}=25\,\rm{Hz}$. Our framework is not fundamentally limited by the exact functional form of the \ac{CGWB}, and we choose a simple flat \ac{CGWB} for demonstration purposes.

We consider a network of \ac{XG} interferometers composed of two 40-km long \ac{CE} detectors located at the  LIGO Livingston and LIGO Hanford sites~\cite{LIGOScientific:2014pky}, and an \ac{ET}
detector located at the Virgo site~\cite{VIRGO:2014yos}. We assume Gaussian and stationary detector noise, modeled via a perfectly known \ac{PSD} and uncorrelated among different detectors. For both \ac{CE} detectors, we adopt the \ac{PSD} curve of Ref.~\cite{Gupta:2023lga}, while for \ac{ET} we assume the \ac{PSD} of the triangular configuration in Ref.~\cite{ET_PSD}. We set the low-frequency sensitivity limit to be $5~\rm{Hz}$ for all detectors in the network.

We use \texttt{Bilby}~\cite{Bilby} to simulate and add detector noise and \ac{CBC} signals into a time series. For \ac{BBH} signals, we adopt the \texttt{IMRPhenomXAS} waveform model~\cite{IMRXAS}, which is a full inspiral-merger-ringdown, non-precessing waveform model for the fundamental mode. For \ac{BNS} signals we use the inspiral-only \texttt{TaylorT4} waveform model~\cite{Buonanno:2002fy}, as we expect most of the \ac{BNS} signals in \ac{XG} detectors to be inspiral-dominated.

We use \texttt{pygwb}~\cite{pygwb} to simulate and add the \ac{CGWB} into the time series and compute cross-correlations for the stochastic search.
With our network of \ac{XG} detectors, Ref.~\cite{Zhong:2024} showed that combining cross-correlation results from all ten possible baselines yields an improvement in the sensitivity of the stochastic search of less than $10\%$ compared to using a single baseline of two \ac{CE} detectors. Using all baselines would require 10 times more computational power. Therefore, for demonstration purposes, here we only cross-correlate time series data from two \ac{CE} interferometers.


\noindent
\textbf{\em Data-analysis method.} We devise a procedure that combines the notching method of Refs.~\cite{Zhong:2022ylh,Zhong:2024} with the joint Bayesian inference of Ref.~\cite{Callister:2020arv}.

We assess the detectability of individual \ac{CBC} signals by computing their network \ac{SNR}~\cite{Sathyaprakash:2009xs}, choosing detection thresholds of $\mathrm{SNR}_\mathrm{thr}^\mathrm{BBH}=8$ for \acp{BBH} and $\mathrm{SNR}_\mathrm{thr}^\mathrm{BNS}=12$ for \acp{BNS}~\cite{Zhong:2024}. With the population model we adopt, we find that 99.7\% of the \ac{BBH} and 50.4\% of the \ac{BNS} signals are detected given two \ac{CE} and one \ac{ET} detectors network. 
We then apply the time-frequency approach of Ref.~\cite{Zhong:2024} to notch out all the detected \ac{BBH} signals from the data. As we show below, the remaining \ac{BBH} foreground, resulting from the combination of the few unresolved \ac{BBH} signals and the residuals from imperfect removal, is weaker than the network sensitivity; hence, we do not include it in the inference studies.

After applying this notching procedure, the data contains only \ac{BNS} signals and the \ac{CGWB}. If we followed Ref.~\cite{Zhong:2024} and notched out the resolved \ac{BNS} events as well, we would still be limited by the foreground from unresolved \ac{BNS} signals. Instead, we adopt the Bayesian framework of Ref.~\cite{Callister:2020arv} to jointly fit the resolved \ac{BNS} signals along with the \ac{BNS} foreground and the \ac{CGWB}. Ref.~\cite{Callister:2020arv} devised this method in the context of current \ac{LVK} detectors, where the \ac{SGWB} complements the hierarchical inference on individually resolved events by providing information on the \ac{CBC} population at high redshift~\cite{KAGRA:2021duu}. Here, instead, the information from detected events helps us tighten the constraints on the \ac{BNS} foreground and distinguish it from the \ac{CGWB}.

Let us consider a set of data segments $\data$ from $N_\mathrm{obs}$ individual \ac{BNS} detections and the cross-correlation spectrum $\hat{C}^\star(f)$ after notching out the resolvable \ac{BBH} signals (see the Supplement~\cite{note1} for the definition of $\hat{C}^\star(f)$). Their joint likelihood can be written as~\cite{Callister:2020arv}
\begin{widetext}
\begin{equation}
    \mathscr{L}(\data,\hat
{C}^\star(f)|\feature_\mathrm{BNS},\feature_\mathrm{CGWB})=\mathscr{L}_\mathrm{BNS}(\data|\feature_\mathrm{BNS})
\times\mathscr{L}_\mathrm{SGWB}(\hat
{C}^\star(f)|\feature_\mathrm{BNS},\feature_\mathrm{CGWB})\,,
\label{eq:Joint_Like}
\end{equation}
\end{widetext}
where $\mathscr{L}_\mathrm{BNS}(\data|\feature_\mathrm{BNS})$ is the usual hierarchical likelihood for population inference on the detected events~\cite{Mandel:2018mve,Taylor:2018iat}, while $\mathscr{L}_\mathrm{SGWB}(\hat
{C}^\star(f)|\feature_\mathrm{BNS},\feature_\mathrm{CGWB})$ is the Gaussian likelihood for stochastic searches~\cite{Mandic:2012pj,Callister:2017ocg}. Here $\feature_\mathrm{CGWB}=\{\Omega_\mathrm{ref},\alpha\}$ encodes the parameters characterizing the \ac{CGWB}, while $\feature_\mathrm{BNS}$ is the set of hyperparameters defining the \ac{BNS} population, which affect both the individually detected events and the \ac{SGWB} from unresolved \ac{BNS} signals. We note that the above joint likelihood might not be valid in general for two reasons: (1) it ignores the \ac{BBH} notching residual, which is a foreground; and (2) we use the BNS data in both the individual event and in the stochastic search, which could result in a bias. Regarding the first concern, since we notch the resolvable \ac{BBH} events instead of subtracting them, the notching residual is rather weak (see the solid gray line in Fig.~\ref{fig:sensitivity}, and also Ref.~\cite{Zhong:2024}). To address the second concern, in Section IV of the Supplemental Material we perform a test to show that the bias is negligible.
To simplify the inference, we assume that the \ac{BNS} mass distribution is perfectly known and that the angular parameters are isotropically distributed. Hence, the only \ac{BNS} hyperparameters $\feature_\mathrm{BNS}$ that we are inferring are the ones characterizing the redshift distribution. In particular, we consider two cases:
\begin{itemize}
\item[1)] A simplistic \emph{known \ac{SFR}} scenario where we assume that the \ac{SFR} is completely known, and all that is left to determine are the local merger rate, $\mathcal{R}_0$, and the power-law index of the time-delay distribution, $p$.
    Then the joint inference involves a total of four parameters $\{\feature_\mathrm{BNS},\feature_\mathrm{CGWB}\}=\{\mathcal{R}_0,p,\Omega_\mathrm{ref},\alpha\}$.
    \item[2)] A more realistic \emph{unkown \ac{SFR}} scenario where we assume no prior knowledge of the redshift distribution, except for its functional form. Following the approach of Ref.~\cite{Ng:2020qpk}, we assume that the merger rate has the same functional form as the \ac{SFR} of Eq.~\eqref{eq:SFR}, then we fit this function to the merger rate $R_m(z)$ computed via Eq.~\eqref{eq:Rm} to obtain the injected values $a_\mathrm{inj}=2.15,~b_\mathrm{inj}=1.40,~z_p^\mathrm{inj}=1.64$. In this case, we directly infer these parameters along with the local merger rate $\mathcal{R}_0$, so the joint inference involves a total of six parameters $\{\feature_\mathrm{BNS},\feature_\mathrm{CGWB}\}=\{\mathcal{R}_0,a,b,z_p,\Omega_\mathrm{ref},\alpha\}$.
    
\end{itemize}

Performing full Bayesian \ac{PE} on thousands of individually resolved \ac{BNS} signals is currently computationally unfeasible. Hence, we generate synthetic samples for the individual \ac{BNS} likelihoods on source parameters using the approximate method of Refs.~\cite{Fishbach:2018edt,Callister:2020arv}. For the exact expressions of the joint likelihood and additional details on the inference procedure, we refer the reader to the Supplemental Material.

\noindent
\textbf{\em Joint inference results.} For demonstration purposes, we perform our analysis assuming an observation time $T_\mathrm{obs}=10$ days. With our population model, this corresponds to roughly $\sim 650$ \ac{BBH} signals that we notch out, and $\sim 7000$ resolvable \ac{BNS} events that enter the hierarchical likelihood $\mathscr{L}_\mathrm{BNS}(\data|\feature_\mathrm{BNS})$.

In vanilla searches for a single, dominant \ac{SGWB}, its \ac{SNR} is straightforward to compute, and it is a good figure of merit to estimate whether the target \ac{SGWB} is strong enough to be detectable relative to the detector noise~\cite{KAGRA:2021duu, Allen:1997ad, O3stoch}. In our case, instead, we are jointly fitting for the \ac{BNS} hyperparameters from individually detected sources plus two distinct \acp{SGWB}, and our target is, in fact, the subdominant \ac{CGWB}. Hence, we turn to a full Bayesian approach and we study the inferred posterior to find evidence for CGWB detection.

\begin{figure}[t]
    \centering
\includegraphics[width=0.45\textwidth]{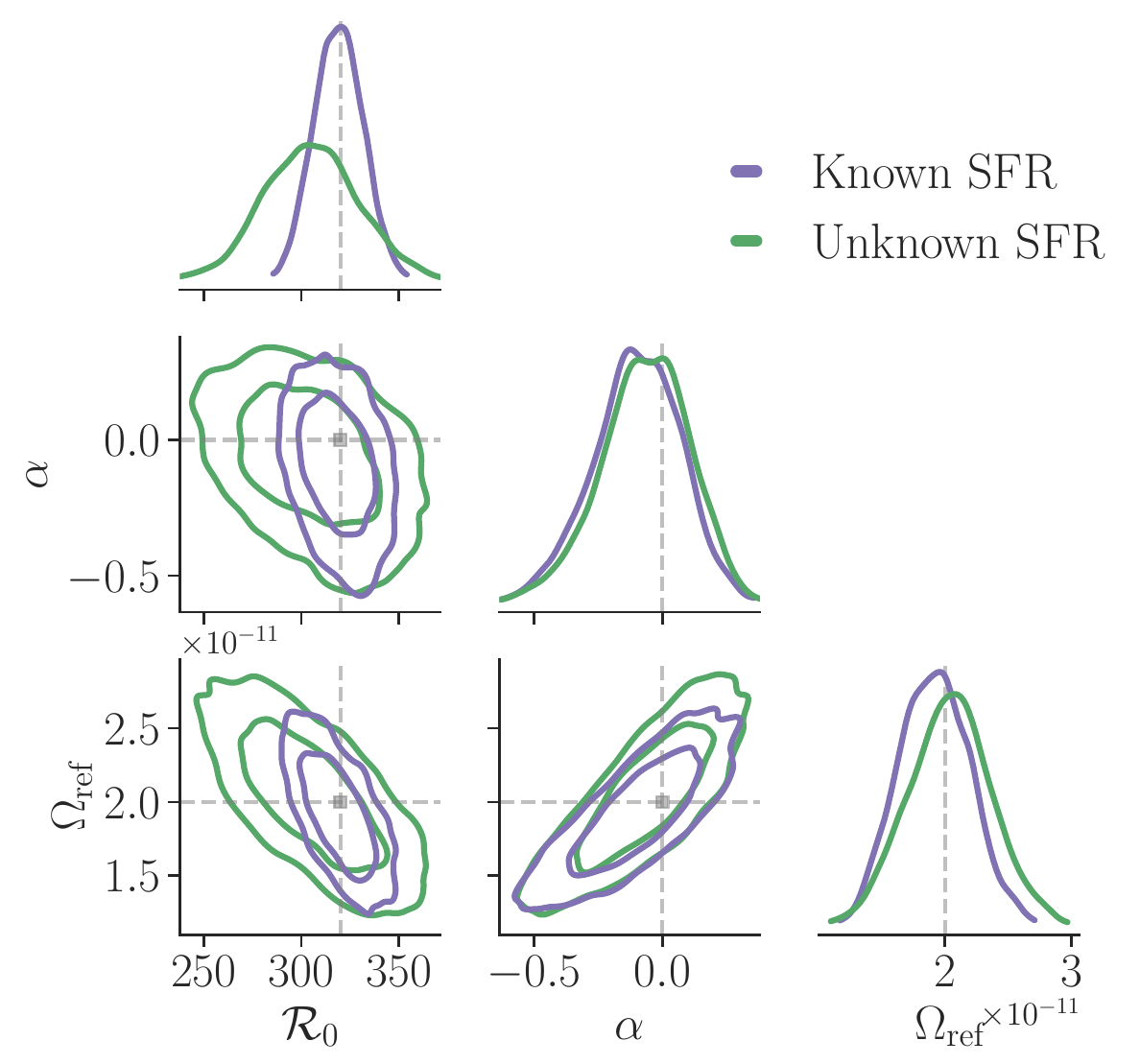}
\caption{Marginalized Joint posterior distributions on $\{\mathcal{R}_0,\alpha,\Omega_\mathrm{ref}\}$ for the models assuming known \ac{SFR} (purple curves) and unknown \ac{SFR} (green curves). Contours enclose $68\%$ and $95\%$ of the probability mass. Gray dashed lines indicate the true simulated values of parameters.}
    \label{fig:comparison}
\end{figure}

In both our scenarios, we find that the posteriors on all the parameters are fully consistent with the simulated values. In Fig.~\ref{fig:comparison}, we compare the joint posteriors on the common parameters for both inferences, namely the \ac{CGWB} parameters $\{\Omega_\mathrm{ref},\alpha\}$ and the local \ac{BNS} merger rate $\mathcal{R}_0$. As expected, we find that assuming better knowledge of the \ac{BNS} redshift distribution has a large impact on inferring the local merger rate. The $95\%$ confidence interval on $\mathcal{R}_0$ roughly doubles going from known (purple curves) to unknown (green curves) \ac{SFR}.

However, the joint posterior on $\{\alpha,~\Omega_\mathrm{ref}\}$ is remarkably similar in both analyses, indicating that adding more parameters to our inference of the \ac{BNS} redshift distribution does not have a large impact on the inference of the \ac{CGWB}. Crucially, we find that we can confidently claim the detection of the \ac{CGWB} in both scenarios: we exclude $\Omega_\mathrm{ref}=0$ at the $6.8\,\sigma$ level assuming known \ac{SFR} and at $6.1\,\sigma$ level with unknown \ac{SFR}. We refer to Table~\ref{tab:posterior} for the marginalized posterior distributions of all parameters in the different cases.
\begin{table*}[]
    \centering
    \caption{Marginalized posterior distribution (median and 95\% credible symmetric interval).}
    \label{tab:posterior}
    \begin{ruledtabular}
    \begin{tabular}{lrrrrrrrr}
    Scenario & $\mathcal{R}_0$ & $p$ & $a$ & $b$ & $z_p$& $\Omega_\mathrm{ref}$ & $\alpha$ \\
    \hline
Joint search: Known SFR & $319^{+25}_{-24}$ & $-1.00^{+0.04}_{-0.04}$ & - & - &-&$1.92^{+0.58}_{-0.52}\times 10^{-11}$&$-0.10^{+0.32}_{-0.37}$\\
Joint search: Unknown SFR & $305^{+48}_{-47}$ &-&$2.24^{+0.19}_{-0.16}$&$1.43^{+0.28}_{-0.24}$&$1.64^{+0.07}_{-0.07}$&$2.04^{+0.66}_{-0.66}\times 10^{-11}$&$-0.06^{+0.31}_{-0.39}$\\
\hline
Ideal search & -& -& -& -& -&$1.95^{+0.35}_{-0.34}\times10^{-11}$&$-0.14^{+0.29}_{-0.31}$\\
    \end{tabular}
    \end{ruledtabular}
\end{table*}

\begin{figure}[h]
    \centering
    \includegraphics[width=1.07\linewidth]{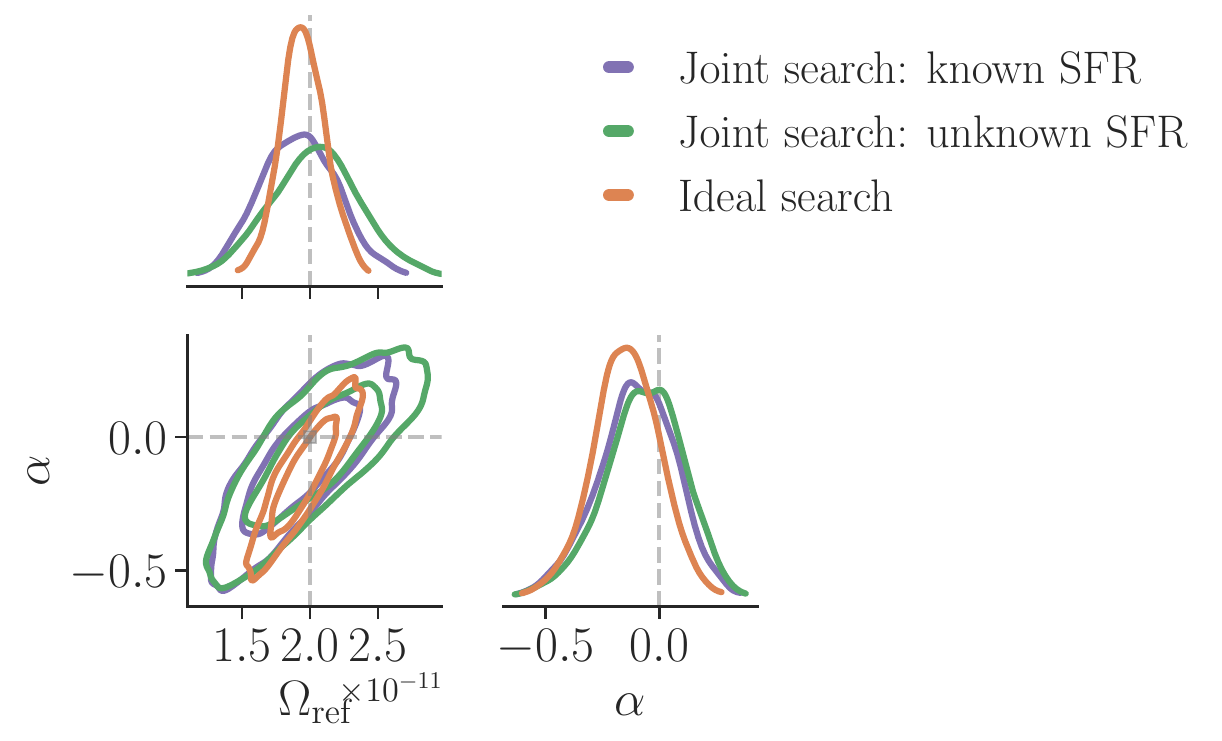}
    \caption{Joint posterior distributions on $\{\Omega_\mathrm{ref},\alpha\}$ across three different scenarios: known \ac{SFR} (purple curves), unknown \ac{SFR} (orange curves), and an ideal search without any \ac{CBC} foreground (green curves). Contours represent  $68\%$ and $95\%$ of the probability.}\label{fig:Comparisons_joint_ideal}
\end{figure}

In Fig.~\ref{fig:Comparisons_joint_ideal}, we further compare the marginalized posterior distributions on $\{\Omega_\mathrm{ref},\alpha\}$ from our two joint-inference scenarios (purple and green curves) with the posterior from an ideal search for the \ac{CGWB} where no \ac{CBC} foreground is present (orange curves). In other words, the ideal search corresponds to a benchmark case where only the \ac{CGWB} is simulated, and the second term of Eq.~\eqref{eq:Joint_Like} in the form $\mathscr{L}_\mathrm{SGWB}(\hat{C}^\star(f)|\feature_\mathrm{CGWB})$ is used for the inference. After applying our notching+joint inference scheme, the recovery of the \ac{CGWB} is not significantly affected by the presence of a \ac{CBC} foreground. We find that the $1\,\sigma$ error on the \ac{CGWB} amplitude $\Omega_\mathrm{ref}$ increases by a factor of $\sim 90\%$ ($\sim60\%$) in the unknown (known) \ac{SFR} case compared to the ideal search with no foreground present. The difference in the marginalized posterior for the power-law index $\alpha$ is even milder, with a 19\% (17\%) increase in the $1\,\sigma$ error for the unknown (known) \ac{SFR} scenario with respect to the ideal search.
 

\begin{figure*}[t]
    \centering
    \includegraphics[width=0.85\textwidth]{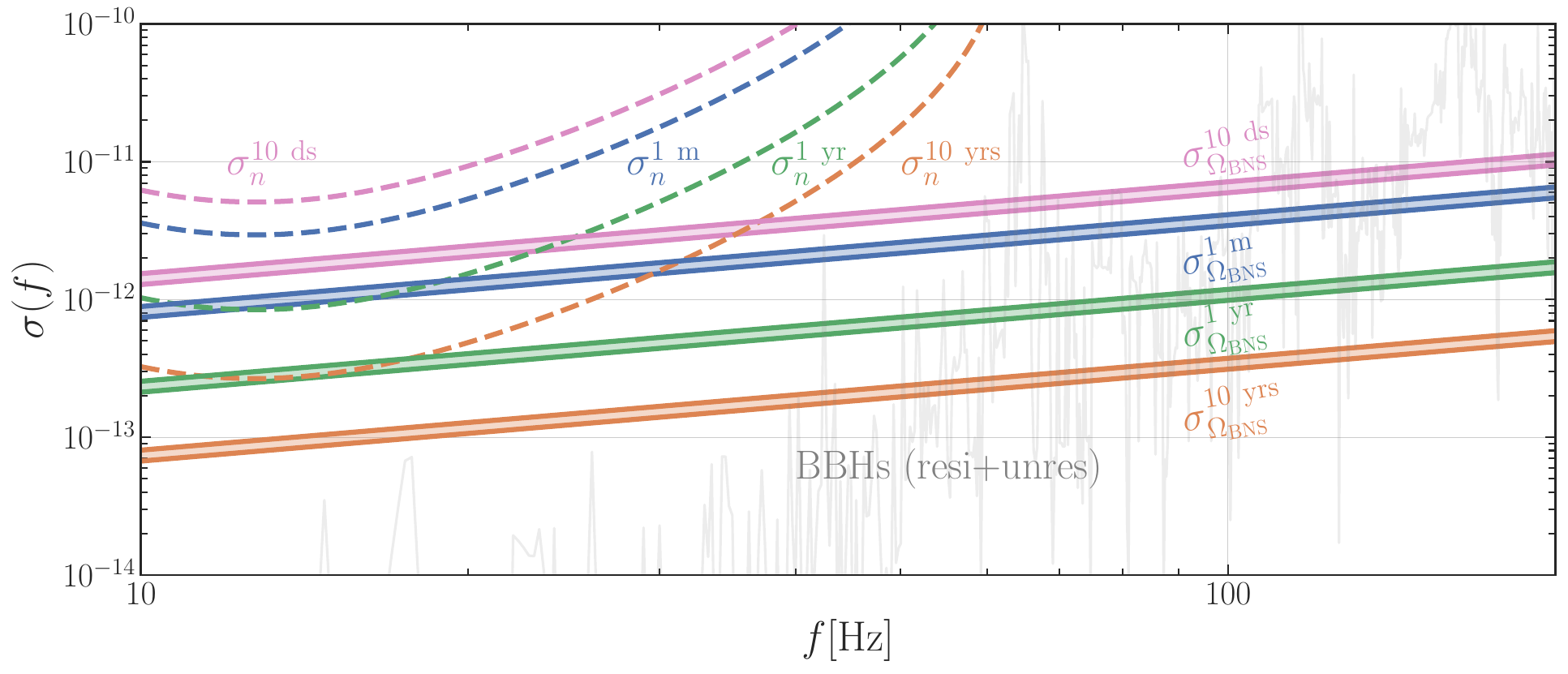}
    \caption{The sensitivity of the \ac{CGWB} search. The top four dashed curves show the baseline sensitivity to stochastic searches given different observation times: ten days (pink curve), one month (blue curve), one year (green curve), and ten years (orange curve). The straight bands represent the uncertainty on the \ac{BNS} foreground for each observation time, with the same color scheme. Within each band, the upper (lower) line corresponds to the unknown (known) \ac{SFR} scenario. The gray curve is the remaining \ac{BBH} foreground after notching with 10 days of observation.}
    \label{fig:sensitivity}
\end{figure*}

\noindent
\textbf{\em Sensitivity of the \ac{CGWB} search.} In order to further assess the constraining power of our procedure, in Fig.~\ref{fig:sensitivity} we compare the noise level for stochastic searches, the uncertainty on the \ac{BNS} foreground, and the residual \ac{BBH} foreground after notching out the resolved signals. The baseline noise level $\sigma_n(f)$ is computed through Eqs.~(S1) and~(S2) of the Supplement. 
The uncertainty on the inferred \ac{BNS} foreground is estimated by computing energy spectra $\Omega_\mathrm{BNS}(f)$ for all posterior samples and the resulting standard deviation $\sigma_{\Omega_\mathrm{BNS}}(f)$ at each frequency (see the Supplement for the description of $\Omega_\mathrm{BNS}(f)$ calculation). 
Increasing the observation time leads to a larger number of individually observed BNS systems, resulting in a better estimate of the BNS population parameters and of $\sigma_{\Omega_\mathrm{BNS}}(f)$. 
By running our full simulation using fewer data, we explicitly verified that the width of the posterior distributions of BNS population parameters, as well as $\sigma_{\Omega_{\mathrm{BNS}}}$, scale as $1/\sqrt{T_\mathrm{obs}}$: $\sigma_{\Omega_\mathrm{BNS}}^\mathrm{T'_{obs}}(f)=\sigma_{\Omega_\mathrm{BNS}}^\mathrm{T_{obs}}(f) \left(\sqrt{T_\mathrm{obs}}/\sqrt{T'_\mathrm{obs}}\right)$. 
The sensitivity of stochastic searches also scales as $\sim 1/\sqrt{T_\mathrm{obs}}$~\cite{Allen:1997ad}. We can, therefore, extrapolate our results from the 10 day simulation to much longer observation times, up to 10 years in Fig.~\ref{fig:sensitivity}, provided that the BBH notching residual remains subdominant. 

By only injecting \acp{BBH} signals into the time series and notching out the resolvable ones in the time-frequency domain as we did in Ref.~\cite{Zhong:2022ylh, Zhong:2024}, we find that the residual \ac{BBH} foreground is $\sim 100$ times lower than the baseline noise $\sigma_n(f)$ at every frequency, even for 10 yrs observation, which justifies neglecting its contribution to the joint likelihood in Eq.~\eqref{eq:Joint_Like}. To confirm that this residual foreground is indeed negligible when the observation time is much longer than 10 days, we also show the \ac{BBH} notching residual considering a one-year-long observation in Fig.~S1, and again find it much weaker than the detector sensitivity; hence, it can be neglected. The uncertainty on the \ac{BNS} foreground differs by about 17\% between the inference with known and unknown \ac{SFR}, and it is well below the baseline noise at all frequencies, for both known and unknown \ac{SFR}, and for all observation times. This is consistent with the results discussed in the previous section: the \ac{BNS} foreground is well constrained in both our inferences, allowing us to decouple it from the \ac{CGWB}. By applying the $\sim 1/\sqrt{T_\mathrm{obs}}$ scaling on the \ac{CGWB} uncertainty, we can project sensitivity estimates  for longer observations than 10 days. For the flat \ac{CGWB}, we find that we can exclude $\Omega_\mathrm{ref}=0$ at $5\,\sigma$ level for $\Omega_\mathrm{ref}=2.7\times 10^{-12}/\sqrt{T_\mathrm{obs}}$ ($2.5\times 10^{-12}/\sqrt{T_\mathrm{obs}}$) in the unknown (known) \ac{SFR} case, where $T_\mathrm{obs}$ is the observation time in years.

\noindent
\textbf{\em Conclusions.} In this work, we combine the notching procedure of Refs.~\cite{Zhong:2022ylh,Zhong:2024} with the joint-likelihood approach of Ref.~\cite{Callister:2020arv} to search for \acp{CGWB} with \ac{XG} detectors. In this context, we demonstrate a methodology capable, for the first time, of simultaneously handling \ac{BBH} signals, \ac{BNS} signals, and a \ac{CGWB}. Considering a simple population model for the hierarchical inference on the detected \ac{BNS} events and a flat \ac{CGWB}, we find that we can successfully claim detection of a sufficiently loud \ac{CGWB} within 10 days of observation. In particular, for a flat \ac{CGWB} with amplitude of $\Omega_\mathrm{ref}=2.0\times10^{-11}$, we find that we can exclude $\Omega_\mathrm{ref}=0$ at at least $\sim 6\,\sigma$ confidence level, depending on the assumptions on the \ac{BNS} population model.
The uncertainty on the recovered \ac{BNS} foreground is well below the detector noise level. If we project the sensitivity to 1 year of observation, then we will reach the $5\sigma$ sensitivity of $\Omega_\mathrm{ref}=2.7\times 10^{-12}$, which is within a factor of $\lesssim2$ from the ideal stochastic search in absence of CBC foregrounds.

This study represents a crucial step forward in developing methods to detect \acp{CGWB} with \ac{XG} detectors, but more work is needed before such methods can be applied to real data:

\noindent
(1) We have assumed that the detector noise is perfectly known, while in practice, it can be challenging to estimate the \ac{PSD} in the presence of glitches, non-stationary noise, cosmological and astrophysical \acp{SGWB}.

\noindent
(2) We have neglected the contribution of \acp{NSBH}, which can give rise to an unresolved foreground that is comparable to the one from \acp{BNS} in \ac{XG} detectors~\cite{Bellie:2023jlq}. Our formalism can be straightforwardly extended to include a \ac{NSBH} population. However, this would inevitably increase the computational cost and
it could worsen the constraints on some of the inferred parameters. 

\noindent
(3) When performing hierarchical inference on the resolved \ac{BNS} signals, we generate mock likelihood samples on the source parameters of individual events by using a Newtonian approximation of the signals and a Gaussian approximation of the likelihood~\cite{Fishbach:2018edt,Callister:2020arv}. This method can underestimate errors on source parameters compared to performing full Bayesian \ac{PE} with more realistic waveform models, although we still expect the population hyperparameters to be tightly constrained when the number of events is large enough~\cite{DeRenzis:2024dvx}. Sampling individual-event likelihoods with full \ac{PE} and state-of-the-art waveform models for several thousands of \ac{BNS} signals in \ac{XG} detectors is currently computationally prohibitive. Several promising techniques to speed up \ac{PE} in preparation for \ac{XG} detectors are being developed~\cite{Green:2020dnx,Dax:2022pxd,Wong:2023lgb,Alvey:2023naa} and can be incorporated in our formalism in the future.

\noindent
(4) We assume that the detected \ac{CBC} signals originate from the same astrophysical population that generates the \ac{CBC} foreground from unresolved signals. The presence of an unmodeled sub-population predominantly at high redshifts (e.g., from Population III stars~\cite{Martinovic:2021fzj} or primordial black holes~\cite{Mukherjee:2021itf}) could potentially give rise to biases in the estimate of the \ac{CBC} foreground, and thus affect the detection of the \ac{CGWB}.

We leave the exploration of these effects to future work.


 \noindent
\textbf{\em Acknowledgements}
    We thank Will M. Farr, Thomas Callister, and Francesco Iacovelli for helpful discussions. We also thank Joseph D. Romano for providing useful comments and suggestions on a first draft of this manuscript. The authors are grateful for computational resources provided by the LIGO Laboratory and supported by National Science Foundation (NSF) Grants PHY-0757058 and PHY-0823459. B.Z. is supported by the Fermi Research Alliance, LLC, acting under Contract No.\ DE-AC02-07CH11359. E.B. and L.R. are supported by NSF Grants No. AST-2307146, PHY-2207502, PHY-090003 and PHY-20043, by NASA Grant No. 21-ATP21-0010, by the John Templeton Foundation Grant 62840, by the Simons Foundation, and by the Italian Ministry of Foreign Affairs and International Cooperation grant No.~PGR01167. H.Z. and V.M. are in part supported by the NSF grant PHY-2409173.
This work was carried out at the Advanced Research Computing at Hopkins (ARCH) core facility (\url{rockfish.jhu.edu}), which is supported by the NSF Grant No.~OAC-1920103.

\bibliography{references}

\begin{thebibliography}{71}%
\makeatletter
\providecommand \@ifxundefined [1]{%
 \@ifx{#1\undefined}
}%
\providecommand \@ifnum [1]{%
 \ifnum #1\expandafter \@firstoftwo
 \else \expandafter \@secondoftwo
 \fi
}%
\providecommand \@ifx [1]{%
 \ifx #1\expandafter \@firstoftwo
 \else \expandafter \@secondoftwo
 \fi
}%
\providecommand \natexlab [1]{#1}%
\providecommand \enquote  [1]{``#1''}%
\providecommand \bibnamefont  [1]{#1}%
\providecommand \bibfnamefont [1]{#1}%
\providecommand \citenamefont [1]{#1}%
\providecommand \href@noop [0]{\@secondoftwo}%
\providecommand \href [0]{\begingroup \@sanitize@url \@href}%
\providecommand \@href[1]{\@@startlink{#1}\@@href}%
\providecommand \@@href[1]{\endgroup#1\@@endlink}%
\providecommand \@sanitize@url [0]{\catcode `\\12\catcode `\$12\catcode
  `\&12\catcode `\#12\catcode `\^12\catcode `\_12\catcode `\%12\relax}%
\providecommand \@@startlink[1]{}%
\providecommand \@@endlink[0]{}%
\providecommand \url  [0]{\begingroup\@sanitize@url \@url }%
\providecommand \@url [1]{\endgroup\@href {#1}{\urlprefix }}%
\providecommand \urlprefix  [0]{URL }%
\providecommand \Eprint [0]{\href }%
\providecommand \doibase [0]{http://dx.doi.org/}%
\providecommand \selectlanguage [0]{\@gobble}%
\providecommand \bibinfo  [0]{\@secondoftwo}%
\providecommand \bibfield  [0]{\@secondoftwo}%
\providecommand \translation [1]{[#1]}%
\providecommand \BibitemOpen [0]{}%
\providecommand \bibitemStop [0]{}%
\providecommand \bibitemNoStop [0]{.\EOS\space}%
\providecommand \EOS [0]{\spacefactor3000\relax}%
\providecommand \BibitemShut  [1]{\csname bibitem#1\endcsname}%
\let\auto@bib@innerbib\@empty
\bibitem [{\citenamefont {Reitze}\ \emph {et~al.}(2019)\citenamefont {Reitze},
  \citenamefont {Adhikari}, \citenamefont {Ballmer}, \citenamefont {Barish},
  \citenamefont {Barsotti}, \citenamefont {Billingsley}, \citenamefont {Brown},
  \citenamefont {Chen}, \citenamefont {Coyne}, \citenamefont {Eisenstein} \emph
  {et~al.}}]{CE}%
  \BibitemOpen
  \bibfield  {author} {\bibinfo {author} {\bibfnamefont {D.}~\bibnamefont
  {Reitze}}, \bibinfo {author} {\bibfnamefont {R.~X.}\ \bibnamefont
  {Adhikari}}, \bibinfo {author} {\bibfnamefont {S.}~\bibnamefont {Ballmer}},
  \bibinfo {author} {\bibfnamefont {B.}~\bibnamefont {Barish}}, \bibinfo
  {author} {\bibfnamefont {L.}~\bibnamefont {Barsotti}}, \bibinfo {author}
  {\bibfnamefont {G.}~\bibnamefont {Billingsley}}, \bibinfo {author}
  {\bibfnamefont {D.~A.}\ \bibnamefont {Brown}}, \bibinfo {author}
  {\bibfnamefont {Y.}~\bibnamefont {Chen}}, \bibinfo {author} {\bibfnamefont
  {D.}~\bibnamefont {Coyne}}, \bibinfo {author} {\bibfnamefont
  {R.}~\bibnamefont {Eisenstein}},  \emph {et~al.},\ }\href
  {https://arxiv.org/abs/1907.04833} {\bibfield  {journal} {\bibinfo  {journal}
  {arXiv:1907.04833}\ } (\bibinfo {year} {2019})}\BibitemShut {NoStop}%
\bibitem [{\citenamefont {Evans}\ \emph {et~al.}(2021)\citenamefont {Evans},
  \citenamefont {Adhikari}, \citenamefont {Afle} \emph {et~al.}}]{CEHorizon}%
  \BibitemOpen
  \bibfield  {author} {\bibinfo {author} {\bibfnamefont {M.}~\bibnamefont
  {Evans}}, \bibinfo {author} {\bibfnamefont {R.~X.}\ \bibnamefont {Adhikari}},
  \bibinfo {author} {\bibfnamefont {C.}~\bibnamefont {Afle}},  \emph {et~al.},\
  }\href
  {https://dcc.cosmicexplorer.org/public/0163/P2100003/007/ce-horizon-study.pdf}
  {\enquote {\bibinfo {title} {A horizon study for cosmic explorer: Science,
  observatories, and community},}\ } (\bibinfo {year} {2021})\BibitemShut
  {NoStop}%
\bibitem [{\citenamefont {Srivastava}\ \emph {et~al.}(2022)\citenamefont
  {Srivastava}, \citenamefont {Davis}, \citenamefont {Kuns}, \citenamefont
  {Landry}, \citenamefont {Ballmer}, \citenamefont {Evans}, \citenamefont
  {Hall}, \citenamefont {Read},\ and\ \citenamefont {Sathyaprakash}}]{CE2_PSD}%
  \BibitemOpen
  \bibfield  {author} {\bibinfo {author} {\bibfnamefont {V.}~\bibnamefont
  {Srivastava}}, \bibinfo {author} {\bibfnamefont {D.}~\bibnamefont {Davis}},
  \bibinfo {author} {\bibfnamefont {K.}~\bibnamefont {Kuns}}, \bibinfo {author}
  {\bibfnamefont {P.}~\bibnamefont {Landry}}, \bibinfo {author} {\bibfnamefont
  {S.}~\bibnamefont {Ballmer}}, \bibinfo {author} {\bibfnamefont
  {M.}~\bibnamefont {Evans}}, \bibinfo {author} {\bibfnamefont {E.~D.}\
  \bibnamefont {Hall}}, \bibinfo {author} {\bibfnamefont {J.}~\bibnamefont
  {Read}}, \ and\ \bibinfo {author} {\bibfnamefont {B.~S.}\ \bibnamefont
  {Sathyaprakash}},\ }\href {\doibase 10.3847/1538-4357/ac5f04} {\bibfield
  {journal} {\bibinfo  {journal} {Astrophys. J.}\ }\textbf {\bibinfo {volume}
  {931}},\ \bibinfo {pages} {22} (\bibinfo {year} {2022})},\ \Eprint
  {http://arxiv.org/abs/2201.10668} {arXiv:2201.10668 [gr-qc]} \BibitemShut
  {NoStop}%
\bibitem [{\citenamefont {Punturo}\ \emph {et~al.}(2010)\citenamefont
  {Punturo}, \citenamefont {Abernathy}, \citenamefont {Acernese} \emph
  {et~al.}}]{ET}%
  \BibitemOpen
  \bibfield  {author} {\bibinfo {author} {\bibfnamefont {M.}~\bibnamefont
  {Punturo}}, \bibinfo {author} {\bibfnamefont {M.}~\bibnamefont {Abernathy}},
  \bibinfo {author} {\bibfnamefont {F.}~\bibnamefont {Acernese}},  \emph
  {et~al.},\ }\href {\doibase 10.1088/0264-9381/27/19/194002} {\bibfield
  {journal} {\bibinfo  {journal} {Classical and Quantum Gravity}\ }\textbf
  {\bibinfo {volume} {27}},\ \bibinfo {pages} {194002} (\bibinfo {year}
  {2010})}\BibitemShut {NoStop}%
\bibitem [{\citenamefont {Branchesi}\ \emph
  {et~al.}(2023{\natexlab{a}})\citenamefont {Branchesi} \emph
  {et~al.}}]{ET_PSD}%
  \BibitemOpen
  \bibfield  {author} {\bibinfo {author} {\bibfnamefont {M.}~\bibnamefont
  {Branchesi}} \emph {et~al.},\ }\href {\doibase 10.1088/1475-7516/2023/07/068}
  {\bibfield  {journal} {\bibinfo  {journal} {JCAP}\ }\textbf {\bibinfo
  {volume} {07}},\ \bibinfo {pages} {068} (\bibinfo {year}
  {2023}{\natexlab{a}})},\ \Eprint {http://arxiv.org/abs/2303.15923}
  {arXiv:2303.15923 [gr-qc]} \BibitemShut {NoStop}%
\bibitem [{\citenamefont {Gupta}\ \emph {et~al.}(2023)\citenamefont {Gupta}
  \emph {et~al.}}]{Gupta:2023lga}%
  \BibitemOpen
  \bibfield  {author} {\bibinfo {author} {\bibfnamefont {I.}~\bibnamefont
  {Gupta}} \emph {et~al.},\ }\href@noop {} {\  (\bibinfo {year} {2023})},\
  \Eprint {http://arxiv.org/abs/2307.10421} {arXiv:2307.10421 [gr-qc]}
  \BibitemShut {NoStop}%
\bibitem [{\citenamefont {Branchesi}\ \emph
  {et~al.}(2023{\natexlab{b}})\citenamefont {Branchesi} \emph
  {et~al.}}]{Branchesi:2023mws}%
  \BibitemOpen
  \bibfield  {author} {\bibinfo {author} {\bibfnamefont {M.}~\bibnamefont
  {Branchesi}} \emph {et~al.},\ }\href {\doibase 10.1088/1475-7516/2023/07/068}
  {\bibfield  {journal} {\bibinfo  {journal} {JCAP}\ }\textbf {\bibinfo
  {volume} {07}},\ \bibinfo {pages} {068} (\bibinfo {year}
  {2023}{\natexlab{b}})},\ \Eprint {http://arxiv.org/abs/2303.15923}
  {arXiv:2303.15923 [gr-qc]} \BibitemShut {NoStop}%
\bibitem [{\citenamefont {Regimbau}\ \emph {et~al.}(2017)\citenamefont
  {Regimbau}, \citenamefont {Evans}, \citenamefont {Christensen}, \citenamefont
  {Katsavounidis}, \citenamefont {Sathyaprakash},\ and\ \citenamefont
  {Vitale}}]{TaniaCE}%
  \BibitemOpen
  \bibfield  {author} {\bibinfo {author} {\bibfnamefont {T.}~\bibnamefont
  {Regimbau}}, \bibinfo {author} {\bibfnamefont {M.}~\bibnamefont {Evans}},
  \bibinfo {author} {\bibfnamefont {N.}~\bibnamefont {Christensen}}, \bibinfo
  {author} {\bibfnamefont {E.}~\bibnamefont {Katsavounidis}}, \bibinfo {author}
  {\bibfnamefont {B.}~\bibnamefont {Sathyaprakash}}, \ and\ \bibinfo {author}
  {\bibfnamefont {S.}~\bibnamefont {Vitale}},\ }\href {\doibase
  10.1103/PhysRevLett.118.151105} {\bibfield  {journal} {\bibinfo  {journal}
  {Phys. Rev. Lett.}\ }\textbf {\bibinfo {volume} {118}},\ \bibinfo {pages}
  {151105} (\bibinfo {year} {2017})}\BibitemShut {NoStop}%
\bibitem [{\citenamefont {Renzini}\ \emph {et~al.}(2022)\citenamefont
  {Renzini}, \citenamefont {Goncharov}, \citenamefont {Jenkins},\ and\
  \citenamefont {Meyers}}]{Renzini:2022alw}%
  \BibitemOpen
  \bibfield  {author} {\bibinfo {author} {\bibfnamefont {A.~I.}\ \bibnamefont
  {Renzini}}, \bibinfo {author} {\bibfnamefont {B.}~\bibnamefont {Goncharov}},
  \bibinfo {author} {\bibfnamefont {A.~C.}\ \bibnamefont {Jenkins}}, \ and\
  \bibinfo {author} {\bibfnamefont {P.~M.}\ \bibnamefont {Meyers}},\ }\href
  {\doibase 10.3390/galaxies10010034} {\bibfield  {journal} {\bibinfo
  {journal} {Galaxies}\ }\textbf {\bibinfo {volume} {10}},\ \bibinfo {pages}
  {34} (\bibinfo {year} {2022})},\ \Eprint {http://arxiv.org/abs/2202.00178}
  {arXiv:2202.00178 [gr-qc]} \BibitemShut {NoStop}%
\bibitem [{\citenamefont {Grishchuk}(1974)}]{Grishchuk:1974ny}%
  \BibitemOpen
  \bibfield  {author} {\bibinfo {author} {\bibfnamefont {L.~P.}\ \bibnamefont
  {Grishchuk}},\ }\href@noop {} {\bibfield  {journal} {\bibinfo  {journal} {Zh.
  Eksp. Teor. Fiz.}\ }\textbf {\bibinfo {volume} {67}},\ \bibinfo {pages} {825}
  (\bibinfo {year} {1974})}\BibitemShut {NoStop}%
\bibitem [{\citenamefont {Starobinsky}(1979)}]{Starobinsky:1979ty}%
  \BibitemOpen
  \bibfield  {author} {\bibinfo {author} {\bibfnamefont {A.~A.}\ \bibnamefont
  {Starobinsky}},\ }\href@noop {} {\bibfield  {journal} {\bibinfo  {journal}
  {JETP Lett.}\ }\textbf {\bibinfo {volume} {30}},\ \bibinfo {pages} {682}
  (\bibinfo {year} {1979})}\BibitemShut {NoStop}%
\bibitem [{\citenamefont {Grishchuk}(1993)}]{Grishchuk:1993te}%
  \BibitemOpen
  \bibfield  {author} {\bibinfo {author} {\bibfnamefont {L.~P.}\ \bibnamefont
  {Grishchuk}},\ }\href {\doibase 10.1103/PhysRevD.48.3513} {\bibfield
  {journal} {\bibinfo  {journal} {Phys. Rev. D}\ }\textbf {\bibinfo {volume}
  {48}},\ \bibinfo {pages} {3513} (\bibinfo {year} {1993})},\ \Eprint
  {http://arxiv.org/abs/gr-qc/9304018} {arXiv:gr-qc/9304018} \BibitemShut
  {NoStop}%
\bibitem [{\citenamefont {Barnaby}\ \emph {et~al.}(2012)\citenamefont
  {Barnaby}, \citenamefont {Pajer},\ and\ \citenamefont
  {Peloso}}]{Barnaby:2011qe}%
  \BibitemOpen
  \bibfield  {author} {\bibinfo {author} {\bibfnamefont {N.}~\bibnamefont
  {Barnaby}}, \bibinfo {author} {\bibfnamefont {E.}~\bibnamefont {Pajer}}, \
  and\ \bibinfo {author} {\bibfnamefont {M.}~\bibnamefont {Peloso}},\ }\href
  {\doibase 10.1103/PhysRevD.85.023525} {\bibfield  {journal} {\bibinfo
  {journal} {Phys. Rev. D}\ }\textbf {\bibinfo {volume} {85}},\ \bibinfo
  {pages} {023525} (\bibinfo {year} {2012})},\ \Eprint
  {http://arxiv.org/abs/1110.3327} {arXiv:1110.3327 [astro-ph.CO]} \BibitemShut
  {NoStop}%
\bibitem [{\citenamefont {Damour}\ and\ \citenamefont
  {Vilenkin}(2005)}]{Damour:2004kw}%
  \BibitemOpen
  \bibfield  {author} {\bibinfo {author} {\bibfnamefont {T.}~\bibnamefont
  {Damour}}\ and\ \bibinfo {author} {\bibfnamefont {A.}~\bibnamefont
  {Vilenkin}},\ }\href {\doibase 10.1103/PhysRevD.71.063510} {\bibfield
  {journal} {\bibinfo  {journal} {Phys. Rev. D}\ }\textbf {\bibinfo {volume}
  {71}},\ \bibinfo {pages} {063510} (\bibinfo {year} {2005})},\ \Eprint
  {http://arxiv.org/abs/hep-th/0410222} {arXiv:hep-th/0410222} \BibitemShut
  {NoStop}%
\bibitem [{\citenamefont {Siemens}\ \emph {et~al.}(2007)\citenamefont
  {Siemens}, \citenamefont {Mandic},\ and\ \citenamefont
  {Creighton}}]{Siemens:2006yp}%
  \BibitemOpen
  \bibfield  {author} {\bibinfo {author} {\bibfnamefont {X.}~\bibnamefont
  {Siemens}}, \bibinfo {author} {\bibfnamefont {V.}~\bibnamefont {Mandic}}, \
  and\ \bibinfo {author} {\bibfnamefont {J.}~\bibnamefont {Creighton}},\ }\href
  {\doibase 10.1103/PhysRevLett.98.111101} {\bibfield  {journal} {\bibinfo
  {journal} {Phys. Rev. Lett.}\ }\textbf {\bibinfo {volume} {98}},\ \bibinfo
  {pages} {111101} (\bibinfo {year} {2007})},\ \Eprint
  {http://arxiv.org/abs/astro-ph/0610920} {arXiv:astro-ph/0610920} \BibitemShut
  {NoStop}%
\bibitem [{\citenamefont {Olmez}\ \emph {et~al.}(2010)\citenamefont {Olmez},
  \citenamefont {Mandic},\ and\ \citenamefont {Siemens}}]{Olmez:2010bi}%
  \BibitemOpen
  \bibfield  {author} {\bibinfo {author} {\bibfnamefont {S.}~\bibnamefont
  {Olmez}}, \bibinfo {author} {\bibfnamefont {V.}~\bibnamefont {Mandic}}, \
  and\ \bibinfo {author} {\bibfnamefont {X.}~\bibnamefont {Siemens}},\ }\href
  {\doibase 10.1103/PhysRevD.81.104028} {\bibfield  {journal} {\bibinfo
  {journal} {Phys. Rev. D}\ }\textbf {\bibinfo {volume} {81}},\ \bibinfo
  {pages} {104028} (\bibinfo {year} {2010})},\ \Eprint
  {http://arxiv.org/abs/1004.0890} {arXiv:1004.0890 [astro-ph.CO]} \BibitemShut
  {NoStop}%
\bibitem [{\citenamefont {Regimbau}\ \emph {et~al.}(2012)\citenamefont
  {Regimbau}, \citenamefont {Giampanis}, \citenamefont {Siemens},\ and\
  \citenamefont {Mandic}}]{Regimbau:2011bm}%
  \BibitemOpen
  \bibfield  {author} {\bibinfo {author} {\bibfnamefont {T.}~\bibnamefont
  {Regimbau}}, \bibinfo {author} {\bibfnamefont {S.}~\bibnamefont {Giampanis}},
  \bibinfo {author} {\bibfnamefont {X.}~\bibnamefont {Siemens}}, \ and\
  \bibinfo {author} {\bibfnamefont {V.}~\bibnamefont {Mandic}},\ }\href
  {\doibase 10.1103/PhysRevD.85.066001} {\bibfield  {journal} {\bibinfo
  {journal} {Phys. Rev. D}\ }\textbf {\bibinfo {volume} {85}},\ \bibinfo
  {pages} {066001} (\bibinfo {year} {2012})},\ \Eprint
  {http://arxiv.org/abs/1111.6638} {arXiv:1111.6638 [astro-ph.CO]} \BibitemShut
  {NoStop}%
\bibitem [{\citenamefont {Callister}\ \emph {et~al.}(2016)\citenamefont
  {Callister}, \citenamefont {Sammut}, \citenamefont {Qiu}, \citenamefont
  {Mandel},\ and\ \citenamefont {Thrane}}]{Callister:2016ewt}%
  \BibitemOpen
  \bibfield  {author} {\bibinfo {author} {\bibfnamefont {T.}~\bibnamefont
  {Callister}}, \bibinfo {author} {\bibfnamefont {L.}~\bibnamefont {Sammut}},
  \bibinfo {author} {\bibfnamefont {S.}~\bibnamefont {Qiu}}, \bibinfo {author}
  {\bibfnamefont {I.}~\bibnamefont {Mandel}}, \ and\ \bibinfo {author}
  {\bibfnamefont {E.}~\bibnamefont {Thrane}},\ }\href {\doibase
  10.1103/PhysRevX.6.031018} {\bibfield  {journal} {\bibinfo  {journal} {Phys.
  Rev. X}\ }\textbf {\bibinfo {volume} {6}},\ \bibinfo {pages} {031018}
  (\bibinfo {year} {2016})},\ \Eprint {http://arxiv.org/abs/1604.02513}
  {arXiv:1604.02513 [gr-qc]} \BibitemShut {NoStop}%
\bibitem [{\citenamefont {Sachdev}\ \emph {et~al.}(2020)\citenamefont
  {Sachdev}, \citenamefont {Regimbau},\ and\ \citenamefont
  {Sathyaprakash}}]{subtractionSurabhi}%
  \BibitemOpen
  \bibfield  {author} {\bibinfo {author} {\bibfnamefont {S.}~\bibnamefont
  {Sachdev}}, \bibinfo {author} {\bibfnamefont {T.}~\bibnamefont {Regimbau}}, \
  and\ \bibinfo {author} {\bibfnamefont {B.~S.}\ \bibnamefont
  {Sathyaprakash}},\ }\href {\doibase 10.1103/PhysRevD.102.024051} {\bibfield
  {journal} {\bibinfo  {journal} {Phys. Rev. D}\ }\textbf {\bibinfo {volume}
  {102}},\ \bibinfo {pages} {024051} (\bibinfo {year} {2020})}\BibitemShut
  {NoStop}%
\bibitem [{\citenamefont {Zhou}\ \emph {et~al.}(2023)\citenamefont {Zhou},
  \citenamefont {Reali}, \citenamefont {Berti}, \citenamefont
  {\c{C}al\i{}\c{s}kan}, \citenamefont {Creque-Sarbinowski}, \citenamefont
  {Kamionkowski},\ and\ \citenamefont {Sathyaprakash}}]{Zhou:2022nmt}%
  \BibitemOpen
  \bibfield  {author} {\bibinfo {author} {\bibfnamefont {B.}~\bibnamefont
  {Zhou}}, \bibinfo {author} {\bibfnamefont {L.}~\bibnamefont {Reali}},
  \bibinfo {author} {\bibfnamefont {E.}~\bibnamefont {Berti}}, \bibinfo
  {author} {\bibfnamefont {M.}~\bibnamefont {\c{C}al\i{}\c{s}kan}}, \bibinfo
  {author} {\bibfnamefont {C.}~\bibnamefont {Creque-Sarbinowski}}, \bibinfo
  {author} {\bibfnamefont {M.}~\bibnamefont {Kamionkowski}}, \ and\ \bibinfo
  {author} {\bibfnamefont {B.~S.}\ \bibnamefont {Sathyaprakash}},\ }\href
  {\doibase 10.1103/PhysRevD.108.064040} {\bibfield  {journal} {\bibinfo
  {journal} {Phys. Rev. D}\ }\textbf {\bibinfo {volume} {108}},\ \bibinfo
  {pages} {064040} (\bibinfo {year} {2023})},\ \Eprint
  {http://arxiv.org/abs/2209.01310} {arXiv:2209.01310 [gr-qc]} \BibitemShut
  {NoStop}%
\bibitem [{\citenamefont {Zhou}\ \emph {et~al.}(2022)\citenamefont {Zhou},
  \citenamefont {Reali}, \citenamefont {Berti}, \citenamefont
  {\c{C}al\i{}\c{s}kan}, \citenamefont {Creque-Sarbinowski}, \citenamefont
  {Kamionkowski},\ and\ \citenamefont {Sathyaprakash}}]{Zhou:2022otw}%
  \BibitemOpen
  \bibfield  {author} {\bibinfo {author} {\bibfnamefont {B.}~\bibnamefont
  {Zhou}}, \bibinfo {author} {\bibfnamefont {L.}~\bibnamefont {Reali}},
  \bibinfo {author} {\bibfnamefont {E.}~\bibnamefont {Berti}}, \bibinfo
  {author} {\bibfnamefont {M.}~\bibnamefont {\c{C}al\i{}\c{s}kan}}, \bibinfo
  {author} {\bibfnamefont {C.}~\bibnamefont {Creque-Sarbinowski}}, \bibinfo
  {author} {\bibfnamefont {M.}~\bibnamefont {Kamionkowski}}, \ and\ \bibinfo
  {author} {\bibfnamefont {B.~S.}\ \bibnamefont {Sathyaprakash}},\ }\href@noop
  {} {\bibfield  {journal} {\bibinfo  {journal} {arXiv}\ } (\bibinfo {year}
  {2022})},\ \Eprint {http://arxiv.org/abs/2209.01221} {arXiv:2209.01221
  [gr-qc]} \BibitemShut {NoStop}%
\bibitem [{\citenamefont {Song}\ \emph {et~al.}(2024)\citenamefont {Song},
  \citenamefont {Liang}, \citenamefont {Wang},\ and\ \citenamefont
  {Shao}}]{Song:2024pnk}%
  \BibitemOpen
  \bibfield  {author} {\bibinfo {author} {\bibfnamefont {H.}~\bibnamefont
  {Song}}, \bibinfo {author} {\bibfnamefont {D.}~\bibnamefont {Liang}},
  \bibinfo {author} {\bibfnamefont {Z.}~\bibnamefont {Wang}}, \ and\ \bibinfo
  {author} {\bibfnamefont {L.}~\bibnamefont {Shao}},\ }\href@noop {} {\bibfield
   {journal} {\bibinfo  {journal} {arXiv}\ } (\bibinfo {year} {2024})},\
  \Eprint {http://arxiv.org/abs/2401.00984} {arXiv:2401.00984 [gr-qc]}
  \BibitemShut {NoStop}%
\bibitem [{\citenamefont {Cutler}\ and\ \citenamefont
  {Harms}(2006)}]{Cutler:2005qq}%
  \BibitemOpen
  \bibfield  {author} {\bibinfo {author} {\bibfnamefont {C.}~\bibnamefont
  {Cutler}}\ and\ \bibinfo {author} {\bibfnamefont {J.}~\bibnamefont {Harms}},\
  }\href {\doibase 10.1103/PhysRevD.73.042001} {\bibfield  {journal} {\bibinfo
  {journal} {Phys. Rev. D}\ }\textbf {\bibinfo {volume} {73}},\ \bibinfo
  {pages} {042001} (\bibinfo {year} {2006})},\ \Eprint
  {http://arxiv.org/abs/gr-qc/0511092} {arXiv:gr-qc/0511092} \BibitemShut
  {NoStop}%
\bibitem [{\citenamefont {Sharma}\ and\ \citenamefont
  {Harms}(2020)}]{Sharma:2020btq}%
  \BibitemOpen
  \bibfield  {author} {\bibinfo {author} {\bibfnamefont {A.}~\bibnamefont
  {Sharma}}\ and\ \bibinfo {author} {\bibfnamefont {J.}~\bibnamefont {Harms}},\
  }\href {\doibase 10.1103/PhysRevD.102.063009} {\bibfield  {journal} {\bibinfo
   {journal} {Phys. Rev. D}\ }\textbf {\bibinfo {volume} {102}},\ \bibinfo
  {pages} {063009} (\bibinfo {year} {2020})},\ \Eprint
  {http://arxiv.org/abs/2006.16116} {arXiv:2006.16116 [gr-qc]} \BibitemShut
  {NoStop}%
\bibitem [{\citenamefont {Pan}\ and\ \citenamefont {Yang}(2023)}]{Pan:2023naq}%
  \BibitemOpen
  \bibfield  {author} {\bibinfo {author} {\bibfnamefont {Z.}~\bibnamefont
  {Pan}}\ and\ \bibinfo {author} {\bibfnamefont {H.}~\bibnamefont {Yang}},\
  }\href {\doibase 10.1103/PhysRevD.107.123036} {\bibfield  {journal} {\bibinfo
   {journal} {Phys. Rev. D}\ }\textbf {\bibinfo {volume} {107}},\ \bibinfo
  {pages} {123036} (\bibinfo {year} {2023})},\ \Eprint
  {http://arxiv.org/abs/2301.04529} {arXiv:2301.04529 [gr-qc]} \BibitemShut
  {NoStop}%
\bibitem [{\citenamefont {Belgacem}\ \emph {et~al.}(2024)\citenamefont
  {Belgacem}, \citenamefont {Iacovelli}, \citenamefont {Maggiore},
  \citenamefont {Mancarella},\ and\ \citenamefont
  {Muttoni}}]{Belgacem:2024ntv}%
  \BibitemOpen
  \bibfield  {author} {\bibinfo {author} {\bibfnamefont {E.}~\bibnamefont
  {Belgacem}}, \bibinfo {author} {\bibfnamefont {F.}~\bibnamefont {Iacovelli}},
  \bibinfo {author} {\bibfnamefont {M.}~\bibnamefont {Maggiore}}, \bibinfo
  {author} {\bibfnamefont {M.}~\bibnamefont {Mancarella}}, \ and\ \bibinfo
  {author} {\bibfnamefont {N.}~\bibnamefont {Muttoni}},\ }\href@noop {} {\
  (\bibinfo {year} {2024})},\ \Eprint {http://arxiv.org/abs/2411.04029}
  {arXiv:2411.04029 [gr-qc]} \BibitemShut {NoStop}%
\bibitem [{\citenamefont {Zhong}\ \emph {et~al.}(2023)\citenamefont {Zhong},
  \citenamefont {Ormiston},\ and\ \citenamefont {Mandic}}]{Zhong:2022ylh}%
  \BibitemOpen
  \bibfield  {author} {\bibinfo {author} {\bibfnamefont {H.}~\bibnamefont
  {Zhong}}, \bibinfo {author} {\bibfnamefont {R.}~\bibnamefont {Ormiston}}, \
  and\ \bibinfo {author} {\bibfnamefont {V.}~\bibnamefont {Mandic}},\ }\href
  {\doibase 10.1103/PhysRevD.107.064048} {\bibfield  {journal} {\bibinfo
  {journal} {Phys. Rev. D}\ }\textbf {\bibinfo {volume} {107}},\ \bibinfo
  {pages} {064048} (\bibinfo {year} {2023})},\ \bibinfo {note} {[Erratum:
  Phys.Rev.D 108, 089902 (2023)]},\ \Eprint {http://arxiv.org/abs/2209.11877}
  {arXiv:2209.11877 [gr-qc]} \BibitemShut {NoStop}%
\bibitem [{\citenamefont {Zhong}\ \emph {et~al.}(2024)\citenamefont {Zhong},
  \citenamefont {Zhou}, \citenamefont {Reali}, \citenamefont {Berti},\ and\
  \citenamefont {Mandic}}]{Zhong:2024}%
  \BibitemOpen
  \bibfield  {author} {\bibinfo {author} {\bibfnamefont {H.}~\bibnamefont
  {Zhong}}, \bibinfo {author} {\bibfnamefont {B.}~\bibnamefont {Zhou}},
  \bibinfo {author} {\bibfnamefont {L.}~\bibnamefont {Reali}}, \bibinfo
  {author} {\bibfnamefont {E.}~\bibnamefont {Berti}}, \ and\ \bibinfo {author}
  {\bibfnamefont {V.}~\bibnamefont {Mandic}},\ }\href {\doibase
  10.1103/PhysRevD.110.064047} {\bibfield  {journal} {\bibinfo  {journal}
  {Phys. Rev. D}\ }\textbf {\bibinfo {volume} {110}},\ \bibinfo {pages}
  {064047} (\bibinfo {year} {2024})}\BibitemShut {NoStop}%
\bibitem [{\citenamefont {Bellie}\ \emph {et~al.}(2023)\citenamefont {Bellie},
  \citenamefont {Banagiri}, \citenamefont {Doctor},\ and\ \citenamefont
  {Kalogera}}]{Bellie:2023jlq}%
  \BibitemOpen
  \bibfield  {author} {\bibinfo {author} {\bibfnamefont {D.~S.}\ \bibnamefont
  {Bellie}}, \bibinfo {author} {\bibfnamefont {S.}~\bibnamefont {Banagiri}},
  \bibinfo {author} {\bibfnamefont {Z.}~\bibnamefont {Doctor}}, \ and\ \bibinfo
  {author} {\bibfnamefont {V.}~\bibnamefont {Kalogera}},\ }\href@noop {}
  {\bibfield  {journal} {\bibinfo  {journal} {arXiv}\ } (\bibinfo {year}
  {2023})},\ \Eprint {http://arxiv.org/abs/2310.02517} {arXiv:2310.02517
  [gr-qc]} \BibitemShut {NoStop}%
\bibitem [{\citenamefont {Smith}\ and\ \citenamefont
  {Thrane}(2018)}]{Smith:2017vfk}%
  \BibitemOpen
  \bibfield  {author} {\bibinfo {author} {\bibfnamefont {R.}~\bibnamefont
  {Smith}}\ and\ \bibinfo {author} {\bibfnamefont {E.}~\bibnamefont {Thrane}},\
  }\href {\doibase 10.1103/PhysRevX.8.021019} {\bibfield  {journal} {\bibinfo
  {journal} {Phys. Rev. X}\ }\textbf {\bibinfo {volume} {8}},\ \bibinfo {pages}
  {021019} (\bibinfo {year} {2018})},\ \Eprint
  {http://arxiv.org/abs/1712.00688} {arXiv:1712.00688 [gr-qc]} \BibitemShut
  {NoStop}%
\bibitem [{\citenamefont {Smith}\ \emph {et~al.}(2020)\citenamefont {Smith},
  \citenamefont {Talbot}, \citenamefont {Hernandez~Vivanco},\ and\
  \citenamefont {Thrane}}]{Smith:2020lkj}%
  \BibitemOpen
  \bibfield  {author} {\bibinfo {author} {\bibfnamefont {R.~J.~E.}\
  \bibnamefont {Smith}}, \bibinfo {author} {\bibfnamefont {C.}~\bibnamefont
  {Talbot}}, \bibinfo {author} {\bibfnamefont {F.}~\bibnamefont
  {Hernandez~Vivanco}}, \ and\ \bibinfo {author} {\bibfnamefont
  {E.}~\bibnamefont {Thrane}},\ }\href {\doibase 10.1093/mnras/staa1642}
  {\bibfield  {journal} {\bibinfo  {journal} {Mon. Not. Roy. Astron. Soc.}\
  }\textbf {\bibinfo {volume} {496}},\ \bibinfo {pages} {3281} (\bibinfo {year}
  {2020})},\ \Eprint {http://arxiv.org/abs/2004.09700} {arXiv:2004.09700
  [astro-ph.HE]} \BibitemShut {NoStop}%
\bibitem [{\citenamefont {Biscoveanu}\ \emph {et~al.}(2020)\citenamefont
  {Biscoveanu}, \citenamefont {Talbot}, \citenamefont {Thrane},\ and\
  \citenamefont {Smith}}]{Biscoveanu:2020gds}%
  \BibitemOpen
  \bibfield  {author} {\bibinfo {author} {\bibfnamefont {S.}~\bibnamefont
  {Biscoveanu}}, \bibinfo {author} {\bibfnamefont {C.}~\bibnamefont {Talbot}},
  \bibinfo {author} {\bibfnamefont {E.}~\bibnamefont {Thrane}}, \ and\ \bibinfo
  {author} {\bibfnamefont {R.}~\bibnamefont {Smith}},\ }\href {\doibase
  10.1103/PhysRevLett.125.241101} {\bibfield  {journal} {\bibinfo  {journal}
  {Phys. Rev. Lett.}\ }\textbf {\bibinfo {volume} {125}},\ \bibinfo {pages}
  {241101} (\bibinfo {year} {2020})},\ \Eprint
  {http://arxiv.org/abs/2009.04418} {arXiv:2009.04418 [astro-ph.HE]}
  \BibitemShut {NoStop}%
\bibitem [{\citenamefont {Li}\ \emph {et~al.}(2024)\citenamefont {Li},
  \citenamefont {Yu},\ and\ \citenamefont {Pan}}]{Li:2024iua}%
  \BibitemOpen
  \bibfield  {author} {\bibinfo {author} {\bibfnamefont {M.}~\bibnamefont
  {Li}}, \bibinfo {author} {\bibfnamefont {J.}~\bibnamefont {Yu}}, \ and\
  \bibinfo {author} {\bibfnamefont {Z.}~\bibnamefont {Pan}},\ }\href@noop {}
  {\bibfield  {journal} {\bibinfo  {journal} {arXiv}\ } (\bibinfo {year}
  {2024})},\ \Eprint {http://arxiv.org/abs/2403.01846} {arXiv:2403.01846
  [gr-qc]} \BibitemShut {NoStop}%
\bibitem [{\citenamefont {Callister}\ \emph {et~al.}(2020)\citenamefont
  {Callister}, \citenamefont {Fishbach}, \citenamefont {Holz},\ and\
  \citenamefont {Farr}}]{Callister:2020arv}%
  \BibitemOpen
  \bibfield  {author} {\bibinfo {author} {\bibfnamefont {T.}~\bibnamefont
  {Callister}}, \bibinfo {author} {\bibfnamefont {M.}~\bibnamefont {Fishbach}},
  \bibinfo {author} {\bibfnamefont {D.}~\bibnamefont {Holz}}, \ and\ \bibinfo
  {author} {\bibfnamefont {W.}~\bibnamefont {Farr}},\ }\href {\doibase
  10.3847/2041-8213/ab9743} {\bibfield  {journal} {\bibinfo  {journal}
  {Astrophys. J. Lett.}\ }\textbf {\bibinfo {volume} {896}},\ \bibinfo {pages}
  {L32} (\bibinfo {year} {2020})},\ \Eprint {http://arxiv.org/abs/2003.12152}
  {arXiv:2003.12152 [astro-ph.HE]} \BibitemShut {NoStop}%
\bibitem [{\citenamefont {Abbott}\ \emph {et~al.}(2023)\citenamefont {Abbott}
  \emph {et~al.}}]{KAGRA:2021duu}%
  \BibitemOpen
  \bibfield  {author} {\bibinfo {author} {\bibfnamefont {R.}~\bibnamefont
  {Abbott}} \emph {et~al.} (\bibinfo {collaboration} {KAGRA, VIRGO, LIGO
  Scientific}),\ }\href {\doibase 10.1103/PhysRevX.13.011048} {\bibfield
  {journal} {\bibinfo  {journal} {Phys. Rev. X}\ }\textbf {\bibinfo {volume}
  {13}},\ \bibinfo {pages} {011048} (\bibinfo {year} {2023})},\ \Eprint
  {http://arxiv.org/abs/2111.03634} {arXiv:2111.03634 [astro-ph.HE]}
  \BibitemShut {NoStop}%
\bibitem [{\citenamefont {Landry}\ and\ \citenamefont
  {Read}(2021)}]{Landry:2021hvl}%
  \BibitemOpen
  \bibfield  {author} {\bibinfo {author} {\bibfnamefont {P.}~\bibnamefont
  {Landry}}\ and\ \bibinfo {author} {\bibfnamefont {J.~S.}\ \bibnamefont
  {Read}},\ }\href {\doibase 10.3847/2041-8213/ac2f3e} {\bibfield  {journal}
  {\bibinfo  {journal} {Astrophys. J. Lett.}\ }\textbf {\bibinfo {volume}
  {921}},\ \bibinfo {pages} {L25} (\bibinfo {year} {2021})},\ \Eprint
  {http://arxiv.org/abs/2107.04559} {arXiv:2107.04559 [astro-ph.HE]}
  \BibitemShut {NoStop}%
\bibitem [{\citenamefont {Finkel}\ \emph {et~al.}(2022)\citenamefont {Finkel},
  \citenamefont {Andresen},\ and\ \citenamefont {Mandic}}]{Finkel:2021zgf}%
  \BibitemOpen
  \bibfield  {author} {\bibinfo {author} {\bibfnamefont {B.}~\bibnamefont
  {Finkel}}, \bibinfo {author} {\bibfnamefont {H.}~\bibnamefont {Andresen}}, \
  and\ \bibinfo {author} {\bibfnamefont {V.}~\bibnamefont {Mandic}},\ }\href
  {\doibase 10.1103/PhysRevD.105.063022} {\bibfield  {journal} {\bibinfo
  {journal} {Phys. Rev. D}\ }\textbf {\bibinfo {volume} {105}},\ \bibinfo
  {pages} {063022} (\bibinfo {year} {2022})},\ \Eprint
  {http://arxiv.org/abs/2110.01478} {arXiv:2110.01478 [gr-qc]} \BibitemShut
  {NoStop}%
\bibitem [{\citenamefont {Aasi}\ \emph {et~al.}(2015)\citenamefont {Aasi} \emph
  {et~al.}}]{LIGOScientific:2014pky}%
  \BibitemOpen
  \bibfield  {author} {\bibinfo {author} {\bibfnamefont {J.}~\bibnamefont
  {Aasi}} \emph {et~al.} (\bibinfo {collaboration} {LIGO Scientific}),\ }\href
  {\doibase 10.1088/0264-9381/32/7/074001} {\bibfield  {journal} {\bibinfo
  {journal} {Class. Quant. Grav.}\ }\textbf {\bibinfo {volume} {32}},\ \bibinfo
  {pages} {074001} (\bibinfo {year} {2015})},\ \Eprint
  {http://arxiv.org/abs/1411.4547} {arXiv:1411.4547 [gr-qc]} \BibitemShut
  {NoStop}%
\bibitem [{\citenamefont {Acernese}\ \emph {et~al.}(2015)\citenamefont
  {Acernese} \emph {et~al.}}]{VIRGO:2014yos}%
  \BibitemOpen
  \bibfield  {author} {\bibinfo {author} {\bibfnamefont {F.}~\bibnamefont
  {Acernese}} \emph {et~al.} (\bibinfo {collaboration} {VIRGO}),\ }\href
  {\doibase 10.1088/0264-9381/32/2/024001} {\bibfield  {journal} {\bibinfo
  {journal} {Class. Quant. Grav.}\ }\textbf {\bibinfo {volume} {32}},\ \bibinfo
  {pages} {024001} (\bibinfo {year} {2015})},\ \Eprint
  {http://arxiv.org/abs/1408.3978} {arXiv:1408.3978 [gr-qc]} \BibitemShut
  {NoStop}%
\bibitem [{\citenamefont {Ashton}\ \emph {et~al.}(2019)\citenamefont {Ashton},
  \citenamefont {H{\"u}bner}, \citenamefont {Lasky}, \citenamefont {Talbot},
  \citenamefont {Ackley}, \citenamefont {Biscoveanu}, \citenamefont {Chu},
  \citenamefont {Divakarla}, \citenamefont {Easter}, \citenamefont {Goncharov}
  \emph {et~al.}}]{Bilby}%
  \BibitemOpen
  \bibfield  {author} {\bibinfo {author} {\bibfnamefont {G.}~\bibnamefont
  {Ashton}}, \bibinfo {author} {\bibfnamefont {M.}~\bibnamefont {H{\"u}bner}},
  \bibinfo {author} {\bibfnamefont {P.~D.}\ \bibnamefont {Lasky}}, \bibinfo
  {author} {\bibfnamefont {C.}~\bibnamefont {Talbot}}, \bibinfo {author}
  {\bibfnamefont {K.}~\bibnamefont {Ackley}}, \bibinfo {author} {\bibfnamefont
  {S.}~\bibnamefont {Biscoveanu}}, \bibinfo {author} {\bibfnamefont
  {Q.}~\bibnamefont {Chu}}, \bibinfo {author} {\bibfnamefont {A.}~\bibnamefont
  {Divakarla}}, \bibinfo {author} {\bibfnamefont {P.~J.}\ \bibnamefont
  {Easter}}, \bibinfo {author} {\bibfnamefont {B.}~\bibnamefont {Goncharov}},
  \emph {et~al.},\ }\href
  {https://iopscience.iop.org/article/10.3847/1538-4365/ab06fc} {\bibfield
  {journal} {\bibinfo  {journal} {The Astrophysical Journal Supplement Series}\
  }\textbf {\bibinfo {volume} {241}},\ \bibinfo {pages} {27} (\bibinfo {year}
  {2019})}\BibitemShut {NoStop}%
\bibitem [{\citenamefont {Pratten}\ \emph {et~al.}(2020)\citenamefont
  {Pratten}, \citenamefont {Husa}, \citenamefont {Garc\'{\i}a-Quir\'os},
  \citenamefont {Colleoni}, \citenamefont {Ramos-Buades}, \citenamefont
  {Estell\'es},\ and\ \citenamefont {Jaume}}]{IMRXAS}%
  \BibitemOpen
  \bibfield  {author} {\bibinfo {author} {\bibfnamefont {G.}~\bibnamefont
  {Pratten}}, \bibinfo {author} {\bibfnamefont {S.}~\bibnamefont {Husa}},
  \bibinfo {author} {\bibfnamefont {C.}~\bibnamefont {Garc\'{\i}a-Quir\'os}},
  \bibinfo {author} {\bibfnamefont {M.}~\bibnamefont {Colleoni}}, \bibinfo
  {author} {\bibfnamefont {A.}~\bibnamefont {Ramos-Buades}}, \bibinfo {author}
  {\bibfnamefont {H.}~\bibnamefont {Estell\'es}}, \ and\ \bibinfo {author}
  {\bibfnamefont {R.}~\bibnamefont {Jaume}},\ }\href {\doibase
  10.1103/PhysRevD.102.064001} {\bibfield  {journal} {\bibinfo  {journal}
  {Phys. Rev. D}\ }\textbf {\bibinfo {volume} {102}},\ \bibinfo {pages}
  {064001} (\bibinfo {year} {2020})}\BibitemShut {NoStop}%
\bibitem [{\citenamefont {Buonanno}\ \emph {et~al.}(2003)\citenamefont
  {Buonanno}, \citenamefont {Chen},\ and\ \citenamefont
  {Vallisneri}}]{Buonanno:2002fy}%
  \BibitemOpen
  \bibfield  {author} {\bibinfo {author} {\bibfnamefont {A.}~\bibnamefont
  {Buonanno}}, \bibinfo {author} {\bibfnamefont {Y.-b.}\ \bibnamefont {Chen}},
  \ and\ \bibinfo {author} {\bibfnamefont {M.}~\bibnamefont {Vallisneri}},\
  }\href {\doibase 10.1103/PhysRevD.67.104025} {\bibfield  {journal} {\bibinfo
  {journal} {Phys. Rev. D}\ }\textbf {\bibinfo {volume} {67}},\ \bibinfo
  {pages} {104025} (\bibinfo {year} {2003})},\ \bibinfo {note} {[Erratum:
  Phys.Rev.D 74, 029904 (2006)]},\ \Eprint {http://arxiv.org/abs/gr-qc/0211087}
  {arXiv:gr-qc/0211087} \BibitemShut {NoStop}%
\bibitem [{\citenamefont {Renzini}\ \emph {et~al.}(2023)\citenamefont
  {Renzini}, \citenamefont {Romero-Rodríguez}, \citenamefont {Talbot} \emph
  {et~al.}}]{pygwb}%
  \BibitemOpen
  \bibfield  {author} {\bibinfo {author} {\bibfnamefont {A.~I.}\ \bibnamefont
  {Renzini}}, \bibinfo {author} {\bibfnamefont {A.}~\bibnamefont
  {Romero-Rodríguez}}, \bibinfo {author} {\bibfnamefont {C.}~\bibnamefont
  {Talbot}},  \emph {et~al.},\ }\href {\doibase 10.3847/1538-4357/acd775}
  {\bibfield  {journal} {\bibinfo  {journal} {The Astrophysical Journal}\
  }\textbf {\bibinfo {volume} {952}},\ \bibinfo {pages} {25} (\bibinfo {year}
  {2023})}\BibitemShut {NoStop}%
\bibitem [{\citenamefont {Sathyaprakash}\ and\ \citenamefont
  {Schutz}(2009)}]{Sathyaprakash:2009xs}%
  \BibitemOpen
  \bibfield  {author} {\bibinfo {author} {\bibfnamefont {B.~S.}\ \bibnamefont
  {Sathyaprakash}}\ and\ \bibinfo {author} {\bibfnamefont {B.~F.}\ \bibnamefont
  {Schutz}},\ }\href {\doibase 10.12942/lrr-2009-2} {\bibfield  {journal}
  {\bibinfo  {journal} {Living Rev. Rel.}\ }\textbf {\bibinfo {volume} {12}},\
  \bibinfo {pages} {2} (\bibinfo {year} {2009})},\ \Eprint
  {http://arxiv.org/abs/0903.0338} {arXiv:0903.0338 [gr-qc]} \BibitemShut
  {NoStop}%
\bibitem [{not()}]{note1}%
  \BibitemOpen
  \href@noop {} {}\bibinfo {note} {See Supplemental Material for additional
  technical details, which includes Refs. [61–71].}\BibitemShut {Stop}%
\bibitem [{\citenamefont {Mandel}\ \emph {et~al.}(2019)\citenamefont {Mandel},
  \citenamefont {Farr},\ and\ \citenamefont {Gair}}]{Mandel:2018mve}%
  \BibitemOpen
  \bibfield  {author} {\bibinfo {author} {\bibfnamefont {I.}~\bibnamefont
  {Mandel}}, \bibinfo {author} {\bibfnamefont {W.~M.}\ \bibnamefont {Farr}}, \
  and\ \bibinfo {author} {\bibfnamefont {J.~R.}\ \bibnamefont {Gair}},\ }\href
  {\doibase 10.1093/mnras/stz896} {\bibfield  {journal} {\bibinfo  {journal}
  {Mon. Not. Roy. Astron. Soc.}\ }\textbf {\bibinfo {volume} {486}},\ \bibinfo
  {pages} {1086} (\bibinfo {year} {2019})},\ \Eprint
  {http://arxiv.org/abs/1809.02063} {arXiv:1809.02063 [physics.data-an]}
  \BibitemShut {NoStop}%
\bibitem [{\citenamefont {Taylor}\ and\ \citenamefont
  {Gerosa}(2018)}]{Taylor:2018iat}%
  \BibitemOpen
  \bibfield  {author} {\bibinfo {author} {\bibfnamefont {S.~R.}\ \bibnamefont
  {Taylor}}\ and\ \bibinfo {author} {\bibfnamefont {D.}~\bibnamefont
  {Gerosa}},\ }\href {\doibase 10.1103/PhysRevD.98.083017} {\bibfield
  {journal} {\bibinfo  {journal} {Phys. Rev. D}\ }\textbf {\bibinfo {volume}
  {98}},\ \bibinfo {pages} {083017} (\bibinfo {year} {2018})},\ \Eprint
  {http://arxiv.org/abs/1806.08365} {arXiv:1806.08365 [astro-ph.HE]}
  \BibitemShut {NoStop}%
\bibitem [{\citenamefont {Mandic}\ \emph {et~al.}(2012)\citenamefont {Mandic},
  \citenamefont {Thrane}, \citenamefont {Giampanis},\ and\ \citenamefont
  {Regimbau}}]{Mandic:2012pj}%
  \BibitemOpen
  \bibfield  {author} {\bibinfo {author} {\bibfnamefont {V.}~\bibnamefont
  {Mandic}}, \bibinfo {author} {\bibfnamefont {E.}~\bibnamefont {Thrane}},
  \bibinfo {author} {\bibfnamefont {S.}~\bibnamefont {Giampanis}}, \ and\
  \bibinfo {author} {\bibfnamefont {T.}~\bibnamefont {Regimbau}},\ }\href
  {\doibase 10.1103/PhysRevLett.109.171102} {\bibfield  {journal} {\bibinfo
  {journal} {Phys. Rev. Lett.}\ }\textbf {\bibinfo {volume} {109}},\ \bibinfo
  {pages} {171102} (\bibinfo {year} {2012})},\ \Eprint
  {http://arxiv.org/abs/1209.3847} {arXiv:1209.3847 [astro-ph.CO]} \BibitemShut
  {NoStop}%
\bibitem [{\citenamefont {Callister}\ \emph {et~al.}(2017)\citenamefont
  {Callister}, \citenamefont {Biscoveanu}, \citenamefont {Christensen},
  \citenamefont {Isi}, \citenamefont {Matas}, \citenamefont {Minazzoli},
  \citenamefont {Regimbau}, \citenamefont {Sakellariadou}, \citenamefont
  {Tasson},\ and\ \citenamefont {Thrane}}]{Callister:2017ocg}%
  \BibitemOpen
  \bibfield  {author} {\bibinfo {author} {\bibfnamefont {T.}~\bibnamefont
  {Callister}}, \bibinfo {author} {\bibfnamefont {A.~S.}\ \bibnamefont
  {Biscoveanu}}, \bibinfo {author} {\bibfnamefont {N.}~\bibnamefont
  {Christensen}}, \bibinfo {author} {\bibfnamefont {M.}~\bibnamefont {Isi}},
  \bibinfo {author} {\bibfnamefont {A.}~\bibnamefont {Matas}}, \bibinfo
  {author} {\bibfnamefont {O.}~\bibnamefont {Minazzoli}}, \bibinfo {author}
  {\bibfnamefont {T.}~\bibnamefont {Regimbau}}, \bibinfo {author}
  {\bibfnamefont {M.}~\bibnamefont {Sakellariadou}}, \bibinfo {author}
  {\bibfnamefont {J.}~\bibnamefont {Tasson}}, \ and\ \bibinfo {author}
  {\bibfnamefont {E.}~\bibnamefont {Thrane}},\ }\href {\doibase
  10.1103/PhysRevX.7.041058} {\bibfield  {journal} {\bibinfo  {journal} {Phys.
  Rev. X}\ }\textbf {\bibinfo {volume} {7}},\ \bibinfo {pages} {041058}
  (\bibinfo {year} {2017})},\ \Eprint {http://arxiv.org/abs/1704.08373}
  {arXiv:1704.08373 [gr-qc]} \BibitemShut {NoStop}%
\bibitem [{\citenamefont {Ng}\ \emph {et~al.}(2021)\citenamefont {Ng},
  \citenamefont {Vitale}, \citenamefont {Farr},\ and\ \citenamefont
  {Rodriguez}}]{Ng:2020qpk}%
  \BibitemOpen
  \bibfield  {author} {\bibinfo {author} {\bibfnamefont {K.~K.~Y.}\
  \bibnamefont {Ng}}, \bibinfo {author} {\bibfnamefont {S.}~\bibnamefont
  {Vitale}}, \bibinfo {author} {\bibfnamefont {W.~M.}\ \bibnamefont {Farr}}, \
  and\ \bibinfo {author} {\bibfnamefont {C.~L.}\ \bibnamefont {Rodriguez}},\
  }\href {\doibase 10.3847/2041-8213/abf8be} {\bibfield  {journal} {\bibinfo
  {journal} {Astrophys. J. Lett.}\ }\textbf {\bibinfo {volume} {913}},\
  \bibinfo {pages} {L5} (\bibinfo {year} {2021})},\ \Eprint
  {http://arxiv.org/abs/2012.09876} {arXiv:2012.09876 [astro-ph.CO]}
  \BibitemShut {NoStop}%
\bibitem [{\citenamefont {Fishbach}\ \emph {et~al.}(2018)\citenamefont
  {Fishbach}, \citenamefont {Holz},\ and\ \citenamefont
  {Farr}}]{Fishbach:2018edt}%
  \BibitemOpen
  \bibfield  {author} {\bibinfo {author} {\bibfnamefont {M.}~\bibnamefont
  {Fishbach}}, \bibinfo {author} {\bibfnamefont {D.~E.}\ \bibnamefont {Holz}},
  \ and\ \bibinfo {author} {\bibfnamefont {W.~M.}\ \bibnamefont {Farr}},\
  }\href {\doibase 10.3847/2041-8213/aad800} {\bibfield  {journal} {\bibinfo
  {journal} {Astrophys. J. Lett.}\ }\textbf {\bibinfo {volume} {863}},\
  \bibinfo {pages} {L41} (\bibinfo {year} {2018})},\ \Eprint
  {http://arxiv.org/abs/1805.10270} {arXiv:1805.10270 [astro-ph.HE]}
  \BibitemShut {NoStop}%
\bibitem [{\citenamefont {Allen}\ and\ \citenamefont
  {Romano}(1999)}]{Allen:1997ad}%
  \BibitemOpen
  \bibfield  {author} {\bibinfo {author} {\bibfnamefont {B.}~\bibnamefont
  {Allen}}\ and\ \bibinfo {author} {\bibfnamefont {J.~D.}\ \bibnamefont
  {Romano}},\ }\href {\doibase 10.1103/PhysRevD.59.102001} {\bibfield
  {journal} {\bibinfo  {journal} {Phys. Rev. D}\ }\textbf {\bibinfo {volume}
  {59}},\ \bibinfo {pages} {102001} (\bibinfo {year} {1999})},\ \Eprint
  {http://arxiv.org/abs/gr-qc/9710117} {arXiv:gr-qc/9710117} \BibitemShut
  {NoStop}%
\bibitem [{\citenamefont {Abbott}\ \emph
  {et~al.}(2021{\natexlab{a}})\citenamefont {Abbott} \emph {et~al.}}]{O3stoch}%
  \BibitemOpen
  \bibfield  {author} {\bibinfo {author} {\bibfnamefont {R.}~\bibnamefont
  {Abbott}} \emph {et~al.} (\bibinfo {collaboration} {LIGO Scientific
  Collaboration, Virgo Collaboration, and KAGRA Collaboration}),\ }\href
  {\doibase 10.1103/PhysRevD.104.022004} {\bibfield  {journal} {\bibinfo
  {journal} {Phys. Rev. D}\ }\textbf {\bibinfo {volume} {104}},\ \bibinfo
  {pages} {022004} (\bibinfo {year} {2021}{\natexlab{a}})}\BibitemShut
  {NoStop}%
\bibitem [{\citenamefont {De~Renzis}\ \emph {et~al.}(2024)\citenamefont
  {De~Renzis}, \citenamefont {Iacovelli}, \citenamefont {Gerosa}, \citenamefont
  {Mancarella},\ and\ \citenamefont {Pacilio}}]{DeRenzis:2024dvx}%
  \BibitemOpen
  \bibfield  {author} {\bibinfo {author} {\bibfnamefont {V.}~\bibnamefont
  {De~Renzis}}, \bibinfo {author} {\bibfnamefont {F.}~\bibnamefont
  {Iacovelli}}, \bibinfo {author} {\bibfnamefont {D.}~\bibnamefont {Gerosa}},
  \bibinfo {author} {\bibfnamefont {M.}~\bibnamefont {Mancarella}}, \ and\
  \bibinfo {author} {\bibfnamefont {C.}~\bibnamefont {Pacilio}},\ }\href@noop
  {} {\  (\bibinfo {year} {2024})},\ \Eprint {http://arxiv.org/abs/2410.17325}
  {arXiv:2410.17325 [astro-ph.HE]} \BibitemShut {NoStop}%
\bibitem [{\citenamefont {Green}\ and\ \citenamefont
  {Gair}(2021)}]{Green:2020dnx}%
  \BibitemOpen
  \bibfield  {author} {\bibinfo {author} {\bibfnamefont {S.~R.}\ \bibnamefont
  {Green}}\ and\ \bibinfo {author} {\bibfnamefont {J.}~\bibnamefont {Gair}},\
  }\href {\doibase 10.1088/2632-2153/abfaed} {\bibfield  {journal} {\bibinfo
  {journal} {Mach. Learn. Sci. Tech.}\ }\textbf {\bibinfo {volume} {2}},\
  \bibinfo {pages} {03LT01} (\bibinfo {year} {2021})},\ \Eprint
  {http://arxiv.org/abs/2008.03312} {arXiv:2008.03312 [astro-ph.IM]}
  \BibitemShut {NoStop}%
\bibitem [{\citenamefont {Dax}\ \emph {et~al.}(2023)\citenamefont {Dax},
  \citenamefont {Green}, \citenamefont {Gair}, \citenamefont {P\"urrer},
  \citenamefont {Wildberger}, \citenamefont {Macke}, \citenamefont {Buonanno},\
  and\ \citenamefont {Sch\"olkopf}}]{Dax:2022pxd}%
  \BibitemOpen
  \bibfield  {author} {\bibinfo {author} {\bibfnamefont {M.}~\bibnamefont
  {Dax}}, \bibinfo {author} {\bibfnamefont {S.~R.}\ \bibnamefont {Green}},
  \bibinfo {author} {\bibfnamefont {J.}~\bibnamefont {Gair}}, \bibinfo {author}
  {\bibfnamefont {M.}~\bibnamefont {P\"urrer}}, \bibinfo {author}
  {\bibfnamefont {J.}~\bibnamefont {Wildberger}}, \bibinfo {author}
  {\bibfnamefont {J.~H.}\ \bibnamefont {Macke}}, \bibinfo {author}
  {\bibfnamefont {A.}~\bibnamefont {Buonanno}}, \ and\ \bibinfo {author}
  {\bibfnamefont {B.}~\bibnamefont {Sch\"olkopf}},\ }\href {\doibase
  10.1103/PhysRevLett.130.171403} {\bibfield  {journal} {\bibinfo  {journal}
  {Phys. Rev. Lett.}\ }\textbf {\bibinfo {volume} {130}},\ \bibinfo {pages}
  {171403} (\bibinfo {year} {2023})},\ \Eprint
  {http://arxiv.org/abs/2210.05686} {arXiv:2210.05686 [gr-qc]} \BibitemShut
  {NoStop}%
\bibitem [{\citenamefont {Wong}\ \emph {et~al.}(2023)\citenamefont {Wong},
  \citenamefont {Isi},\ and\ \citenamefont {Edwards}}]{Wong:2023lgb}%
  \BibitemOpen
  \bibfield  {author} {\bibinfo {author} {\bibfnamefont {K.~W.~K.}\
  \bibnamefont {Wong}}, \bibinfo {author} {\bibfnamefont {M.}~\bibnamefont
  {Isi}}, \ and\ \bibinfo {author} {\bibfnamefont {T.~D.~P.}\ \bibnamefont
  {Edwards}},\ }\href {\doibase 10.3847/1538-4357/acf5cd} {\bibfield  {journal}
  {\bibinfo  {journal} {Astrophys. J.}\ }\textbf {\bibinfo {volume} {958}},\
  \bibinfo {pages} {129} (\bibinfo {year} {2023})},\ \Eprint
  {http://arxiv.org/abs/2302.05333} {arXiv:2302.05333 [astro-ph.IM]}
  \BibitemShut {NoStop}%
\bibitem [{\citenamefont {Alvey}\ \emph {et~al.}(2023)\citenamefont {Alvey},
  \citenamefont {Bhardwaj}, \citenamefont {Nissanke},\ and\ \citenamefont
  {Weniger}}]{Alvey:2023naa}%
  \BibitemOpen
  \bibfield  {author} {\bibinfo {author} {\bibfnamefont {J.}~\bibnamefont
  {Alvey}}, \bibinfo {author} {\bibfnamefont {U.}~\bibnamefont {Bhardwaj}},
  \bibinfo {author} {\bibfnamefont {S.}~\bibnamefont {Nissanke}}, \ and\
  \bibinfo {author} {\bibfnamefont {C.}~\bibnamefont {Weniger}},\ }\href@noop
  {} {\  (\bibinfo {year} {2023})},\ \Eprint {http://arxiv.org/abs/2308.06318}
  {arXiv:2308.06318 [gr-qc]} \BibitemShut {NoStop}%
\bibitem [{\citenamefont {Martinovic}\ \emph {et~al.}(2022)\citenamefont
  {Martinovic}, \citenamefont {Perigois}, \citenamefont {Regimbau},\ and\
  \citenamefont {Sakellariadou}}]{Martinovic:2021fzj}%
  \BibitemOpen
  \bibfield  {author} {\bibinfo {author} {\bibfnamefont {K.}~\bibnamefont
  {Martinovic}}, \bibinfo {author} {\bibfnamefont {C.}~\bibnamefont
  {Perigois}}, \bibinfo {author} {\bibfnamefont {T.}~\bibnamefont {Regimbau}},
  \ and\ \bibinfo {author} {\bibfnamefont {M.}~\bibnamefont {Sakellariadou}},\
  }\href {\doibase 10.3847/1538-4357/ac9840} {\bibfield  {journal} {\bibinfo
  {journal} {Astrophys. J.}\ }\textbf {\bibinfo {volume} {940}},\ \bibinfo
  {pages} {29} (\bibinfo {year} {2022})},\ \Eprint
  {http://arxiv.org/abs/2109.09779} {arXiv:2109.09779 [astro-ph.SR]}
  \BibitemShut {NoStop}%
\bibitem [{\citenamefont {Mukherjee}\ \emph {et~al.}(2022)\citenamefont
  {Mukherjee}, \citenamefont {Meinema},\ and\ \citenamefont
  {Silk}}]{Mukherjee:2021itf}%
  \BibitemOpen
  \bibfield  {author} {\bibinfo {author} {\bibfnamefont {S.}~\bibnamefont
  {Mukherjee}}, \bibinfo {author} {\bibfnamefont {M.~S.~P.}\ \bibnamefont
  {Meinema}}, \ and\ \bibinfo {author} {\bibfnamefont {J.}~\bibnamefont
  {Silk}},\ }\href {\doibase 10.1093/mnras/stab3756} {\bibfield  {journal}
  {\bibinfo  {journal} {Mon. Not. Roy. Astron. Soc.}\ }\textbf {\bibinfo
  {volume} {510}},\ \bibinfo {pages} {6218} (\bibinfo {year} {2022})},\ \Eprint
  {http://arxiv.org/abs/2107.02181} {arXiv:2107.02181 [astro-ph.CO]}
  \BibitemShut {NoStop}%
\bibitem [{\citenamefont {Leyde}\ \emph {et~al.}(2024)\citenamefont {Leyde},
  \citenamefont {Green}, \citenamefont {Toubiana},\ and\ \citenamefont
  {Gair}}]{Leyde:2023iof}%
  \BibitemOpen
  \bibfield  {author} {\bibinfo {author} {\bibfnamefont {K.}~\bibnamefont
  {Leyde}}, \bibinfo {author} {\bibfnamefont {S.~R.}\ \bibnamefont {Green}},
  \bibinfo {author} {\bibfnamefont {A.}~\bibnamefont {Toubiana}}, \ and\
  \bibinfo {author} {\bibfnamefont {J.}~\bibnamefont {Gair}},\ }\href {\doibase
  10.1103/PhysRevD.109.064056} {\bibfield  {journal} {\bibinfo  {journal}
  {Phys. Rev. D}\ }\textbf {\bibinfo {volume} {109}},\ \bibinfo {pages}
  {064056} (\bibinfo {year} {2024})},\ \Eprint
  {http://arxiv.org/abs/2311.12093} {arXiv:2311.12093 [gr-qc]} \BibitemShut
  {NoStop}%
\bibitem [{\citenamefont {Mancarella}\ \emph {et~al.}(2024)\citenamefont
  {Mancarella}, \citenamefont {Iacovelli}, \citenamefont {Foffa}, \citenamefont
  {Muttoni},\ and\ \citenamefont {Maggiore}}]{Mancarella:2024qle}%
  \BibitemOpen
  \bibfield  {author} {\bibinfo {author} {\bibfnamefont {M.}~\bibnamefont
  {Mancarella}}, \bibinfo {author} {\bibfnamefont {F.}~\bibnamefont
  {Iacovelli}}, \bibinfo {author} {\bibfnamefont {S.}~\bibnamefont {Foffa}},
  \bibinfo {author} {\bibfnamefont {N.}~\bibnamefont {Muttoni}}, \ and\
  \bibinfo {author} {\bibfnamefont {M.}~\bibnamefont {Maggiore}},\ }\href@noop
  {} {\  (\bibinfo {year} {2024})},\ \Eprint {http://arxiv.org/abs/2405.02286}
  {arXiv:2405.02286 [astro-ph.HE]} \BibitemShut {NoStop}%
\bibitem [{\citenamefont {Finn}\ and\ \citenamefont
  {Chernoff}(1993)}]{Finn:1992xs}%
  \BibitemOpen
  \bibfield  {author} {\bibinfo {author} {\bibfnamefont {L.~S.}\ \bibnamefont
  {Finn}}\ and\ \bibinfo {author} {\bibfnamefont {D.~F.}\ \bibnamefont
  {Chernoff}},\ }\href {\doibase 10.1103/PhysRevD.47.2198} {\bibfield
  {journal} {\bibinfo  {journal} {Phys. Rev. D}\ }\textbf {\bibinfo {volume}
  {47}},\ \bibinfo {pages} {2198} (\bibinfo {year} {1993})},\ \Eprint
  {http://arxiv.org/abs/gr-qc/9301003} {arXiv:gr-qc/9301003} \BibitemShut
  {NoStop}%
\bibitem [{\citenamefont {Maggiore}(2007)}]{Maggiore:2007ulw}%
  \BibitemOpen
  \bibfield  {author} {\bibinfo {author} {\bibfnamefont {M.}~\bibnamefont
  {Maggiore}},\ }\href@noop {} {\emph {\bibinfo {title} {{Gravitational Waves.
  Vol. 1: Theory and Experiments}}}},\ Oxford Master Series in Physics\
  (\bibinfo  {publisher} {Oxford University Press},\ \bibinfo {year}
  {2007})\BibitemShut {NoStop}%
\bibitem [{\citenamefont {Iacovelli}\ \emph {et~al.}(2022)\citenamefont
  {Iacovelli}, \citenamefont {Mancarella}, \citenamefont {Foffa},\ and\
  \citenamefont {Maggiore}}]{Iacovelli:2022bbs}%
  \BibitemOpen
  \bibfield  {author} {\bibinfo {author} {\bibfnamefont {F.}~\bibnamefont
  {Iacovelli}}, \bibinfo {author} {\bibfnamefont {M.}~\bibnamefont
  {Mancarella}}, \bibinfo {author} {\bibfnamefont {S.}~\bibnamefont {Foffa}}, \
  and\ \bibinfo {author} {\bibfnamefont {M.}~\bibnamefont {Maggiore}},\ }\href
  {\doibase 10.3847/1538-4357/ac9cd4} {\bibfield  {journal} {\bibinfo
  {journal} {Astrophys. J.}\ }\textbf {\bibinfo {volume} {941}},\ \bibinfo
  {pages} {208} (\bibinfo {year} {2022})},\ \Eprint
  {http://arxiv.org/abs/2207.02771} {arXiv:2207.02771 [gr-qc]} \BibitemShut
  {NoStop}%
\bibitem [{\citenamefont {Vallisneri}(2008)}]{Vallisneri:2007ev}%
  \BibitemOpen
  \bibfield  {author} {\bibinfo {author} {\bibfnamefont {M.}~\bibnamefont
  {Vallisneri}},\ }\href {\doibase 10.1103/PhysRevD.77.042001} {\bibfield
  {journal} {\bibinfo  {journal} {Phys. Rev. D}\ }\textbf {\bibinfo {volume}
  {77}},\ \bibinfo {pages} {042001} (\bibinfo {year} {2008})},\ \Eprint
  {http://arxiv.org/abs/gr-qc/0703086} {arXiv:gr-qc/0703086} \BibitemShut
  {NoStop}%
\bibitem [{\citenamefont {Aghanim}\ \emph {et~al.}(2020)\citenamefont {Aghanim}
  \emph {et~al.}}]{Planck:2018vyg}%
  \BibitemOpen
  \bibfield  {author} {\bibinfo {author} {\bibfnamefont {N.}~\bibnamefont
  {Aghanim}} \emph {et~al.} (\bibinfo {collaboration} {Planck}),\ }\href
  {\doibase 10.1051/0004-6361/201833910} {\bibfield  {journal} {\bibinfo
  {journal} {Astron. Astrophys.}\ }\textbf {\bibinfo {volume} {641}},\ \bibinfo
  {pages} {A6} (\bibinfo {year} {2020})},\ \bibinfo {note} {[Erratum:
  Astron.Astrophys. 652, C4 (2021)]},\ \Eprint
  {http://arxiv.org/abs/1807.06209} {arXiv:1807.06209 [astro-ph.CO]}
  \BibitemShut {NoStop}%
\bibitem [{\citenamefont {Phinney}(2001)}]{Phinney:2001di}%
  \BibitemOpen
  \bibfield  {author} {\bibinfo {author} {\bibfnamefont {E.~S.}\ \bibnamefont
  {Phinney}},\ }\href@noop {} {\  (\bibinfo {year} {2001})},\ \Eprint
  {http://arxiv.org/abs/astro-ph/0108028} {arXiv:astro-ph/0108028} \BibitemShut
  {NoStop}%
\bibitem [{\citenamefont {Abbott}\ \emph
  {et~al.}(2021{\natexlab{b}})\citenamefont {Abbott} \emph
  {et~al.}}]{KAGRA:2021kbb}%
  \BibitemOpen
  \bibfield  {author} {\bibinfo {author} {\bibfnamefont {R.}~\bibnamefont
  {Abbott}} \emph {et~al.} (\bibinfo {collaboration} {KAGRA, Virgo, LIGO
  Scientific}),\ }\href {\doibase 10.1103/PhysRevD.104.022004} {\bibfield
  {journal} {\bibinfo  {journal} {Phys. Rev. D}\ }\textbf {\bibinfo {volume}
  {104}},\ \bibinfo {pages} {022004} (\bibinfo {year} {2021}{\natexlab{b}})},\
  \Eprint {http://arxiv.org/abs/2101.12130} {arXiv:2101.12130 [gr-qc]}
  \BibitemShut {NoStop}%
\bibitem [{\citenamefont {Dupletsa}\ \emph {et~al.}(2023)\citenamefont
  {Dupletsa}, \citenamefont {Harms}, \citenamefont {Banerjee}, \citenamefont
  {Branchesi}, \citenamefont {Goncharov}, \citenamefont {Maselli},
  \citenamefont {Oliveira}, \citenamefont {Ronchini},\ and\ \citenamefont
  {Tissino}}]{Dupletsa:2022scg}%
  \BibitemOpen
  \bibfield  {author} {\bibinfo {author} {\bibfnamefont {U.}~\bibnamefont
  {Dupletsa}}, \bibinfo {author} {\bibfnamefont {J.}~\bibnamefont {Harms}},
  \bibinfo {author} {\bibfnamefont {B.}~\bibnamefont {Banerjee}}, \bibinfo
  {author} {\bibfnamefont {M.}~\bibnamefont {Branchesi}}, \bibinfo {author}
  {\bibfnamefont {B.}~\bibnamefont {Goncharov}}, \bibinfo {author}
  {\bibfnamefont {A.}~\bibnamefont {Maselli}}, \bibinfo {author} {\bibfnamefont
  {A.~C.~S.}\ \bibnamefont {Oliveira}}, \bibinfo {author} {\bibfnamefont
  {S.}~\bibnamefont {Ronchini}}, \ and\ \bibinfo {author} {\bibfnamefont
  {J.}~\bibnamefont {Tissino}},\ }\href {\doibase 10.1016/j.ascom.2022.100671}
  {\bibfield  {journal} {\bibinfo  {journal} {Astron. Comput.}\ }\textbf
  {\bibinfo {volume} {42}},\ \bibinfo {pages} {100671} (\bibinfo {year}
  {2023})},\ \Eprint {http://arxiv.org/abs/2205.02499} {arXiv:2205.02499
  [gr-qc]} \BibitemShut {NoStop}%
\bibitem [{\citenamefont {Borhanian}(2021)}]{Borhanian:2020ypi}%
  \BibitemOpen
  \bibfield  {author} {\bibinfo {author} {\bibfnamefont {S.}~\bibnamefont
  {Borhanian}},\ }\href {\doibase 10.1088/1361-6382/ac1618} {\bibfield
  {journal} {\bibinfo  {journal} {Class. Quant. Grav.}\ }\textbf {\bibinfo
  {volume} {38}},\ \bibinfo {pages} {175014} (\bibinfo {year} {2021})},\
  \Eprint {http://arxiv.org/abs/2010.15202} {arXiv:2010.15202 [gr-qc]}
  \BibitemShut {NoStop}%
\end{thebibliography}%


\begin{thebibliography}{25}%
\makeatletter
\providecommand \@ifxundefined [1]{%
 \@ifx{#1\undefined}
}%
\providecommand \@ifnum [1]{%
 \ifnum #1\expandafter \@firstoftwo
 \else \expandafter \@secondoftwo
 \fi
}%
\providecommand \@ifx [1]{%
 \ifx #1\expandafter \@firstoftwo
 \else \expandafter \@secondoftwo
 \fi
}%
\providecommand \natexlab [1]{#1}%
\providecommand \enquote  [1]{``#1''}%
\providecommand \bibnamefont  [1]{#1}%
\providecommand \bibfnamefont [1]{#1}%
\providecommand \citenamefont [1]{#1}%
\providecommand \href@noop [0]{\@secondoftwo}%
\providecommand \href [0]{\begingroup \@sanitize@url \@href}%
\providecommand \@href[1]{\@@startlink{#1}\@@href}%
\providecommand \@@href[1]{\endgroup#1\@@endlink}%
\providecommand \@sanitize@url [0]{\catcode `\\12\catcode `\$12\catcode
  `\&12\catcode `\#12\catcode `\^12\catcode `\_12\catcode `\%12\relax}%
\providecommand \@@startlink[1]{}%
\providecommand \@@endlink[0]{}%
\providecommand \url  [0]{\begingroup\@sanitize@url \@url }%
\providecommand \@url [1]{\endgroup\@href {#1}{\urlprefix }}%
\providecommand \urlprefix  [0]{URL }%
\providecommand \Eprint [0]{\href }%
\providecommand \doibase [0]{http://dx.doi.org/}%
\providecommand \selectlanguage [0]{\@gobble}%
\providecommand \bibinfo  [0]{\@secondoftwo}%
\providecommand \bibfield  [0]{\@secondoftwo}%
\providecommand \translation [1]{[#1]}%
\providecommand \BibitemOpen [0]{}%
\providecommand \bibitemStop [0]{}%
\providecommand \bibitemNoStop [0]{.\EOS\space}%
\providecommand \EOS [0]{\spacefactor3000\relax}%
\providecommand \BibitemShut  [1]{\csname bibitem#1\endcsname}%
\let\auto@bib@innerbib\@empty
\bibitem [{\citenamefont {Zhong}\ \emph {et~al.}(2024)\citenamefont {Zhong},
  \citenamefont {Zhou}, \citenamefont {Reali}, \citenamefont {Berti},\ and\
  \citenamefont {Mandic}}]{Zhong:2024dss}%
  \BibitemOpen
  \bibfield  {author} {\bibinfo {author} {\bibfnamefont {H.}~\bibnamefont
  {Zhong}}, \bibinfo {author} {\bibfnamefont {B.}~\bibnamefont {Zhou}},
  \bibinfo {author} {\bibfnamefont {L.}~\bibnamefont {Reali}}, \bibinfo
  {author} {\bibfnamefont {E.}~\bibnamefont {Berti}}, \ and\ \bibinfo {author}
  {\bibfnamefont {V.}~\bibnamefont {Mandic}},\ }\href {\doibase
  10.1103/PhysRevD.110.064047} {\bibfield  {journal} {\bibinfo  {journal}
  {Phys. Rev. D}\ }\textbf {\bibinfo {volume} {110}},\ \bibinfo {pages}
  {064047} (\bibinfo {year} {2024})},\ \Eprint
  {http://arxiv.org/abs/2406.10757} {arXiv:2406.10757 [gr-qc]} \BibitemShut
  {NoStop}%
\bibitem [{\citenamefont {Abbott}\ \emph {et~al.}(2021)\citenamefont {Abbott}
  \emph {et~al.}}]{KAGRA:2021kbb}%
  \BibitemOpen
  \bibfield  {author} {\bibinfo {author} {\bibfnamefont {R.}~\bibnamefont
  {Abbott}} \emph {et~al.} (\bibinfo {collaboration} {KAGRA, Virgo, LIGO
  Scientific}),\ }\href {\doibase 10.1103/PhysRevD.104.022004} {\bibfield
  {journal} {\bibinfo  {journal} {Phys. Rev. D}\ }\textbf {\bibinfo {volume}
  {104}},\ \bibinfo {pages} {022004} (\bibinfo {year} {2021})},\ \Eprint
  {http://arxiv.org/abs/2101.12130} {arXiv:2101.12130 [gr-qc]} \BibitemShut
  {NoStop}%
\bibitem [{\citenamefont {Allen}\ and\ \citenamefont
  {Romano}(1999)}]{Allen:1997ad}%
  \BibitemOpen
  \bibfield  {author} {\bibinfo {author} {\bibfnamefont {B.}~\bibnamefont
  {Allen}}\ and\ \bibinfo {author} {\bibfnamefont {J.~D.}\ \bibnamefont
  {Romano}},\ }\href {\doibase 10.1103/PhysRevD.59.102001} {\bibfield
  {journal} {\bibinfo  {journal} {Phys. Rev. D}\ }\textbf {\bibinfo {volume}
  {59}},\ \bibinfo {pages} {102001} (\bibinfo {year} {1999})},\ \Eprint
  {http://arxiv.org/abs/gr-qc/9710117} {arXiv:gr-qc/9710117} \BibitemShut
  {NoStop}%
\bibitem [{\citenamefont {Finn}(1992)}]{Finn:1992wt}%
  \BibitemOpen
  \bibfield  {author} {\bibinfo {author} {\bibfnamefont {L.~S.}\ \bibnamefont
  {Finn}},\ }\href {\doibase 10.1103/PhysRevD.46.5236} {\bibfield  {journal}
  {\bibinfo  {journal} {Phys. Rev. D}\ }\textbf {\bibinfo {volume} {46}},\
  \bibinfo {pages} {5236} (\bibinfo {year} {1992})},\ \Eprint
  {http://arxiv.org/abs/gr-qc/9209010} {arXiv:gr-qc/9209010} \BibitemShut
  {NoStop}%
\bibitem [{\citenamefont {Borhanian}(2021)}]{Borhanian:2020ypi}%
  \BibitemOpen
  \bibfield  {author} {\bibinfo {author} {\bibfnamefont {S.}~\bibnamefont
  {Borhanian}},\ }\href {\doibase 10.1088/1361-6382/ac1618} {\bibfield
  {journal} {\bibinfo  {journal} {Class. Quant. Grav.}\ }\textbf {\bibinfo
  {volume} {38}},\ \bibinfo {pages} {175014} (\bibinfo {year} {2021})},\
  \Eprint {http://arxiv.org/abs/2010.15202} {arXiv:2010.15202 [gr-qc]}
  \BibitemShut {NoStop}%
\bibitem [{\citenamefont {Zhong}\ \emph {et~al.}(2023)\citenamefont {Zhong},
  \citenamefont {Ormiston},\ and\ \citenamefont {Mandic}}]{Zhong:2022ylh}%
  \BibitemOpen
  \bibfield  {author} {\bibinfo {author} {\bibfnamefont {H.}~\bibnamefont
  {Zhong}}, \bibinfo {author} {\bibfnamefont {R.}~\bibnamefont {Ormiston}}, \
  and\ \bibinfo {author} {\bibfnamefont {V.}~\bibnamefont {Mandic}},\ }\href
  {\doibase 10.1103/PhysRevD.107.064048} {\bibfield  {journal} {\bibinfo
  {journal} {Phys. Rev. D}\ }\textbf {\bibinfo {volume} {107}},\ \bibinfo
  {pages} {064048} (\bibinfo {year} {2023})},\ \bibinfo {note} {[Erratum:
  Phys.Rev.D 108, 089902 (2023)]},\ \Eprint {http://arxiv.org/abs/2209.11877}
  {arXiv:2209.11877 [gr-qc]} \BibitemShut {NoStop}%
\bibitem [{\citenamefont {Callister}\ \emph {et~al.}(2020)\citenamefont
  {Callister}, \citenamefont {Fishbach}, \citenamefont {Holz},\ and\
  \citenamefont {Farr}}]{Callister:2020arv}%
  \BibitemOpen
  \bibfield  {author} {\bibinfo {author} {\bibfnamefont {T.}~\bibnamefont
  {Callister}}, \bibinfo {author} {\bibfnamefont {M.}~\bibnamefont {Fishbach}},
  \bibinfo {author} {\bibfnamefont {D.}~\bibnamefont {Holz}}, \ and\ \bibinfo
  {author} {\bibfnamefont {W.}~\bibnamefont {Farr}},\ }\href {\doibase
  10.3847/2041-8213/ab9743} {\bibfield  {journal} {\bibinfo  {journal}
  {Astrophys. J. Lett.}\ }\textbf {\bibinfo {volume} {896}},\ \bibinfo {pages}
  {L32} (\bibinfo {year} {2020})},\ \Eprint {http://arxiv.org/abs/2003.12152}
  {arXiv:2003.12152 [astro-ph.HE]} \BibitemShut {NoStop}%
\bibitem [{\citenamefont {Mandel}\ \emph {et~al.}(2019)\citenamefont {Mandel},
  \citenamefont {Farr},\ and\ \citenamefont {Gair}}]{Mandel:2018mve}%
  \BibitemOpen
  \bibfield  {author} {\bibinfo {author} {\bibfnamefont {I.}~\bibnamefont
  {Mandel}}, \bibinfo {author} {\bibfnamefont {W.~M.}\ \bibnamefont {Farr}}, \
  and\ \bibinfo {author} {\bibfnamefont {J.~R.}\ \bibnamefont {Gair}},\ }\href
  {\doibase 10.1093/mnras/stz896} {\bibfield  {journal} {\bibinfo  {journal}
  {Mon. Not. Roy. Astron. Soc.}\ }\textbf {\bibinfo {volume} {486}},\ \bibinfo
  {pages} {1086} (\bibinfo {year} {2019})},\ \Eprint
  {http://arxiv.org/abs/1809.02063} {arXiv:1809.02063 [physics.data-an]}
  \BibitemShut {NoStop}%
\bibitem [{\citenamefont {Taylor}\ and\ \citenamefont
  {Gerosa}(2018)}]{Taylor:2018iat}%
  \BibitemOpen
  \bibfield  {author} {\bibinfo {author} {\bibfnamefont {S.~R.}\ \bibnamefont
  {Taylor}}\ and\ \bibinfo {author} {\bibfnamefont {D.}~\bibnamefont
  {Gerosa}},\ }\href {\doibase 10.1103/PhysRevD.98.083017} {\bibfield
  {journal} {\bibinfo  {journal} {Phys. Rev. D}\ }\textbf {\bibinfo {volume}
  {98}},\ \bibinfo {pages} {083017} (\bibinfo {year} {2018})},\ \Eprint
  {http://arxiv.org/abs/1806.08365} {arXiv:1806.08365 [astro-ph.HE]}
  \BibitemShut {NoStop}%
\bibitem [{\citenamefont {Mandic}\ \emph {et~al.}(2012)\citenamefont {Mandic},
  \citenamefont {Thrane}, \citenamefont {Giampanis},\ and\ \citenamefont
  {Regimbau}}]{Mandic:2012pj}%
  \BibitemOpen
  \bibfield  {author} {\bibinfo {author} {\bibfnamefont {V.}~\bibnamefont
  {Mandic}}, \bibinfo {author} {\bibfnamefont {E.}~\bibnamefont {Thrane}},
  \bibinfo {author} {\bibfnamefont {S.}~\bibnamefont {Giampanis}}, \ and\
  \bibinfo {author} {\bibfnamefont {T.}~\bibnamefont {Regimbau}},\ }\href
  {\doibase 10.1103/PhysRevLett.109.171102} {\bibfield  {journal} {\bibinfo
  {journal} {Phys. Rev. Lett.}\ }\textbf {\bibinfo {volume} {109}},\ \bibinfo
  {pages} {171102} (\bibinfo {year} {2012})},\ \Eprint
  {http://arxiv.org/abs/1209.3847} {arXiv:1209.3847 [astro-ph.CO]} \BibitemShut
  {NoStop}%
\bibitem [{\citenamefont {Callister}\ \emph {et~al.}(2017)\citenamefont
  {Callister}, \citenamefont {Biscoveanu}, \citenamefont {Christensen},
  \citenamefont {Isi}, \citenamefont {Matas}, \citenamefont {Minazzoli},
  \citenamefont {Regimbau}, \citenamefont {Sakellariadou}, \citenamefont
  {Tasson},\ and\ \citenamefont {Thrane}}]{Callister:2017ocg}%
  \BibitemOpen
  \bibfield  {author} {\bibinfo {author} {\bibfnamefont {T.}~\bibnamefont
  {Callister}}, \bibinfo {author} {\bibfnamefont {A.~S.}\ \bibnamefont
  {Biscoveanu}}, \bibinfo {author} {\bibfnamefont {N.}~\bibnamefont
  {Christensen}}, \bibinfo {author} {\bibfnamefont {M.}~\bibnamefont {Isi}},
  \bibinfo {author} {\bibfnamefont {A.}~\bibnamefont {Matas}}, \bibinfo
  {author} {\bibfnamefont {O.}~\bibnamefont {Minazzoli}}, \bibinfo {author}
  {\bibfnamefont {T.}~\bibnamefont {Regimbau}}, \bibinfo {author}
  {\bibfnamefont {M.}~\bibnamefont {Sakellariadou}}, \bibinfo {author}
  {\bibfnamefont {J.}~\bibnamefont {Tasson}}, \ and\ \bibinfo {author}
  {\bibfnamefont {E.}~\bibnamefont {Thrane}},\ }\href {\doibase
  10.1103/PhysRevX.7.041058} {\bibfield  {journal} {\bibinfo  {journal} {Phys.
  Rev. X}\ }\textbf {\bibinfo {volume} {7}},\ \bibinfo {pages} {041058}
  (\bibinfo {year} {2017})},\ \Eprint {http://arxiv.org/abs/1704.08373}
  {arXiv:1704.08373 [gr-qc]} \BibitemShut {NoStop}%
\bibitem [{\citenamefont {Phinney}(2001)}]{Phinney:2001di}%
  \BibitemOpen
  \bibfield  {author} {\bibinfo {author} {\bibfnamefont {E.~S.}\ \bibnamefont
  {Phinney}},\ }\href@noop {} {\  (\bibinfo {year} {2001})},\ \Eprint
  {http://arxiv.org/abs/astro-ph/0108028} {arXiv:astro-ph/0108028} \BibitemShut
  {NoStop}%
\bibitem [{\citenamefont {Fishbach}\ \emph {et~al.}(2018)\citenamefont
  {Fishbach}, \citenamefont {Holz},\ and\ \citenamefont
  {Farr}}]{Fishbach:2018edt}%
  \BibitemOpen
  \bibfield  {author} {\bibinfo {author} {\bibfnamefont {M.}~\bibnamefont
  {Fishbach}}, \bibinfo {author} {\bibfnamefont {D.~E.}\ \bibnamefont {Holz}},
  \ and\ \bibinfo {author} {\bibfnamefont {W.~M.}\ \bibnamefont {Farr}},\
  }\href {\doibase 10.3847/2041-8213/aad800} {\bibfield  {journal} {\bibinfo
  {journal} {Astrophys. J. Lett.}\ }\textbf {\bibinfo {volume} {863}},\
  \bibinfo {pages} {L41} (\bibinfo {year} {2018})},\ \Eprint
  {http://arxiv.org/abs/1805.10270} {arXiv:1805.10270 [astro-ph.HE]}
  \BibitemShut {NoStop}%
\bibitem [{\citenamefont {Maggiore}(2007)}]{Maggiore:2007ulw}%
  \BibitemOpen
  \bibfield  {author} {\bibinfo {author} {\bibfnamefont {M.}~\bibnamefont
  {Maggiore}},\ }\href {\doibase 10.1093/acprof:oso/9780198570745.001.0001}
  {\emph {\bibinfo {title} {{Gravitational Waves. Vol. 1: Theory and
  Experiments}}}}\ (\bibinfo  {publisher} {Oxford University Press},\ \bibinfo
  {year} {2007})\BibitemShut {NoStop}%
\bibitem [{\citenamefont {Finn}\ and\ \citenamefont
  {Chernoff}(1993)}]{Finn:1992xs}%
  \BibitemOpen
  \bibfield  {author} {\bibinfo {author} {\bibfnamefont {L.~S.}\ \bibnamefont
  {Finn}}\ and\ \bibinfo {author} {\bibfnamefont {D.~F.}\ \bibnamefont
  {Chernoff}},\ }\href {\doibase 10.1103/PhysRevD.47.2198} {\bibfield
  {journal} {\bibinfo  {journal} {Phys. Rev. D}\ }\textbf {\bibinfo {volume}
  {47}},\ \bibinfo {pages} {2198} (\bibinfo {year} {1993})},\ \Eprint
  {http://arxiv.org/abs/gr-qc/9301003} {arXiv:gr-qc/9301003} \BibitemShut
  {NoStop}%
\bibitem [{\citenamefont {Aghanim}\ \emph {et~al.}(2020)\citenamefont {Aghanim}
  \emph {et~al.}}]{Planck:2018vyg}%
  \BibitemOpen
  \bibfield  {author} {\bibinfo {author} {\bibfnamefont {N.}~\bibnamefont
  {Aghanim}} \emph {et~al.} (\bibinfo {collaboration} {Planck}),\ }\href
  {\doibase 10.1051/0004-6361/201833910} {\bibfield  {journal} {\bibinfo
  {journal} {Astron. Astrophys.}\ }\textbf {\bibinfo {volume} {641}},\ \bibinfo
  {pages} {A6} (\bibinfo {year} {2020})},\ \bibinfo {note} {[Erratum:
  Astron.Astrophys. 652, C4 (2021)]},\ \Eprint
  {http://arxiv.org/abs/1807.06209} {arXiv:1807.06209 [astro-ph.CO]}
  \BibitemShut {NoStop}%
\bibitem [{\citenamefont {Iacovelli}\ \emph {et~al.}(2022)\citenamefont
  {Iacovelli}, \citenamefont {Mancarella}, \citenamefont {Foffa},\ and\
  \citenamefont {Maggiore}}]{Iacovelli:2022bbs}%
  \BibitemOpen
  \bibfield  {author} {\bibinfo {author} {\bibfnamefont {F.}~\bibnamefont
  {Iacovelli}}, \bibinfo {author} {\bibfnamefont {M.}~\bibnamefont
  {Mancarella}}, \bibinfo {author} {\bibfnamefont {S.}~\bibnamefont {Foffa}}, \
  and\ \bibinfo {author} {\bibfnamefont {M.}~\bibnamefont {Maggiore}},\ }\href
  {\doibase 10.3847/1538-4357/ac9cd4} {\bibfield  {journal} {\bibinfo
  {journal} {Astrophys. J.}\ }\textbf {\bibinfo {volume} {941}},\ \bibinfo
  {pages} {208} (\bibinfo {year} {2022})},\ \Eprint
  {http://arxiv.org/abs/2207.02771} {arXiv:2207.02771 [gr-qc]} \BibitemShut
  {NoStop}%
\bibitem [{\citenamefont {Vallisneri}(2008)}]{Vallisneri:2007ev}%
  \BibitemOpen
  \bibfield  {author} {\bibinfo {author} {\bibfnamefont {M.}~\bibnamefont
  {Vallisneri}},\ }\href {\doibase 10.1103/PhysRevD.77.042001} {\bibfield
  {journal} {\bibinfo  {journal} {Phys. Rev. D}\ }\textbf {\bibinfo {volume}
  {77}},\ \bibinfo {pages} {042001} (\bibinfo {year} {2008})},\ \Eprint
  {http://arxiv.org/abs/gr-qc/0703086} {arXiv:gr-qc/0703086} \BibitemShut
  {NoStop}%
\bibitem [{\citenamefont {Mancarella}\ \emph {et~al.}(2024)\citenamefont
  {Mancarella}, \citenamefont {Iacovelli}, \citenamefont {Foffa}, \citenamefont
  {Muttoni},\ and\ \citenamefont {Maggiore}}]{Mancarella:2024qle}%
  \BibitemOpen
  \bibfield  {author} {\bibinfo {author} {\bibfnamefont {M.}~\bibnamefont
  {Mancarella}}, \bibinfo {author} {\bibfnamefont {F.}~\bibnamefont
  {Iacovelli}}, \bibinfo {author} {\bibfnamefont {S.}~\bibnamefont {Foffa}},
  \bibinfo {author} {\bibfnamefont {N.}~\bibnamefont {Muttoni}}, \ and\
  \bibinfo {author} {\bibfnamefont {M.}~\bibnamefont {Maggiore}},\ }\href@noop
  {} {\  (\bibinfo {year} {2024})},\ \Eprint {http://arxiv.org/abs/2405.02286}
  {arXiv:2405.02286 [astro-ph.HE]} \BibitemShut {NoStop}%
\bibitem [{\citenamefont {Dupletsa}\ \emph {et~al.}(2023)\citenamefont
  {Dupletsa}, \citenamefont {Harms}, \citenamefont {Banerjee}, \citenamefont
  {Branchesi}, \citenamefont {Goncharov}, \citenamefont {Maselli},
  \citenamefont {Oliveira}, \citenamefont {Ronchini},\ and\ \citenamefont
  {Tissino}}]{Dupletsa:2022scg}%
  \BibitemOpen
  \bibfield  {author} {\bibinfo {author} {\bibfnamefont {U.}~\bibnamefont
  {Dupletsa}}, \bibinfo {author} {\bibfnamefont {J.}~\bibnamefont {Harms}},
  \bibinfo {author} {\bibfnamefont {B.}~\bibnamefont {Banerjee}}, \bibinfo
  {author} {\bibfnamefont {M.}~\bibnamefont {Branchesi}}, \bibinfo {author}
  {\bibfnamefont {B.}~\bibnamefont {Goncharov}}, \bibinfo {author}
  {\bibfnamefont {A.}~\bibnamefont {Maselli}}, \bibinfo {author} {\bibfnamefont
  {A.~C.~S.}\ \bibnamefont {Oliveira}}, \bibinfo {author} {\bibfnamefont
  {S.}~\bibnamefont {Ronchini}}, \ and\ \bibinfo {author} {\bibfnamefont
  {J.}~\bibnamefont {Tissino}},\ }\href {\doibase 10.1016/j.ascom.2022.100671}
  {\bibfield  {journal} {\bibinfo  {journal} {Astron. Comput.}\ }\textbf
  {\bibinfo {volume} {42}},\ \bibinfo {pages} {100671} (\bibinfo {year}
  {2023})},\ \Eprint {http://arxiv.org/abs/2205.02499} {arXiv:2205.02499
  [gr-qc]} \BibitemShut {NoStop}%
\bibitem [{\citenamefont {Green}\ and\ \citenamefont
  {Gair}(2021)}]{Green:2020dnx}%
  \BibitemOpen
  \bibfield  {author} {\bibinfo {author} {\bibfnamefont {S.~R.}\ \bibnamefont
  {Green}}\ and\ \bibinfo {author} {\bibfnamefont {J.}~\bibnamefont {Gair}},\
  }\href {\doibase 10.1088/2632-2153/abfaed} {\bibfield  {journal} {\bibinfo
  {journal} {Mach. Learn. Sci. Tech.}\ }\textbf {\bibinfo {volume} {2}},\
  \bibinfo {pages} {03LT01} (\bibinfo {year} {2021})},\ \Eprint
  {http://arxiv.org/abs/2008.03312} {arXiv:2008.03312 [astro-ph.IM]}
  \BibitemShut {NoStop}%
\bibitem [{\citenamefont {Dax}\ \emph {et~al.}(2023)\citenamefont {Dax},
  \citenamefont {Green}, \citenamefont {Gair}, \citenamefont {P\"urrer},
  \citenamefont {Wildberger}, \citenamefont {Macke}, \citenamefont {Buonanno},\
  and\ \citenamefont {Sch\"olkopf}}]{Dax:2022pxd}%
  \BibitemOpen
  \bibfield  {author} {\bibinfo {author} {\bibfnamefont {M.}~\bibnamefont
  {Dax}}, \bibinfo {author} {\bibfnamefont {S.~R.}\ \bibnamefont {Green}},
  \bibinfo {author} {\bibfnamefont {J.}~\bibnamefont {Gair}}, \bibinfo {author}
  {\bibfnamefont {M.}~\bibnamefont {P\"urrer}}, \bibinfo {author}
  {\bibfnamefont {J.}~\bibnamefont {Wildberger}}, \bibinfo {author}
  {\bibfnamefont {J.~H.}\ \bibnamefont {Macke}}, \bibinfo {author}
  {\bibfnamefont {A.}~\bibnamefont {Buonanno}}, \ and\ \bibinfo {author}
  {\bibfnamefont {B.}~\bibnamefont {Sch\"olkopf}},\ }\href {\doibase
  10.1103/PhysRevLett.130.171403} {\bibfield  {journal} {\bibinfo  {journal}
  {Phys. Rev. Lett.}\ }\textbf {\bibinfo {volume} {130}},\ \bibinfo {pages}
  {171403} (\bibinfo {year} {2023})},\ \Eprint
  {http://arxiv.org/abs/2210.05686} {arXiv:2210.05686 [gr-qc]} \BibitemShut
  {NoStop}%
\bibitem [{\citenamefont {Wong}\ \emph {et~al.}(2023)\citenamefont {Wong},
  \citenamefont {Isi},\ and\ \citenamefont {Edwards}}]{Wong:2023lgb}%
  \BibitemOpen
  \bibfield  {author} {\bibinfo {author} {\bibfnamefont {K.~W.~K.}\
  \bibnamefont {Wong}}, \bibinfo {author} {\bibfnamefont {M.}~\bibnamefont
  {Isi}}, \ and\ \bibinfo {author} {\bibfnamefont {T.~D.~P.}\ \bibnamefont
  {Edwards}},\ }\href {\doibase 10.3847/1538-4357/acf5cd} {\bibfield  {journal}
  {\bibinfo  {journal} {Astrophys. J.}\ }\textbf {\bibinfo {volume} {958}},\
  \bibinfo {pages} {129} (\bibinfo {year} {2023})},\ \Eprint
  {http://arxiv.org/abs/2302.05333} {arXiv:2302.05333 [astro-ph.IM]}
  \BibitemShut {NoStop}%
\bibitem [{\citenamefont {Alvey}\ \emph {et~al.}(2023)\citenamefont {Alvey},
  \citenamefont {Bhardwaj}, \citenamefont {Nissanke},\ and\ \citenamefont
  {Weniger}}]{Alvey:2023naa}%
  \BibitemOpen
  \bibfield  {author} {\bibinfo {author} {\bibfnamefont {J.}~\bibnamefont
  {Alvey}}, \bibinfo {author} {\bibfnamefont {U.}~\bibnamefont {Bhardwaj}},
  \bibinfo {author} {\bibfnamefont {S.}~\bibnamefont {Nissanke}}, \ and\
  \bibinfo {author} {\bibfnamefont {C.}~\bibnamefont {Weniger}},\ }\href@noop
  {} {\  (\bibinfo {year} {2023})},\ \Eprint {http://arxiv.org/abs/2308.06318}
  {arXiv:2308.06318 [gr-qc]} \BibitemShut {NoStop}%
\bibitem [{\citenamefont {Leyde}\ \emph {et~al.}(2024)\citenamefont {Leyde},
  \citenamefont {Green}, \citenamefont {Toubiana},\ and\ \citenamefont
  {Gair}}]{Leyde:2023iof}%
  \BibitemOpen
  \bibfield  {author} {\bibinfo {author} {\bibfnamefont {K.}~\bibnamefont
  {Leyde}}, \bibinfo {author} {\bibfnamefont {S.~R.}\ \bibnamefont {Green}},
  \bibinfo {author} {\bibfnamefont {A.}~\bibnamefont {Toubiana}}, \ and\
  \bibinfo {author} {\bibfnamefont {J.}~\bibnamefont {Gair}},\ }\href {\doibase
  10.1103/PhysRevD.109.064056} {\bibfield  {journal} {\bibinfo  {journal}
  {Phys. Rev. D}\ }\textbf {\bibinfo {volume} {109}},\ \bibinfo {pages}
  {064056} (\bibinfo {year} {2024})},\ \Eprint
  {http://arxiv.org/abs/2311.12093} {arXiv:2311.12093 [gr-qc]} \BibitemShut
  {NoStop}%
\end{thebibliography}%
\nocite{Leyde:2023iof,Mancarella:2024qle,Finn:1992xs,Maggiore:2007ulw, Iacovelli:2022bbs,Vallisneri:2007ev,Planck:2018vyg,Phinney:2001di,KAGRA:2021kbb,Dupletsa:2022scg,Borhanian:2020ypi}
\end{document}


\maketitle
\begin{acronym}
    \acro{GW}{gravitational wave}
    \acro{PSD}{power spectral density}
    \acro{GR}{general relativity}
    \acro{CBC}{compact binary coalescence}
    \acro{BH}{black hole}
    \acro{BBH}{binary black hole}
    \acro{BNS}{binary neutron star}
    \acro{NSBH}{neutron star-black hole}
    \acro{SFR}{star formation rate}
    \acro{LVK}{LIGO-Virgo-KAGRA}
    \acro{ET}{Einstein Telescope}
    \acro{CE}{Cosmic Explorer}
    \acro{PE}{parameter estimation}
    \acro{SGWB}{stochastic gravitational wave background}
    \acro{CGWB}{cosmological gravitational wave background}
    \acro{XG}{next generation}
    \acro{ML}{Machine Learning}
    \acro{DL}{Deep Learning}
    \acro{NN}{Neural Network}
    \acro{SNR}{signal-to-noise ratio}
    \acro{PDF}{probability density function}
    \acro{CDF}{cumulative density function}
\end{acronym}
\begin{center}
  \textbf{\large Supplemental Material:\\Two-Step Procedure to Detect Cosmological Gravitational Wave Backgrounds\\with Next-Generation Terrestrial Gravitational-Wave Detectors}
\end{center}
\setcounter{equation}{0}
\setcounter{figure}{0}
\setcounter{table}{0}
\setcounter{page}{1}
\makeatletter
\renewcommand{\theequation}{S\arabic{equation}}
\renewcommand{\thefigure}{S\arabic{figure}}
\renewcommand{\bibnumfmt}[1]{[S#1]}
\renewcommand{\citenumfont}[1]{S#1}

\section{Notching procedure}

In this section, we define the cross-correlation statistics we adopt for the stochastic search and briefly outline the notching procedure of Ref.~\cite{Zhong:2024dss}, which we use to remove all the detected \ac{BBH} signals from data.

\subsection{Stochastic search}
\label{subseq:stochsearch}

Let us consider two \ac{GW} detectors $I, J$ with strain time series $h_I(t)$, $h_J(t)$. To perform the cross-correlation search for the \ac{SGWB}, we split the strains into time segments $t_i$ of duration $T=4\,\rm{s}$. For each frequency bin $f_j$, we denote the Fourier transforms of these segments by $\tilde{h}_{I,J}(t_i;f_j)$ and their complex conjugates with an asterisk. For the baseline of these two detectors, one can define the cross-correlation statistics $\hat{C}_{IJ}(t_i;f_j)$ and the corresponding variance estimator $\hat{\sigma}_{IJ}^2$ as~\cite{KAGRA:2021kbb, Allen:1997ad}
\begin{eqnarray}
    \hat{C}_{IJ}(t_i;f_j) & = & \Big( \frac{20\pi^2f_j^3}{3H_0^2T} \Big) \frac{\Re[{\tilde{h}_{\rm{I}}^*(t_i;f_j)\tilde{h}_{\rm{J}}(t_i;f_j)}]} {\gamma_{IJ}(f_j)}\,, \nonumber \\
    \hat{\sigma}_{IJ}^2(t_i;f_j) & = & \Big( \frac{20\pi^2f_j^3}{3H_0^2} \Big)^2 \frac{P_{n_{\rm{I}}}(t_i;f_j)P_{n_{\rm{J}}}(t_i;f_j)} {8T\Delta f\gamma_{\rm{IJ}}^2(f_j)}\,.
    \label{eq:stoch}
\end{eqnarray}
Here $\gamma_{\mathrm{IJ}}(f_j)$ is the overlap reduction function between the two detectors~\cite{Allen:1997ad}, $\Delta f=0.25$\,Hz is the Fourier transform resolution, $P_{nI}(t_i;f_j)$ is the \ac{PSD} of the Ith detector at time $t_i$, and $H_0$ is the Hubble constant. The time-averaged frequency-domain spectrum  $\hat{C}_{IJ}(f_j)$ and its variance $\hat{\sigma}_{IJ}^2(f_j)$ can then be obtained by performing a weighted average over all time segments:
\begin{eqnarray}
\hat{C}_{IJ}(f_j) & = & \frac{\displaystyle{\sum_{i} \hat{C}_{IJ}(t_i;f_j) \hat{\sigma}_{IJ}^{-2}(t_i;f_j)}} {\displaystyle{\sum_{i} \hat{\sigma}_{IJ}^{-2}(t_i;f_j)}}\,,\nonumber\\
\hat{\sigma}_{IJ}^{-2}(f_j) & = & \sum_{i} \frac{1}{\hat{\sigma}^2_{IJ}(t_i;f_j)}\,.
\label{eq:combine_t}
\end{eqnarray}
With a network of more than two detectors, these spectra can be further combined by computing weighted averages among all the possible detector pairs. As mentioned in the main text however, in this study we perform the cross-correlation search using only a baseline of two CE detectors, since including all the other possible baselines would only lead to a $\lesssim 10\%$ improvement~\cite{Zhong:2024dss}. 
Throughout this work, we denote the frequency domain cross-correlation spectrum and its variance from this baseline {\em before notching} with $\hat{C}(f)$ and $\sigma^2(f)$, respectively.

\subsection{Mask definition}
\label{subsec:notching}

We notch out all the resolved \ac{BBH} signals in our dataset by applying the time-frequency approach of Ref.~\cite{Zhong:2024dss}. We consider a signal $h$ to be individually resolved if its network \ac{SNR}
\begin{equation}
    \rho = \sqrt{\sum_{J=1}^{N_\mathrm{det}} (h|h)_J}
\end{equation}
is above a certain threshold, where $(\cdot|\cdot)$ denotes the usual signal inner product
\begin{equation}
\left(a|b\right) = 4\,\mathrm{Re} \int_0^{\infty}\frac{a(f) b^*(f)}{P_n(f)}\,\D f \,,
\label{eq_fisher}
\end{equation}
and the sum index $J$ runs over all the $N_\mathrm{det}$ detectors in the network. For \acp{BBH}, we choose a detection threshold of $\rho_\mathrm{thr}^\mathrm{BBH}=8$~\cite{Zhong:2024dss}. We estimate the impact of the imperfect recovery of the signals on the notching procedure via their Fisher matrix~\cite{Finn:1992wt}
\begin{equation}
\Gamma_{\alpha\beta} = \sum_{J=1}^{N_\mathrm{det}}\left(\frac{\partial h}{\partial\theta^\alpha} \middle| \frac{\partial h}{\partial\theta^\beta}\right)_J \,,
\label{eq:fisher}
\end{equation}
with $\theta^\alpha$ denoting the source parameters. We consider a set of 9 parameters ignoring spin of all \acp{CBC}
\begin{equation}
\bm{\theta}=
\left\{ 
\ln\left(\frac{\mathcal{M}^c_z}{M_\odot}\right), 
\eta, 
\ln\left(\frac{d_L}{\text{Mpc}}\right), 
\cos\iota, 
\cos\delta,
\alpha, 
\psi,
\phi_{c}, 
t_{c} 
\right\} \,.
\label{eq:info_paras}
\end{equation}
Here, $\mathcal{M}_c^z = \mathcal{M}_c (1+z)$ is the detector-frame chirp mass, with $\mathcal{M}_c=(m_1m_2)^{3/5}/(m_1+m_2)^{1/5}$, and $m_{1,2}$ the component masses; $\eta=(m_1m_2)/(m_1+m_2)^2$ is the symmetric mass ratio; $d_{\rm L}$ is the luminosity distance; $\iota,\psi$ are the inclination and polarization angles; $\alpha,\delta$ are the right ascension and declination angles; $\phi_c, t_c$ are the phase and time of coalescence. We compute the Fisher matrices with the public package \texttt{GWBENCH}~\cite{Borhanian:2020ypi}.

For each detected \ac{BBH} event ($\rho^\mathrm{BBH}>8$), we sample its source parameters from a multivariate normal distribution $\mathcal{N}(\widehat{\bm{\theta}},\Gamma^{-1})$, with mean located at the true values of the parameters $\widehat{\bm{\theta}}$. We determine the mask in the time-frequency domain on zero-noise data containing these sampled signals only. We use Eqs.~\eqref{eq:stoch} to define \ac{SNR} maps as
\begin{equation}
    \mathrm{SNR}_\mathrm{IJ}(t_i;f_j)=\abs{\hat{C}_\mathrm{IJ}(t_i;f_j)/\hat{\sigma}_\mathrm{IJ}(t_i;f_j)}\,.
\end{equation}
We adopt an \ac{SNR} threshold of $5\times10^{-4}$ to identify the tracks associated to the signals~\cite{Zhong:2022ylh,Zhong:2024dss}. Our mask comprises all the pixels that pass this threshold. Given these masks, we can notch out \ac{CBC} tracks. We denote the cross-correlation spectra and the detector noise after notching by $\hat{C}^\star(f)$ and $(\sigma^\star)^2(f)$, respectively.
\begin{figure}[!htbp]
    \centering
    \includegraphics[width=0.8\linewidth]{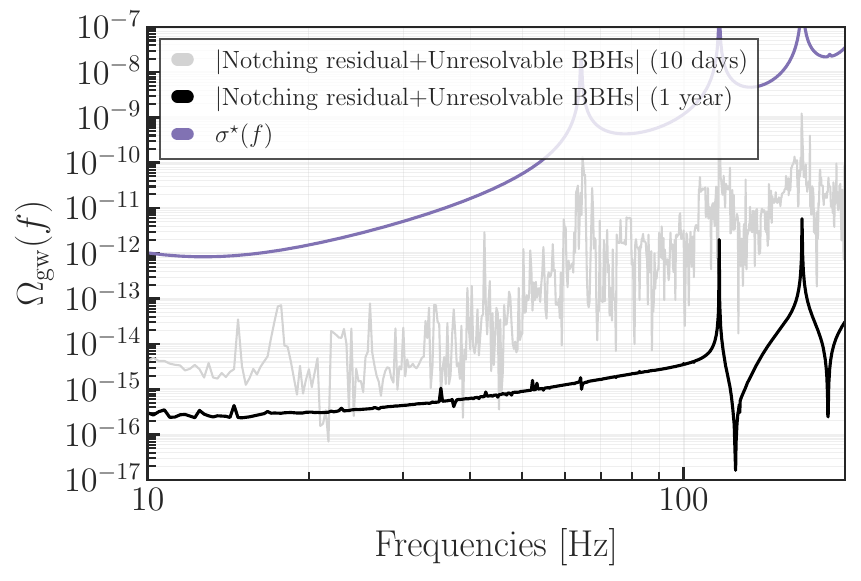}
    \caption{
    Notching residual plus the unresolvable BBH foreground (black), assuming the same underlying BBH population, but given one-year-long observation, and adopting a more aggressive notching strategy by setting a threshold for $|C(t_i;f_j)|$ directly. The purple curve represents the detector sensitivity, and the light gray curve corresponds to the same curve shown in Fig. 3 of the main text.}
    
    \label{fig:notching}
\end{figure}

We highlight that the notching strategy we introduced above is a conservative choice, and one could instead directly set a threshold on the measured $|C(t_i;f_j)|$ spectrograms as the choice we adopted in our very first work~\cite{Zhong:2022ylh}. 
By doing so, one can obtain an even better control of the notching residual foreground. In Fig.~\ref{fig:notching}, we show the corresponding notching residual plus the unresolvable \ac{BBH} foreground for a one-year-long observation considering the same \ac{BBH} population as the main text of this work, and using the notching strategy we introduced above. The foreground is well below the detector sensitivity ($\sigma^\star(f)$), again indicating that we can ignore this foreground in the search.
\section{Joint inference method}

After removing all the resolved \ac{BBH} signals, we perform joint Bayesian inference on the remaining data, which contain \ac{BNS} signals, the \ac{CGWB}, and the notching residual. In this section, we provide details on the calculation of the joint likelihood we use for the inference. 

\subsection{Joint likelihood}

Let us consider a \ac{BNS} population characterized by hyperparameters $\feature_\mathrm{BNS}$ and a \ac{CGWB} with parameters $\feature_\mathrm{CGWB}$. The joint likelihood for the individual \ac{BNS} detections $\data$ and the \ac{SGWB} cross-correlation spectrum ${C}^\star(f)$ can be factorized as~\cite{Callister:2020arv}
\begin{equation}
    \mathscr{L}(\data,\hat
{C}^\star(f)|\feature_\mathrm{BNS},\feature_\mathrm{CGWB})=\mathscr{L}_\mathrm{BNS}(\data|\feature_\mathrm{BNS})
\times\mathscr{L}_\mathrm{SGWB}(\hat
{C}^\star(f)|\feature_\mathrm{BNS},\feature_\mathrm{CGWB})\,.
\label{eq:jointlikelihood}
\end{equation}
The first term on the right-hand side is the usual hierarchical likelihood for population inference with individually resolved signals~\cite{Mandel:2018mve,Taylor:2018iat}
\begin{equation}
     \mathscr{L}(\data|\feature_\mathrm{BNS})\propto [N(\feature_\mathrm{BNS})\xi(\feature_\mathrm{BNS})]^{N_\mathrm{obs}}e^{-N(\feature_\mathrm{BNS})\xi(\feature_\mathrm{BNS})} \Bigg[\prod_{i=1}^{N_\mathrm{obs}}\frac{\displaystyle{\int\D\bm{\theta}~p(\bm{d}_i|\bm{\theta})p_\mathrm{pop}(\bm{\theta}|\feature_\mathrm{BNS})}}{\xi(\feature_\mathrm{BNS})}\Bigg]\,.
     \label{eq:Like_BNS}
\end{equation}
Here, $p(\bm{d}_i|\bm{\theta})$ is the likelihood for the individually detected event $i$ given the source parameters $\bm{\theta}$; $p_\mathrm{pop}(\bm{\theta}|\feature_\mathrm{BNS})$ is the ensemble distribution of the source parameters given population hyperparameters $\feature_\mathrm{BNS}$; $N_\mathrm{obs}$ is the number of events actually detected; and $\xi(\feature_\mathrm{BNS})$ is the expected fraction of resolvable events, given a population model with hyperparameters $\feature_\mathrm{BNS}$. We highlight that in this work, we treat \textit{detectable} and \textit{resolvable} as synonymous and use the terms interchangeably.

If the probability of detecting a signal with source parameters $\bm{\theta}$ is $p_\mathrm{det}(\bm{\theta})$, the fraction $\xi(\feature_\mathrm{BNS})$ is given by
\begin{equation}
    \xi(\feature_\mathrm{BNS}) = \int p_\mathrm{det}(\bm{\theta}) p_\mathrm{pop}(\bm{\theta}|\feature_\mathrm{BNS})\,\mathrm{d}\bm{\theta}\,.
\end{equation}
We compute $p_\mathrm{det}(\bm{\theta})$ on a grid of masses and redshifts with a hard \ac{SNR} cutoff. For \acp{BNS}, we choose a detection \ac{SNR} threshold of $\rho_\mathrm{thr}^\mathrm{BNS}=12$~\cite{Zhong:2024dss}.The expected total number of events (both resolved and unresolved) given population hyperparameters $\feature_\mathrm{BNS}$, $N(\feature_\mathrm{BNS})$, can be calculated as follows:
\begin{equation}
N(\feature_\mathrm{BNS})=T_\mathrm{obs}\int_0^{z_{\max}}\D z\frac{1}{1+z}\mathcal{R}_m(z|\feature_\mathrm{BNS})\frac{\D V_c}{\D z}\Big|_z\,,
\label{eq:ppop}
\end{equation}
where $T_\mathrm{obs}$ is the observation time, $\mathcal{R}_m(z|\feature_\mathrm{BNS})$ is the source-frame merger rate per unit comoving volume, and $\D V_c/\D z$ is the comoving volume element per redshift slice $\D z$. 

Regarding the \ac{BNS} population model $p_\mathrm{pop}(\bm{\theta}|\feature_\mathrm{BNS})$, we assume that all of the angular parameters are isotropically distributed and that the mass distribution is uniform, redshift-independent, and perfectly known. In other words, the only hyperparameters left to determine are the ones characterizing the redshift distribution
\begin{equation}
    p_\mathrm{pop}(\bm{\theta}|\feature_\mathrm{BNS}) \propto p(z|\feature_\mathrm{BNS})\,.
\end{equation}

The second term on the right-hand side of Eq.~\eqref{eq:jointlikelihood} is the usual Gaussian likelihood for the \ac{SGWB} cross-correlation spectrum~\cite{Mandic:2012pj,Callister:2017ocg}
\begin{equation}
    \mathscr{L}_\mathrm{SGWB}(\hat{C}^\star(f)|\feature_\mathrm{BNS},\feature_\mathrm{CGWB})\propto\prod_i\exp\Big(-\frac{[\hat{C}^\star(f_i)-\Omega_\mathrm{SGWB}(f_i;\feature_\mathrm{BNS},\feature_\mathrm{CGWB})]^2}{2(\sigma^{\star})^2(f_i)}\Big)\,,
    \label{eq:sgwblikelihood}
\end{equation}
where $\Omega_\mathrm{SGWB}(f)$ is the expected energy density spectrum of the total \ac{SGWB}, and the product is over all frequency bins $f_i$. After notching out all the resolved \ac{BBH} signals, the expected energy density of the total \ac{SGWB} consists of four contributions:
\begin{equation}
\Omega_\mathrm{SGWB}(f)=\Omega_\mathrm{BNS}(f|\feature_\mathrm{BNS})+\Omega_\mathrm{CGWB}(f|\feature_\mathrm{CGWB})+\Omega_\mathrm{BBH,resi}(f)+\Omega_\mathrm{BBH, unres}(f)\,.
\end{equation}
Here $\Omega_\mathrm{BNS}(f)$ is the foreground generated by all \ac{BNS} events (both resolved and unresolved), $\Omega_\mathrm{CGWB}(f)$ is the cosmological background, $\Omega_\mathrm{BBH,resi}(f)$ is the residual foreground after notching out the resolved \ac{BBH} signals, and $\Omega_\mathrm{BBH,unres}(f)$ is the foreground from the few unresolvable \ac{BBH} signals. We find that the \ac{BBH} contribution $\Omega_\mathrm{BBH,resi}(f)+\Omega_\mathrm{BBH, unres}(f)$ after notching is well below noise level (see Fig.~4 of the main text). For this reason we neglect it in our inference, and we only take $\Omega_\mathrm{BNS}(f|\feature_\mathrm{BNS})$ and $\Omega_\mathrm{CGWB}(f|\feature_\mathrm{CGWB})$ into account. The expected \ac{BNS} foreground is given by~\cite{Phinney:2001di}
\begin{equation}
    \Omega_\mathrm{BNS}(f|\feature_\mathrm{BNS})= \frac{f}{\rho_cH_0}\int_0^{z_{\max}}\D z\frac{\mathcal{R}_m(z|\feature_\mathrm{BNS})}{(1+z)E(z)}\left\langle\frac{\D E}{\D f}\Big|_{f(1+z)}\right\rangle_{\feature_\mathrm{BNS}}\,.
    \label{eq:Omega_BNS}
\end{equation}
Here, $\rho_c$ is the critical energy density required to close the universe; $E(z)=\sqrt{\Omega_\Lambda+(1+z)^3\Omega_\mathrm{M}}$ is the Hubble parameter at redshift $z$ neglecting radiation density, where we take the energy densities of matter and dark energy to be $\Omega_\mathrm{M}=0.3$ and $\Omega_\Lambda=0.7$, respectively; and $\langle \D E/\D f\rangle_{\feature_\mathrm{BNS}}$ is the expectation value over the \ac{BNS} population of the source-frame energy spectrum radiated by a single \ac{BNS} source. We estimate $\D E/\D f$ via the Newtonian approximation~\cite{Phinney:2001di}
\begin{equation}
    \left\langle\frac{\D E}{\D f}\Big|_{f(1+z)}\right\rangle_{\feature_\mathrm{BNS}}=\frac{(\pi G)^{2/3}\langle \mathcal{M}_c\rangle_{\feature_\mathrm{BNS}}^{5/3}}{3}[f(1+z)]^{-1/3}\,,
    \label{eq:dEdf}
\end{equation}
where $G$ is the gravitational constant, and $\langle \mathcal{M}_c\rangle_{\feature_\mathrm{BNS}}$ is the expectation value of the source-frame chirp mass over the \ac{BNS} population.

For the expected \ac{CGWB}, we instead consider a power-law background
\begin{equation}
    \Omega_\mathrm{CGWB}(f|\feature_\mathrm{CGWB})=\Omega_\mathrm{ref}\Big(\frac{f}{f_\mathrm{ref}}\Big)^\alpha\,,
    \label{eq:Omega_CGWB}
\end{equation}
where $\feature_\mathrm{CGWB}=\{\Omega_\mathrm{ref},\alpha\}$. In our simulations, we will assume $\alpha=0$ and $\Omega_{\mathrm{ref}}= 2\times 10^{-11}$ at $f_\mathrm{ref}=25\,\mathrm{Hz}$.

\subsection{Mock likelihoods for resolved BNSs}
\label{subseq:mymethod}

For each detected \ac{BNS} event ($\rho^\mathrm{BNS}>12$), the integral in the hierarchical likelihood of Eq.~\eqref{eq:Like_BNS} can be approximated as an average over discrete likelihood samples
\begin{equation}
\int\D\bm{\theta}~p(\bm{d}_i|\bm{\theta})p_\mathrm{pop}(\bm{\theta}|\feature_\mathrm{BNS})\approx\Big\langle p_\mathrm{pop}(\bm{\theta}_i|\feature_\mathrm{BNS})\Big\rangle_\mathrm{Likelihood~samples}.
\label{eq: samp_avg}
\end{equation}
We emphasize that the hierarchical likelihood — the probability of observing a particular \ac{BNS} event $\bm{d}_i$ given population parameters $\feature_\mathrm{BNS}$ — can be interpreted as an integral over the population distribution $ p_\mathrm{pop}(\bm{\theta}|\feature_\mathrm{BNS})$, which represents the probability of a \ac{BNS} event characterized by parameters $\bm{\theta}$ under the given population model, weighted by the event-specific likelihood.
In a real inference, one typically has access to samples drawn from the individual events' posterior distributions $p(\bm{\theta}|\bm{d}_i)$, which then need to be reweighted by the adopted \ac{PE} prior $p(\bm{\theta})$ to obtain the likelihood samples. By performing the average in Eq.~\eqref{eq: samp_avg}, we have already taken the width of the posterior and of the likelihood into account.Performing full Bayesian \ac{PE} for thousands of \ac{BNS} events in \ac{XG} detectors is, however, currently computationally unfeasible. Hence, we generate synthetic likelihood samples for each detected event following a procedure similar to Refs.~\cite{Fishbach:2018edt,Callister:2020arv}. 

Given our assumptions on the \ac{BNS} population model (see Eq.~\eqref{eq:ppop}), for each detected event, we only need to generate mock likelihood samples on the redshift $z$. The redshift (or rather, the luminosity distance $d_\mathrm{L}$) is mainly determined by the waveform amplitude. For \ac{BNS} signals, the polarization amplitudes $A_+$ and $A_\times$ can be well described by the Newtonian approximation~\cite{Maggiore:2007ulw}
\begin{equation}
\begin{aligned}
     &A_+(\mathcal{M}_c^z,d_L,\cos\iota)=\sqrt{\frac{5}{24}}\frac{(G\mathcal{M}_c^z)^{5/6}}{2\pi^{2/3}c^{3/2}d_L}(1+\cos^2\iota)\,,\\
     &A_\times(\mathcal{M}_c^z,d_L,\cos\iota)=\sqrt{\frac{5}{24}}\frac{(G\mathcal{M}_c^z)^{5/6}}{\pi^{2/3}c^{3/2}d_L}\cos\iota\,.
\label{eq:newtwaveform}
\end{aligned}
\end{equation}
The resulting strain amplitude is thus
\begin{equation}
A(\mathcal{M}_c^z,d_L,\cos\iota)=\sqrt{A_\times^2+A_+^2}=\frac{\mathcal{K}^2\mathcal{M}_c^{z,5/3}}{4d_L^2}\mathcal{F}(\cos\iota)\,,
\label{eq:A}
\end{equation}
where $\mathcal{K}$ encodes all the constants and 
\begin{equation}
\mathcal{F}(\cos\iota)=\sqrt{(1+\cos^2\iota)^2+4\cos^2\iota}\,.
\label{eq:form_factor}
\end{equation}
For each resolved \ac{BNS}, we draw likelihood samples for the strain amplitude A from a Gaussian distribution
\begin{equation}
    \{A\}\sim\mathcal{N}(\bar{A},\sigma_A),
\end{equation}
where $\bar{A}$ is the true value determined via the injected parameters. In this formalism, $\sigma_\mathrm{A}$ represents the detector noise level, and the ratio $\rho^* = \bar{A}/\sigma_{A}$ is a proxy for \ac{SNR}. We fix the value $\sigma_A=1.09\times 10^{-24}$ such that the number of signals that satisfy $\rho^\star\geqslant \rho_\mathrm{thr}^\mathrm{BNS}=12$ in this approximation matches the number of resolved \ac{BNS} signals in our pipeline. We then draw likelihood samples for the detector-frame chirp mass $\mathcal{M}_c^z$ and the factor $\mathcal{F}(\cos\iota)$ from Gaussian distributions centered at the injected values $\overline{\mathcal{M}}_c^z$, $\overline{\mathcal{F}}(\cos\iota)$:
\begin{equation}
    \{\mathcal{M}_c^z\}\sim\mathcal{N}\Big(\overline{\mathcal{M}}_c^z,\sigma_\mathcal{M}\frac{12}{\rho^\star}\Big)\,,\quad \{\mathcal{F}(\cos\iota)\}\sim\mathcal{N}\Big(\overline{\mathcal{F}}(\cos\iota),\sigma_\mathcal{F}\frac{12}{\rho^\star}\Big)\,.
\end{equation}
The values of $\sigma_\mathcal{M}$ and $\sigma_\mathcal{F}$ are derived from Ref.~\cite{Callister:2020arv}. In their work, they provide a reference error $\sigma_\Theta=0.15$ for the projection factor~\cite{Finn:1992xs}
\begin{equation}
    \Theta=4\sqrt{\frac{F^2_+(\alpha,\delta,\psi)}{4}(1+\cos\iota^2)^2+F^2_\times(\alpha,\delta,\psi)\cos^2\iota}\,, 
\end{equation}
where $F_{+,\times}(\alpha,\delta,\psi)$ are the antenna patterns of a given \ac{GW} detector. By fixing $F_+$ and $F_\times$ to their sky-averaged values $\langle F_+^2\rangle_\mathrm{sky}=\langle F^2_\times\rangle_\mathrm{sky}=1/5$~\cite{Maggiore:2007ulw} and comparing with Eq.~\eqref{eq:form_factor}, we obtain
\begin{equation}
\sigma_\mathcal{F}\approx\frac{\sqrt{5}}{2}\sigma_\Theta=0.167\,.
\end{equation}
Regarding the chirp mass, Ref.~\cite{Callister:2020arv} provides a value of $\sigma_\mathcal{M}=0.08 M_\odot$ for \acp{BBH}. Since in this work we apply the formalism to \ac{BNS} signals, we empirically scale this value to $\sigma_\mathcal{M}=0.01\,M_\odot$. We check the validity of this approximation in the next section.

Given likelihood samples for $A,~\mathcal{M}_c^z$ and $\mathcal{F}(\cos\iota)$, we finally obtain likelihood samples for $d_L$ by inverting Eq.~\eqref{eq:A}, and we convert them into redshift samples assuming the Planck 2018 $\Lambda$CDM cosmological model~\cite{Planck:2018vyg}. Albeit simplistic, this method allows us to capture some of the key correlations that drive the uncertainty on the redshift, namely \ac{SNR}, inclination, and chirp mass~\cite{Callister:2020arv}. 
\subsection{Comparison with Fisher matrix}

As an alternative, one could generate mock likelihood samples for individually resolved \acp{BNS} via the Fisher approximation~\cite{Finn:1992wt}. For instance, this is the approach we used to estimate errors on \ac{BBH} source parameters in the notching procedure (see Sec.~\ref{subsec:notching}). However, the approximation is strictly valid only in the high-\ac{SNR} regime, and a large fraction of the resolved \ac{BNS} signals have \ac{SNR} barely above the detection threshold. Furthermore, several events are expected to produce ill-conditioned Fisher matrices~\cite{Iacovelli:2022bbs}, which are notoriously unreliable to invert and can significantly mispredict the likelihood~\cite{Vallisneri:2007ev}. While these effects do not cause significant issues in the notching procedure~\cite{Zhong:2024dss}, they can significantly impact the estimate of the error on luminosity distance (and thus redshift)~\cite{Mancarella:2024qle}.

For these reasons, and given the demonstrative purpose of this work, we choose to generate mock likelihood samples with the approach described in the previous Section.
\begin{figure}[!htbp]
    \centering
    \includegraphics[width=0.8\linewidth]{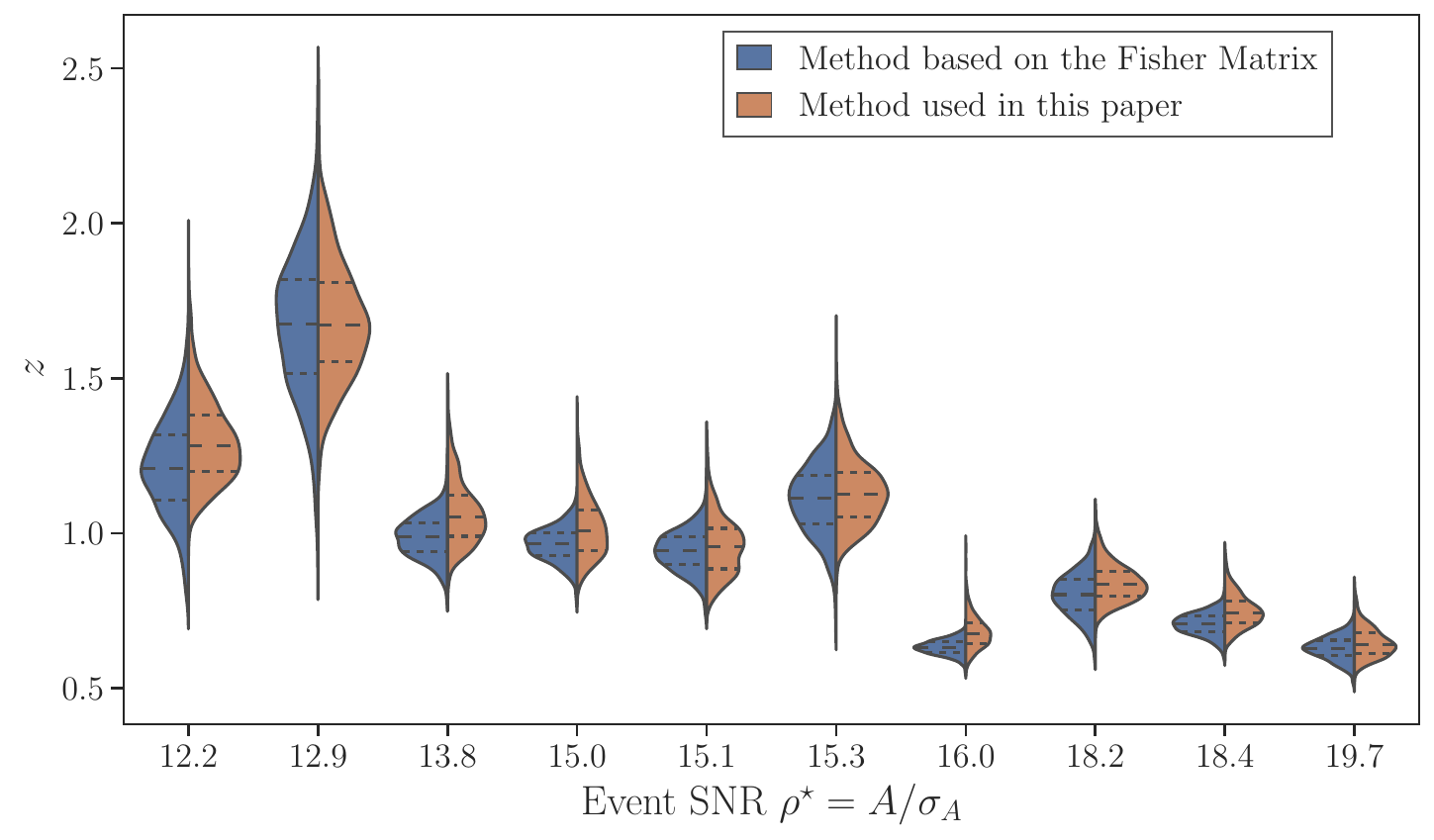}
    \caption{Comparison between mock likelihoods on redshift obtained either via Fisher matrix (blue distributions) or via the approach of Sec.~\ref{subseq:mymethod} (orange distributions). Results are shown for several low-\ac{SNR} \ac{BNS} events and with well-conditioned Fisher matrices drawn from our population. Dashed lines indicate 25\%, 50\%, and 75\% quantiles.}
    \label{fig:Fisher_mine_comp}
\end{figure}

In Fig.~\ref{fig:Fisher_mine_comp}, we compare mock likelihood redshift samples generated with a Fisher-matrix approximation and with the approach described in Sec.~\ref{subseq:mymethod}. We show results for several \ac{BNS} events with low effective \ac{SNR} $\rho^*$ (which are expected to provide the dominant contribution to the uncertainty in the population inference) and well-conditioned Fisher matrices. The Fisher matrices are generated with the public package \texttt{GWFish}~\cite{Dupletsa:2022scg} considering all nine parameters listed in the Eq.~\eqref{eq:info_paras} and \texttt{IMRPhenomXAS} waveform model.
The estimated marginalized likelihoods with the two methods are in good agreement for all the events shown, validating the robustness of our approach. 

In an actual inference, one would need to perform full Bayesian \ac{PE} on all the resolved \ac{BNS} events, which is currently computationally unfeasible in the context of \ac{XG} detectors. In practice, we expect that considering more realistic waveform models with more parameters compared to Eqs.~\eqref{eq:newtwaveform} and performing full \ac{PE} would lead to wider redshift likelihoods compared to our approach. This is due both to the correlations introduced by the additional parameters, and to the potentially non-Gaussian nature of the likelihoods for low-\ac{SNR} events. In this sense, our results represent an optimistic forecast of the capability of our framework in the context of \ac{XG} detectors. However, we argue that the analysis performed in this work only makes use of 10 days of data, and tight constraints on the population parameters can always be achieved by stacking a larger number of events together. Several promising methods to accelerate \ac{PE} in preparation for \ac{XG} detectors are being developed (see e.g.~\cite{Green:2020dnx,Dax:2022pxd,Wong:2023lgb,Alvey:2023naa,Leyde:2023iof}) and can potentially be incorporated in our framework in the future. 
\section{Additional results}

In this section, we provide more details on the results of our joint inference and compare them with the constraints one would obtain from (i) individual \ac{BNS} detections only, or (ii) the \ac{SGWB} search only.

\subsection{Joint inference results}

\begin{figure}[t]
     \centering
     \includegraphics[width=0.45\textwidth]{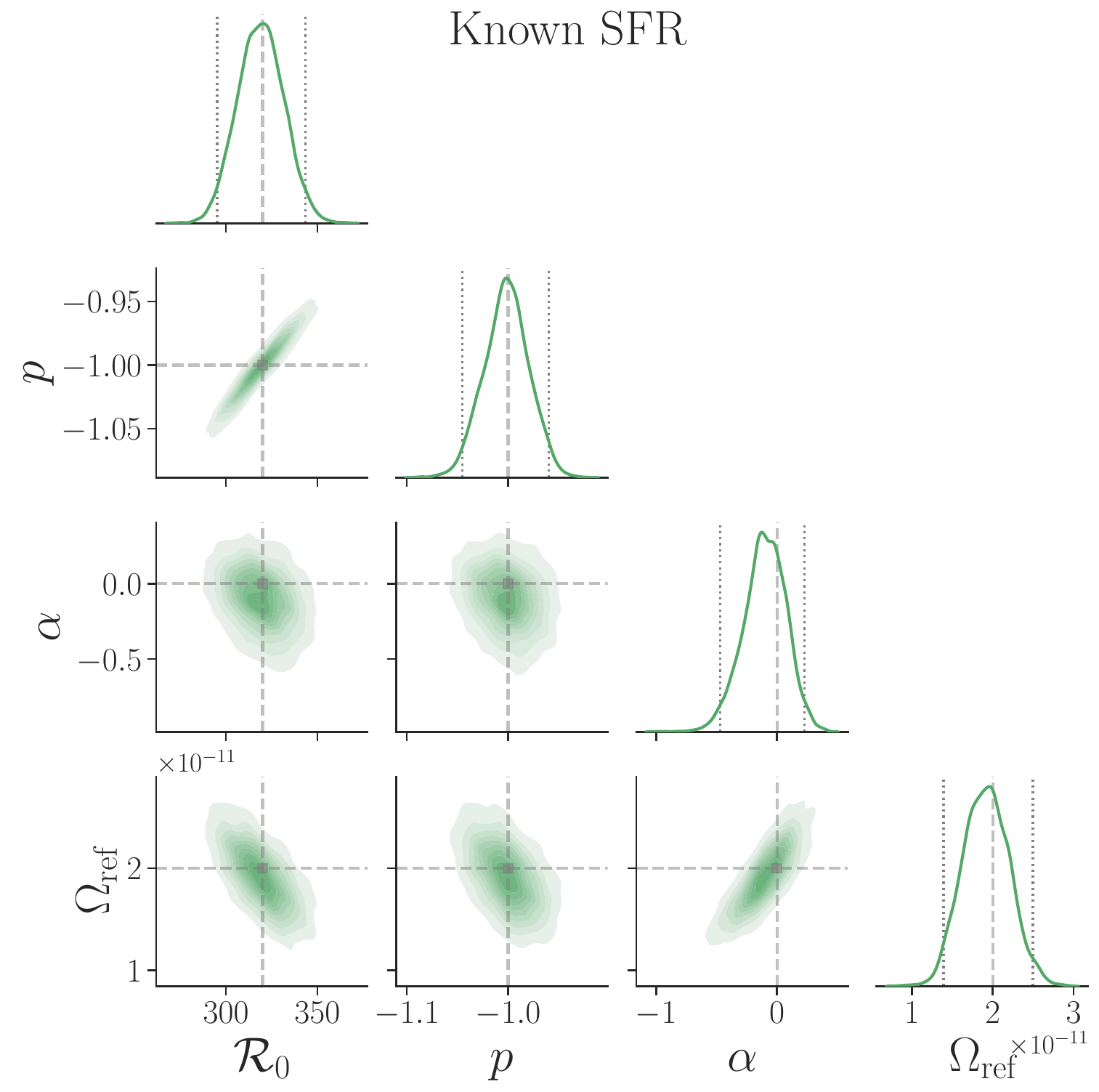}
     \includegraphics[width=0.45\textwidth]{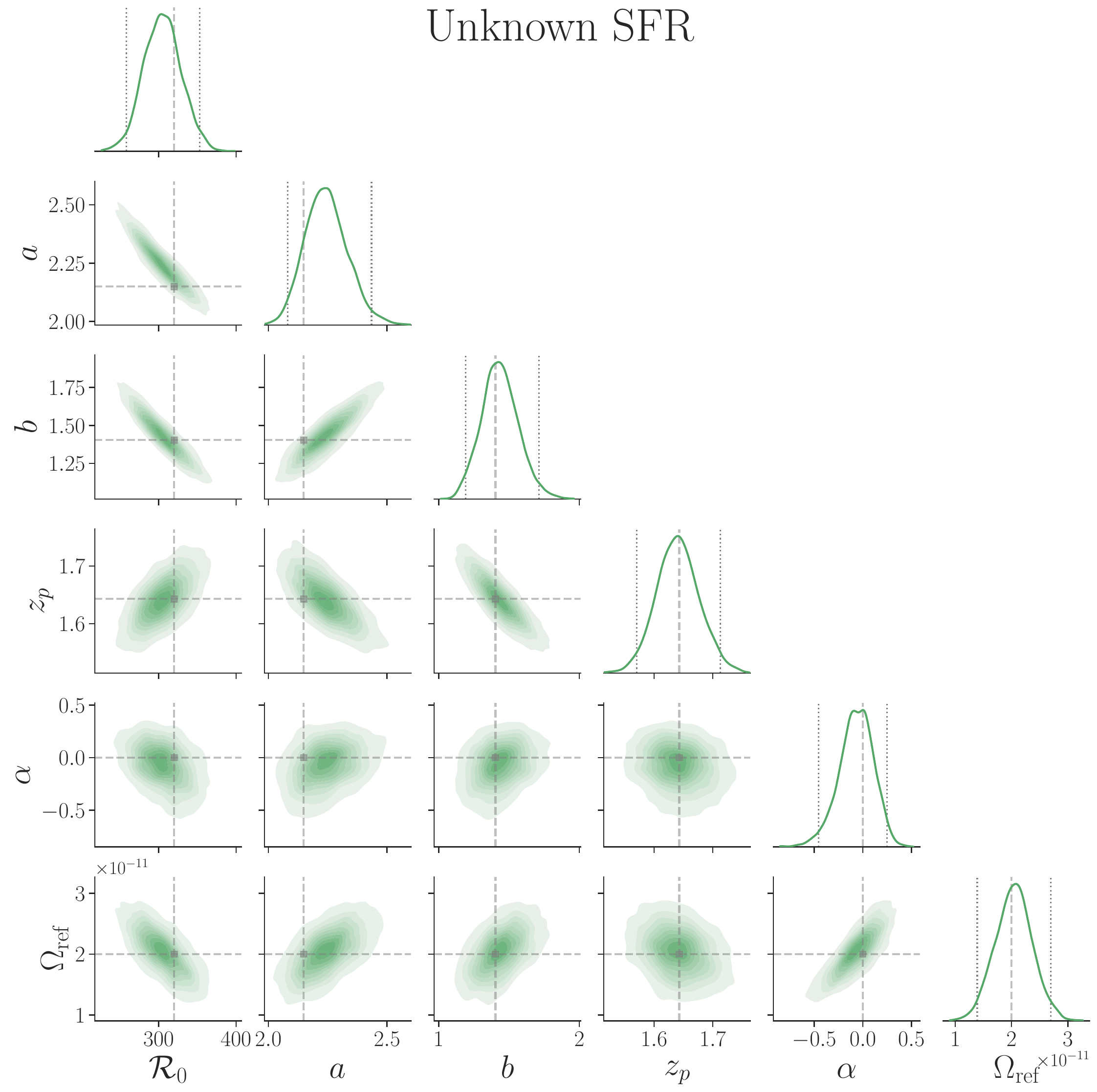}
     \caption{Inferred posterior distributions on the \ac{BNS} hyperparameters and the \ac{CGWB} parameters for the scenarios with known (left corner plot) and unknown (right corner plot) \ac{SFR}. The gray dashed lines indicate the injected values, while the gray dotted lines show the $95\%$ credible intervals in the marginalized 1D posterior distributions.
     }
     \label{fig:full_posteriors}
\end{figure}

\begin{table}[h]
    \centering
    \caption{Priors placed on the hyperparameters describing the \ac{BNS} merger rate and population and the \ac{CGWB}.
    }
    \begin{tabular}{>{\centering\arraybackslash}m{2cm} >{\centering\arraybackslash}m{2cm} >{\centering\arraybackslash}m{2cm} >{\centering\arraybackslash}m{2cm}}
        \toprule
        \textbf{Parameter} & \textbf{Prior} & \textbf{Minimum} & \textbf{Maximum} \\
        \hline
        $\mathcal{R}_0$ & Log-uniform & 10 & 1700 \\
        $p$ & Uniform & -3 & 0 \\
        \hline
        \hline
        $a$ & Uniform & 0 & 10 \\
        $b$ & Uniform & 0 & 10 \\
        $z_p$ & Uniform & 0 & 4 \\
        \hline
        \hline
        $\Omega_\mathrm{ref}$ & Log-uniform & $10^{-15}$ & $10^{-10}$ \\
        $\alpha$ & Uniform & -3 & 3\\
        \bottomrule
    \end{tabular}
    \label{tab:prior}
\end{table}

In Fig.~\ref{fig:full_posteriors}, we show the posterior distributions on all the parameters of our joint inference for both scenarios we consider. The priors we impose are listed in Table~\ref{tab:prior}. As discussed in the main text, we find that the \ac{CGWB} parameters are well constrained in both cases, allowing us to exclude a $\Omega_\mathrm{ref}=0$ amplitude at least $6\,\sigma$ level. With only $10$ days of observation, corresponding to roughly $\sim 7000$ detections of individual \ac{BNS} events, we find that the hyperparameters of the \ac{BNS} redshift distribution are also well constrained. The $95\%$ errors on all the parameters describing the shape of the merger rate are of order $\sim10\%$ or less. The median and 95\% credible interval of the posterior distributions of all parameters are shown in Table~I of the main text.

\begin{figure}[!htbp]
    \centering
    \includegraphics[width=1.0\textwidth]{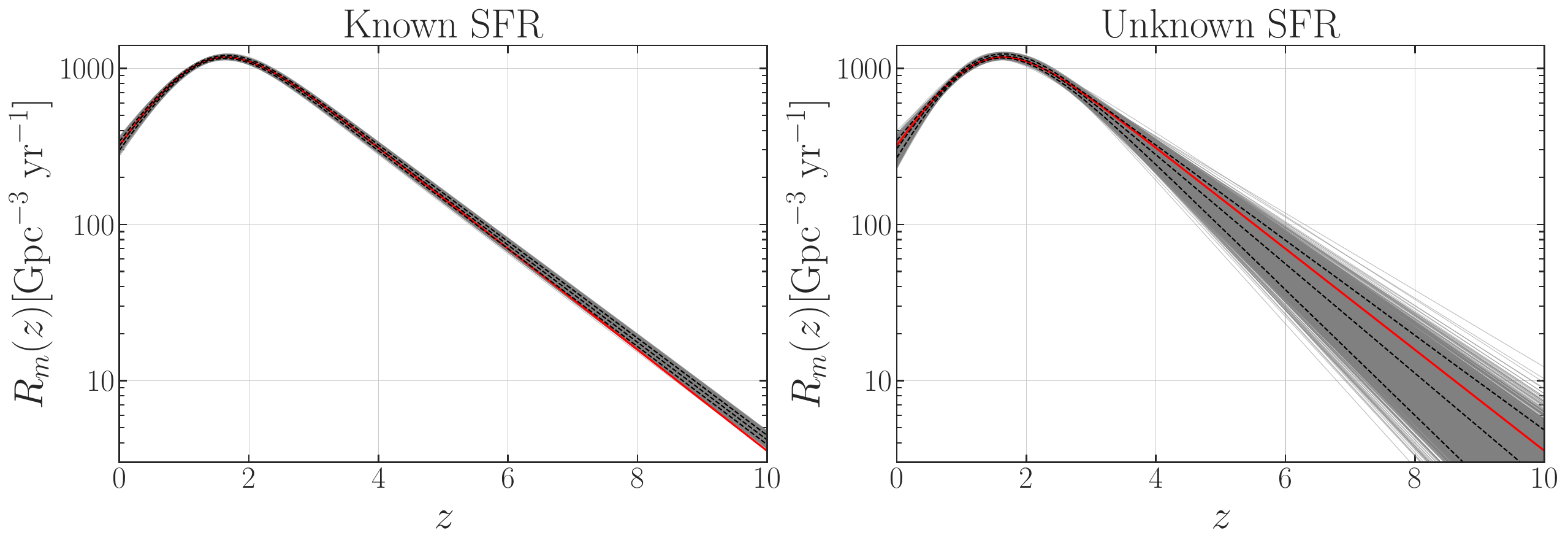}
    \caption{Inferred source-frame \ac{BNS} merger rate $R_m(z)$ as a function of redshift in the two scenarios we consider. Each gray curve corresponds to the merger rate obtained with a different posterior sample on \ac{BNS} hyperparameters. The three dashed black curves denote the median and $90\%$ quantiles. The red curve depicts the injected $R_m(z)$. }
    \label{fig:Rm_com}
\end{figure}
\begin{figure}[!htbp]
    \centering
\includegraphics[width=0.4\linewidth]{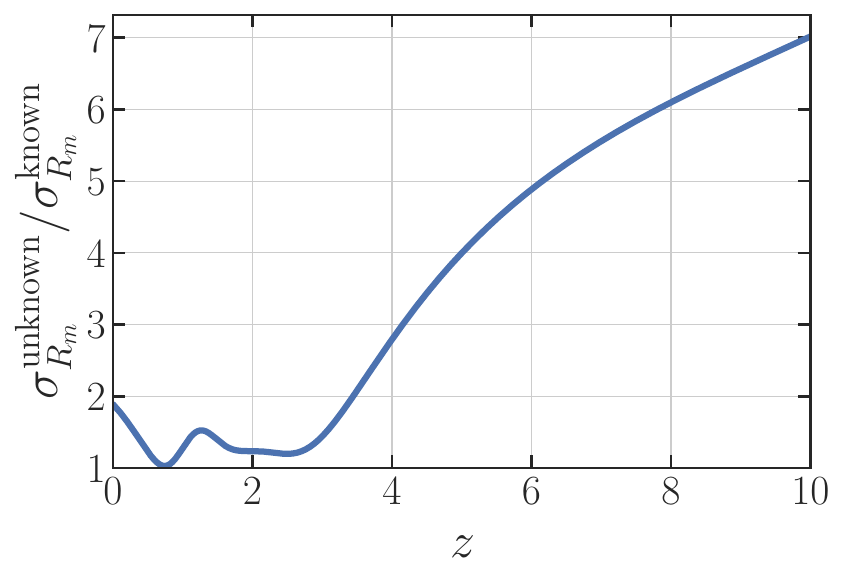}
    \caption{Ratio of the standard deviation in the inferred \ac{BNS} merger rate $R_m(z)$ as a function of redshift between the unknown SFR and the known SFR cases.}
\label{fig:sigma_Rm_ratio}
\end{figure}
The constraints on the source-frame \ac{BNS} merger rate as a function of redshift are shown in Fig.~\ref{fig:Rm_com}. The grey curves are obtained with different posterior samples on the \ac{BNS} hyperparameters, namely $\{\mathcal{R}_0, p\}$ ($\{\mathcal{R}_0, a, b, z_p\}$) in the case with known (unknown) \ac{SFR}. Assuming prior knowledge of the \ac{SFR}, and thus including fewer parameters in the inference, allows for more stringent constraints in $R_m(z)$ in all redshift bins. This effect is particularly prominent at redshifts $z\gtrsim 3$, where less \ac{BNS} detections are available. In particular, we show the ratio of the standard deviations in the inferred merger rate between the two cases in Fig.~\ref{fig:sigma_Rm_ratio}. The standard deviation in the unknown \ac{SFR} case is larger by a factor of $\lesssim 2$ at low redshifts, and then starts increasing monotonically at $z\gtrsim 3$. 
\begin{figure}[t]
    \centering
    \includegraphics[width=1.0\textwidth]{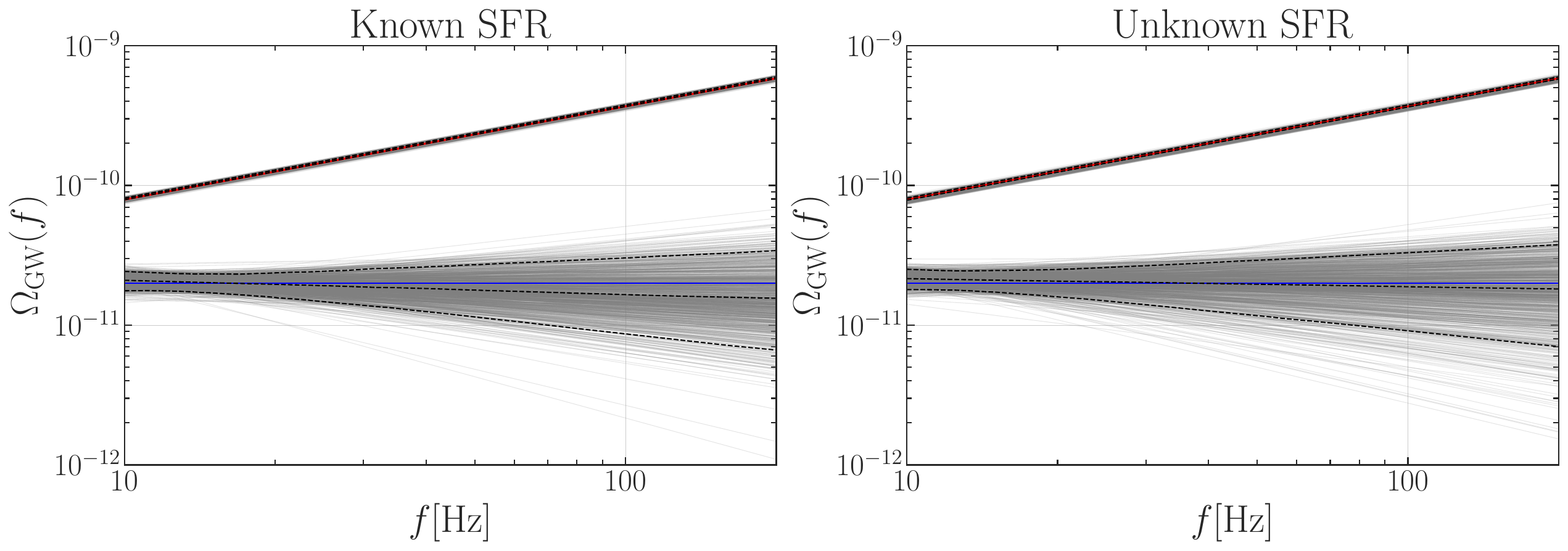}
    \caption{Inferred energy density $\Omega_\mathrm{GW}$ of the \ac{BNS} foreground and \ac{CGWB} as a function of frequency in the two scenarios we consider. 
    Each gray curve represents the resulting $\Omega_\mathrm{GW}(f)$ from a different posterior sample. 
    The red straight line indicates the injected $\Omega_\mathrm{GW}(f)$ for the \ac{BNS} foreground, while the blue line represents the injected $\Omega_\mathrm{GW}(f)$ for the \ac{CGWB}. The dashed black curves depict the median and $90\%$ quantiles.}
    \label{fig:spec_com}
\end{figure}

In Fig.~\ref{fig:spec_com}, we show the inferred energy spectra $\Omega_\mathrm{GW}(f)$ 
for both the \ac{BNS} foreground and the \ac{CGWB} using Eq.~\eqref{eq:Omega_BNS} to  Eq.~\eqref{eq:Omega_CGWB}. We find that the \ac{BNS} foreground spectrum can be tightly constrained in both scenarios we consider. The uncertainty on the \ac{BNS} foreground is larger in the more realistic case with unknown \ac{SFR}, reflecting the larger uncertainty on the merger rate shown in Fig.~\ref{fig:Rm_com}. In particular, we find that the standard deviation on $\Omega_\mathrm{GW}(f)$ is roughly $\sim 20\%$ smaller in every frequency bin in the case with known \ac{SFR}, namely $\sigma^{\mathrm{Known~SFR}}_\mathrm{BNS}(f)=0.83\,\sigma_\mathrm{BNS}^\mathrm{Unknown~SFR}(f)$. 
The recovered \ac{CGWB} is remarkably similar in both cases, reflecting the similarity in the joint posterior distribution of $(\Omega_\mathrm{ref},\alpha)$. Constraints get significantly worse at higher frequencies, as the baseline sensitivity to stochastic searches decreases.

\subsection{Joint inference vs individual BNS only}
\begin{figure}[t]
    \centering
    \includegraphics[width=0.4\textwidth]{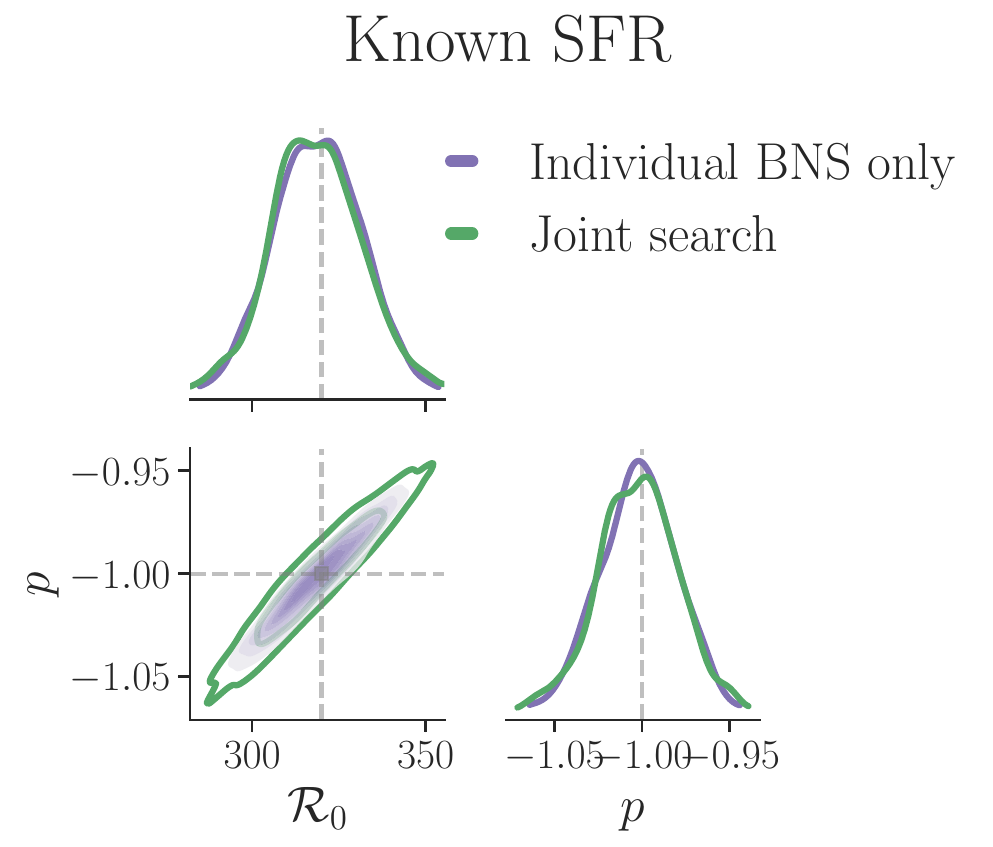}\includegraphics[width=0.55\textwidth]{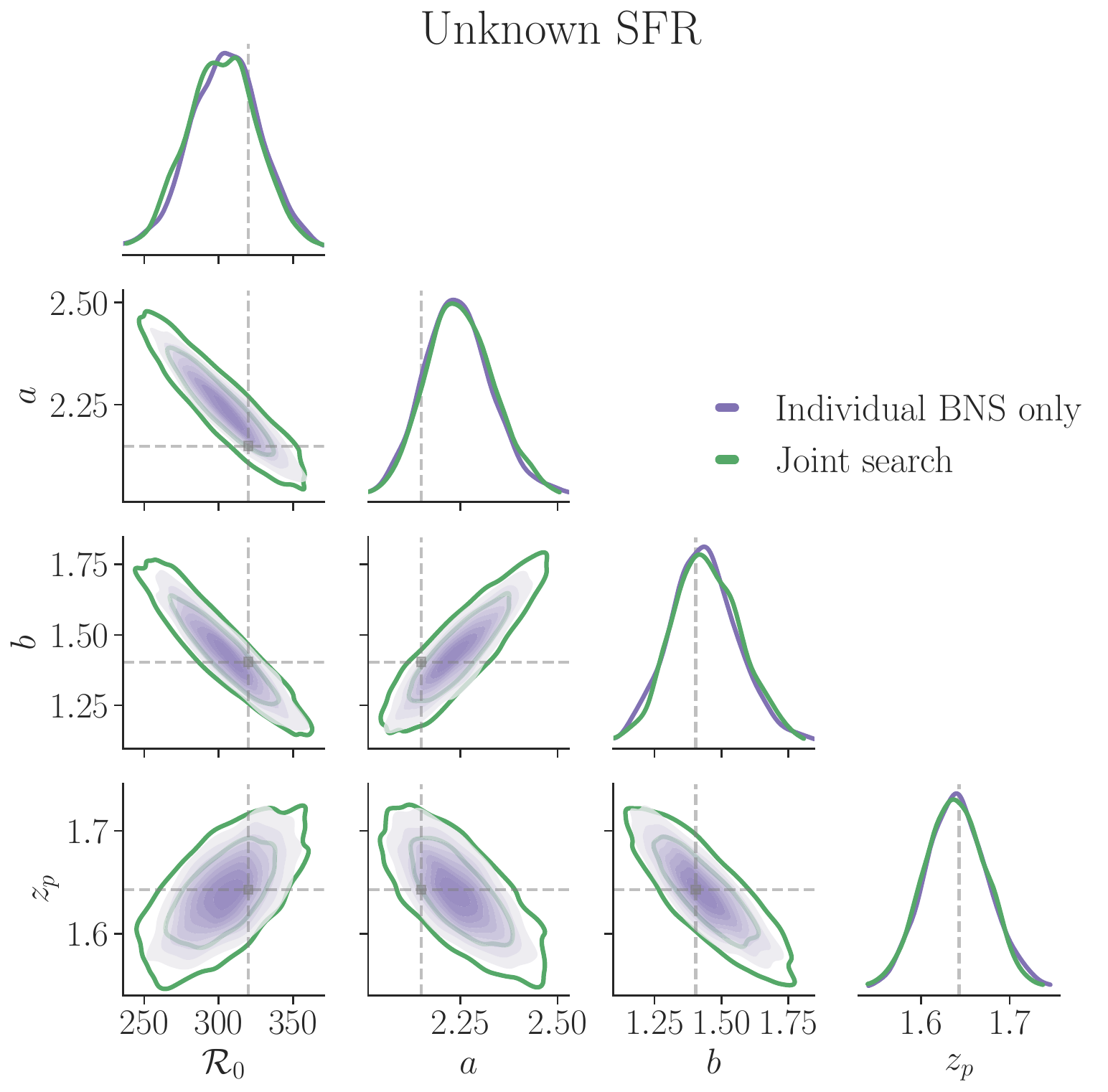}
    \caption{Comparison between posteriors on the \ac{BNS} hyperparameters resulting from the joint inference (green contours) and with individual \ac{BNS} detection only (purple contours). The contours correspond to $68\%$ and $95\%$ of the probability, respectively. Results are shown for both the known (left corner plot) and unknown (right corner plot) \ac{SFR} scenario.}
    \label{fig:BNS_Joint_com}
\end{figure}

In Fig.~\ref{fig:BNS_Joint_com}, we compare the posterior distributions on \ac{BNS} hyperparameters from the joint inference with the posteriors resulting only from the hierarchical inference on the individually detected \ac{BNS} signals. In other words, we compare inference results obtained from the full joint likelihood $\mathscr{L}(\data,\hat{C}^\star(f)|\feature_\mathrm{BNS},\feature_\mathrm{CGWB})$ of Eq.~\eqref{eq:jointlikelihood} with results obtained from just the first term of the same equation, $\mathscr{L}_\mathrm{BNS}(\data|\feature_\mathrm{BNS})$.  The priors on the hyperparameters are the same as in Table~\ref{tab:prior}.

We find that the two posterior distributions are remarkably similar. This is a crucially different conclusion from the results shown in Ref.~\cite{Callister:2020arv} in the context of current \ac{GW} detectors \ac{LVK}. Putting together data from the first two \ac{LVK} observing campaigns (O1/O2), they found that constraints on the \ac{BBH} merger rate significantly improve when including \ac{SGWB} data in the population inference. In particular, constraints on the behavior of the merger rate at redshifts $z\gtrsim 0.5$ were dominated by the stochastic search results. Their findings were primarily caused by the limited number of individual detections available ($\sim 10$) and the fact that all of the detected events originate from \acp{BBH} in the local Universe. After increasing the number of individual \ac{BBH} detections to 44 events by the first half of the third observing run (O3a), the \ac{LVK} collaboration already found that constraints from direct detections begin to dominate over bounds from the stochastic foreground~\cite{KAGRA:2021kbb}. In both of these cases, the information comes from the \emph{non-detection} of the astrophysical \ac{SGWB}. The expected \emph{detection} of the \ac{SGWB} from \ac{CBC} with \ac{LVK} at design sensitivity could once again provide improvements in the \ac{BBH} merger rate constraints compared to individual detections~\cite{KAGRA:2021kbb}. With \ac{XG} detectors, however, our results suggest that the sheer number of detections, the higher detection fraction, and the larger redshift horizon imply that the constraints from individual detections completely dominate over the bounds from the \ac{SGWB}, even for \acp{BNS} and with just $10$ days of observation.
\subsection{Joint inference vs SGWB only}
\begin{figure}
    \centering
\includegraphics[width=0.8\textwidth]{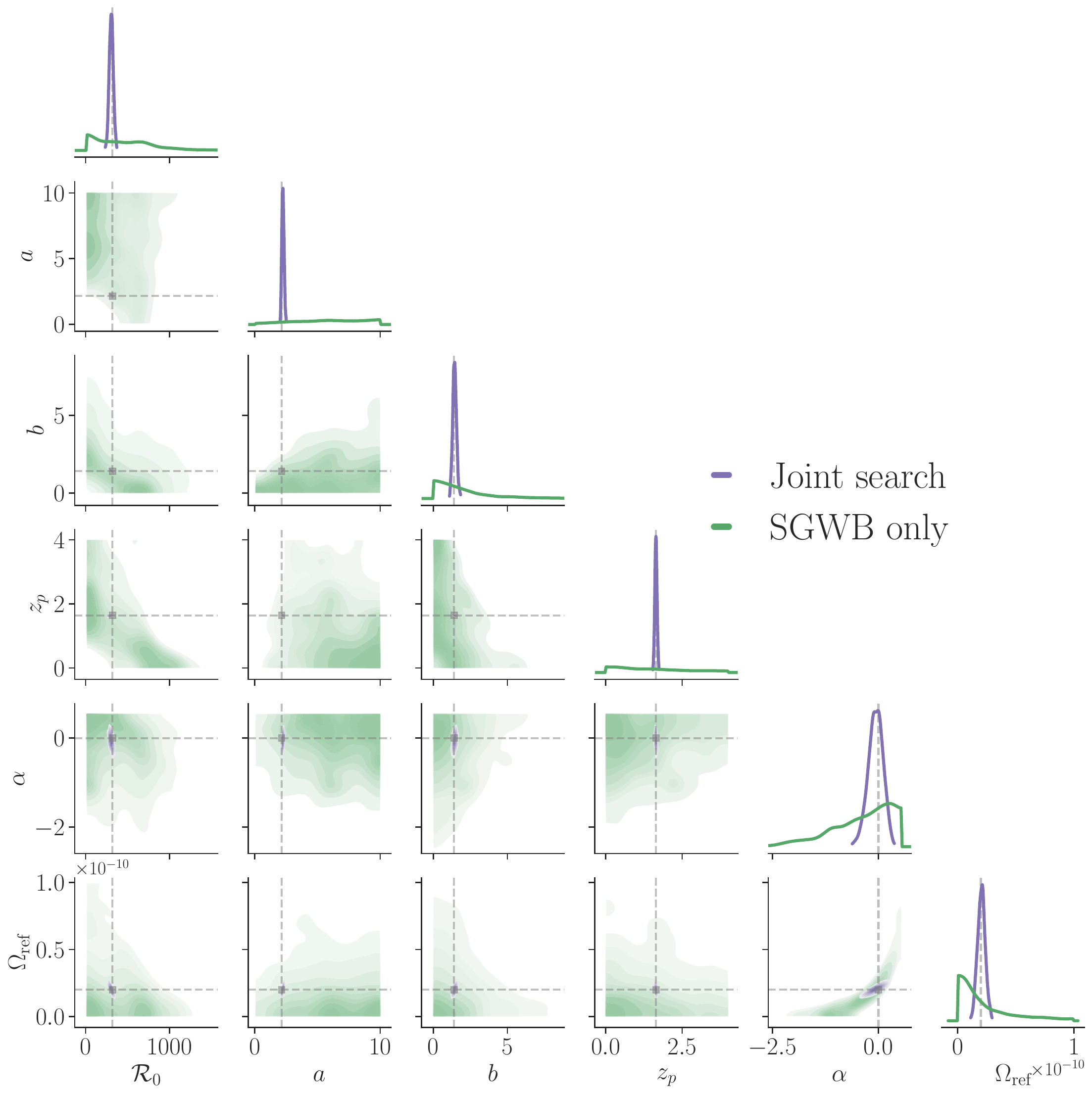}
    \caption{Comparison between posteriors on the \ac{BNS} hyperparameters and the \ac{CGWB} parameters resulting from the joint inference (purple contours) and from \ac{SGWB} data only (green contours). The contours correspond to $68\%$, and $95\%$ of the probability mass, respectively. No prior knowledge of the \ac{SFR} is assumed, other than its functional form.
    }
    \label{fig:SGWB_joint}
\end{figure}
In Fig.~\ref{fig:SGWB_joint} we compare posteriors on the \ac{BNS} hyperparameters and \ac{CGWB} parameters from the joint inference with posteriors resulting from an inference with \ac{SGWB} data only. In other words, this time we compare inference results obtained from the full joint likelihood $\mathscr{L}(\data,\hat{C}^\star(f)|\feature_\mathrm{BNS},\feature_\mathrm{CGWB})$ of Eq.~\eqref{eq:jointlikelihood} with results obtained just with the second term of the same equation, $\mathscr{L}_\mathrm{SGWB}(\hat{C}^\star(f)|\feature_\mathrm{BNS},\feature_\mathrm{CGWB})$. For simplicity, we only show results for the more realistic case with unknown \ac{SFR}. Once again, we adopt the same priors as in Table~\ref{tab:prior}. 

We find that the constraints on all parameters are orders of magnitudes weaker when considering only \ac{SGWB} data than with the joint inference. In particular, with a \ac{SGWB}-only analysis, the marginalized 1D posterior on the \ac{CGWB} amplitude $\Omega_\mathrm{ref}$ becomes fully consistent with $\Omega_\mathrm{ref}=0$, preventing us from claiming detection of the \ac{CGWB}. This is caused by the fact that the \ac{BNS} foreground completely dominates over the \ac{CGWB}, as expected. The stringent constraints on the \ac{BNS} merger rate from individual detections in the joint inference allow us to pin down the \ac{BNS} foreground and to disentangle it from the \ac{CGWB}, consistently with the discussion in the previous sections.
\section{Sanity check: assumptions on the joint likelihood}

Formally, the joint likelihood can be factorized in the form of Eq.~\eqref{eq:jointlikelihood} if and only if the estimators of individual \ac{BNS} signals and of the cross-correlation spectrum are independent random variables. Although these two analyses process raw data very differently, the cross correlation spectrum $\hat{C}^\star(f)$ also contains information from the individually resolved \ac{BNS} signals, meaning that the two are not independent. Therefore, strictly speaking, Eq.~\eqref{eq:jointlikelihood} can only be considered an approximation to the true joint likelihood for $\{\bm{d}_i\}_{i=1}^{N_\mathrm{obs}}$ and $\hat{C}^\star(f)$. 

To our knowledge, this issue has not been discussed in the literature before, neither in the paper that originally proposed this expression for the joint likelihood~\cite{Callister:2020arv}, nor in subsequent applications by the \ac{LVK} collaboration~\cite{KAGRA:2021kbb}. With current \ac{GW} detectors, the number of detected events $\{\bm{d}_i\}_{i=1}^{N_\mathrm{obs}}$ is so small compared to the total number of mergers that happen in the entire Universe (and thus enter $\hat{C}^\star(f)$), that the approximation is likely justified. As detectors keep improving and the fraction of individually resolved signals increases, it is unclear when Eq.~\eqref{eq:jointlikelihood} will start to break down. In particular, with our population model and choice of \ac{XG} detector network, we find that $\sim 50\%$ of all the \ac{BNS} signals are detected, meaning that the “double counting” of such signals in both $\{\bm{d}_i\}_{i=1}^{N_\mathrm{obs}}$ and $\hat{C}^\star(f)$ can potentially bias the inference. 

To circumvent the issue and test the validity of the assumption in Eq.~\eqref{eq:jointlikelihood}, we propose a straightforward alternative approach that consists of partitioning the data into two independent sets. For example, if data from $20$ days of observation are available, one can use data from the first $10$ days exclusively for hierarchical inference on individual \ac{BNS} detections $\{\bm{d}_i\}_{i=1}^{N_\mathrm{obs}}$, and data from the remaining $10$ days exclusively for the stochastic search to calculate $\hat{C}^\star(f)$. This method is certainly suboptimal, as it does not make use of all the information in the data, but it ensures the joint likelihood can now safely be written in the form of Eq.~\eqref{eq:jointlikelihood} without any approximation. 

\begin{figure}
    \centering
    \includegraphics[width=0.6\linewidth]{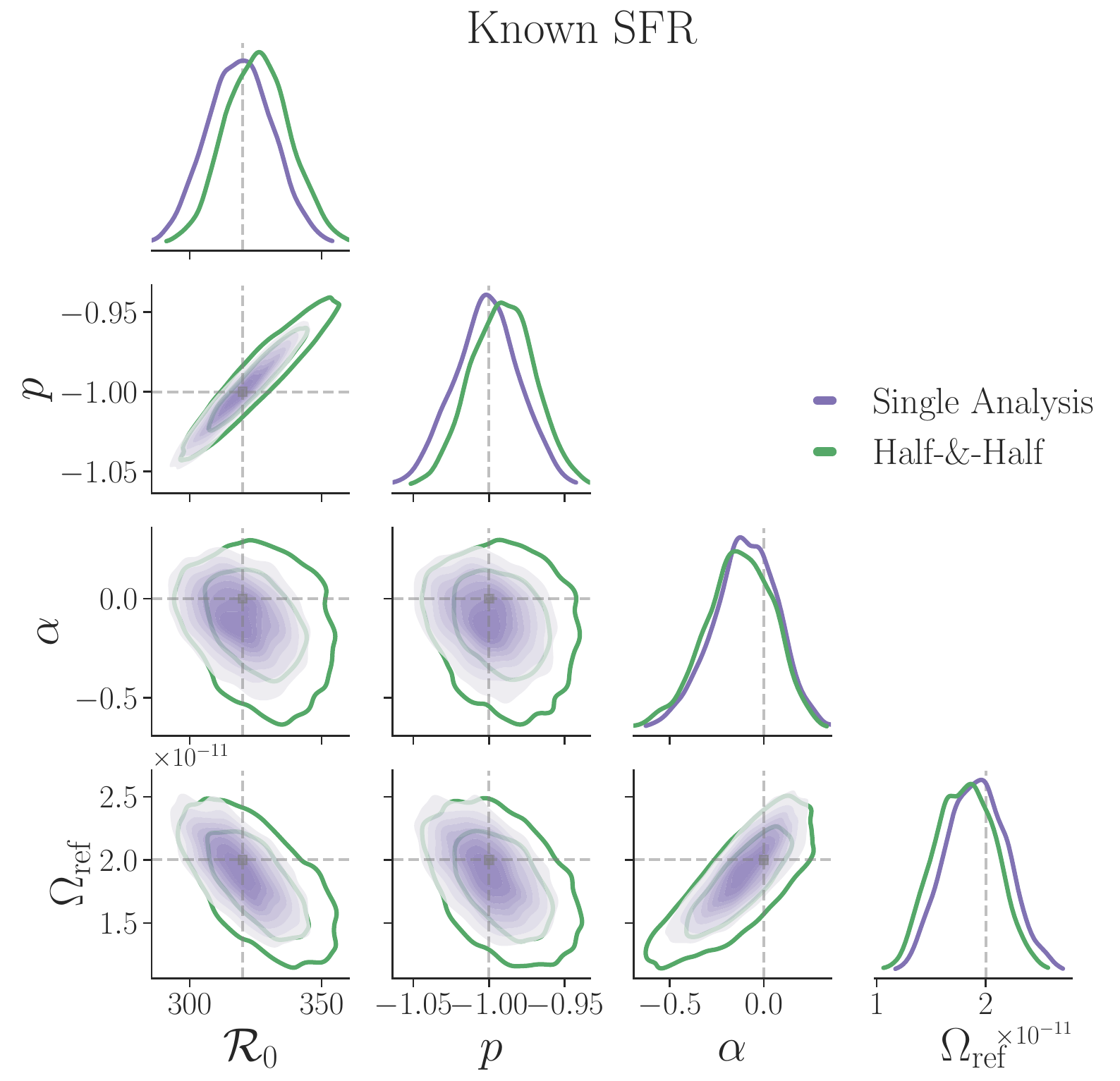}
    \caption{Comparison of posterior distributions from \emph{Single Analysis} (blue curves) and \emph{Half-\&-Half} (green curves) analyses for the known \ac{SFR} scenario. The green contours correspond to $68\%$, and $95\%$ of the probability mass, respectively.}
    \label{fig:Single-HalfnHalf}
\end{figure}

\begin{figure}
    \centering
    \includegraphics[width=0.8\linewidth]{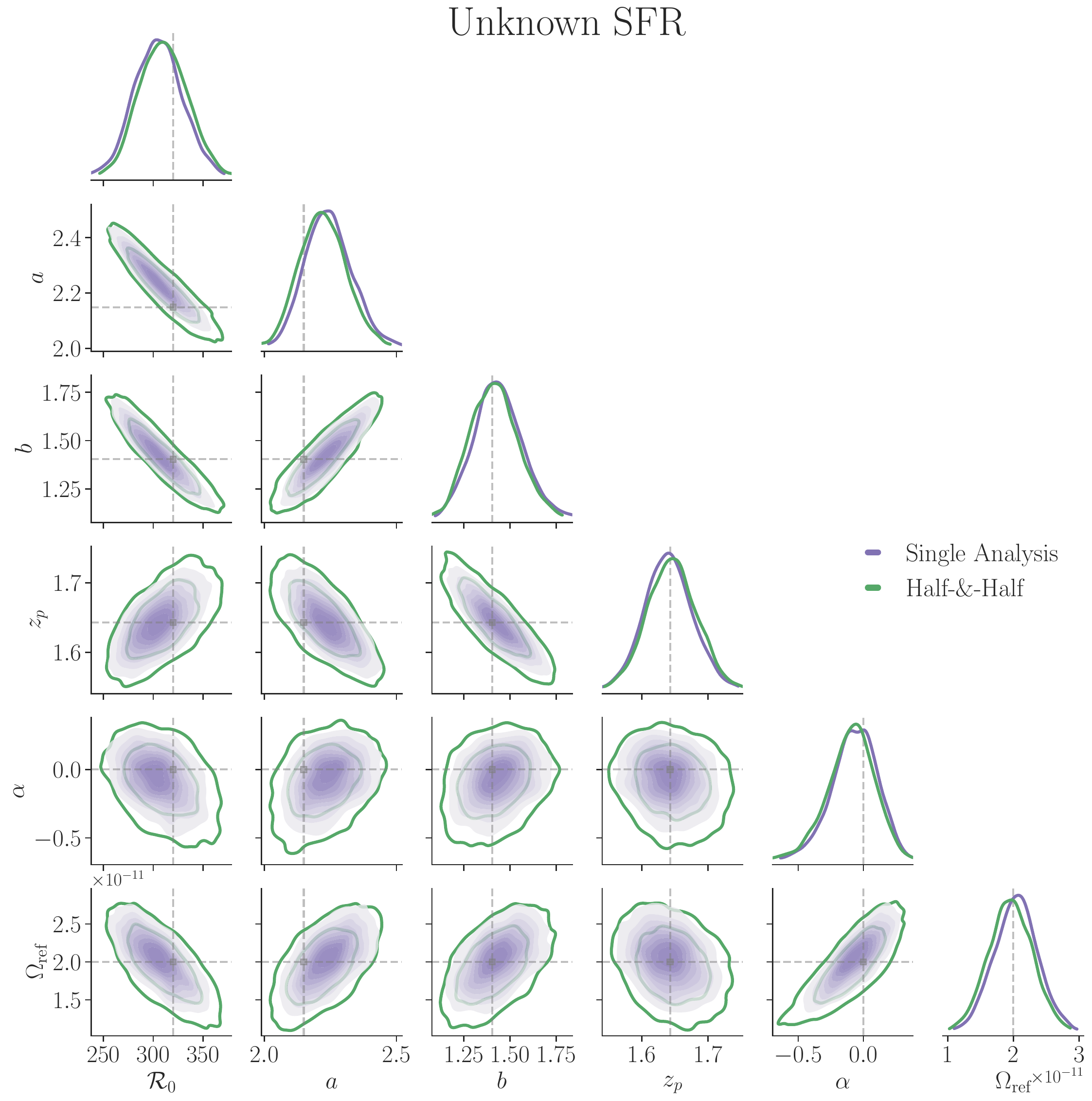}
    \caption{Same as Fig.~\ref{fig:Single-HalfnHalf}, but for the unknown \ac{SFR} scenario.}
    \label{fig:Single-HalfnHalf_unknown}
\end{figure}

In Fig.~\ref{fig:Single-HalfnHalf} and Fig.~\ref{fig:Single-HalfnHalf_unknown}, we compare results obtained by using the same dataset for both the hierarchical inference and stochastic search (i.e., the approach used in the rest of this study and in previous literature~\cite{Callister:2020arv,KAGRA:2021kbb}) with results obtained with the novel approach just discussed, where the dataset is partitioned into two independent chunks. In practice, in the former case (labeled as \emph{Single Analysis} in the figure) we simulate $10$ days of data and use them both to calculate the cross-correlation spectrum $\hat{C}^\star(f)$ and to extract resolved \ac{BNS} signals $\{\bm{d}_i\}_{i=1}^{N_\mathrm{obs}}$. In the latter case (labeled \emph{Half-\&-Half}), we generate two independent realizations of $10$ days of data and use one to compute $\hat{C}^\star(f)$, and one to extract $\{\bm{d}_i\}_{i=1}^{N_\mathrm{obs}}$. We then apply the same factorized joint likelihood of Eq.~\eqref{eq:jointlikelihood} to both cases. Except small variations, the posterior distributions obtained via the two analyses are largely consistent with each other, both for the known and unknown \ac{SFR} scenario. This suggests that using the same dataset for both terms in Eq.~\eqref{eq:jointlikelihood} is a robust approximation for the purposes of our study, and any potential double counting does not significantly bias the inference.

We stress that this conclusion might change for different detector networks or population models, if the relative constraining power of the two likelihood factors in Eq.~\eqref{eq:jointlikelihood} differs significantly from our study. In such a case, the partitioning method proposed in this Section provides a viable and immediate solution. Different partition strategies can also be explored to identify the optimal split of the dataset, i.e., how much observation time to consider for the individual detections $\{\bm{d}_i\}_{i=1}^{N_\mathrm{obs}}$ versus the cross-correlation spectrum $\hat{C}^\star(f)$. Finally, another way to ensure that the joint likelihood can be factorized into two independent contributions, as in Eq.~\eqref{eq:jointlikelihood}, is to further subtract the individually resolved \ac{BNS} signals before computing $\hat{C}^\star(f)$. If the subtraction residuals are negligible, this implies that only unresolved signals and the \ac{CGWB} contribute to $\hat{C}^\star(f)$, ensuring that it is independent from $\{\bm{d}_i\}_{i=1}^{N_\mathrm{obs}}$. In this case, however, the expression for the \ac{SGWB} likelihood of Eq.~\eqref{eq:sgwblikelihood} needs to be modified to account for selection effects, since the unresolved signals no longer represent a fair draw from the \ac{BNS} population. We leave a more detailed exploration of these avenues to future work. 
\bibliography{references_sm}